\documentclass[12pt]{article}

\usepackage{epsf}

\begin{document}

\title{BFKL Pomeron in string models}
\author{G. S. Danilov\thanks{
E-mail address: danilov@thd.pnpi.spb.ru} $\,\,^1$\ and L. N. Lipatov\thanks{
E-mail address: lipatov@thd.pnpi.spb.ru}\thanks{\it Marie Curie Excellence Chair} 
$\,\,^{1,\,2}$ \\
$^1\,$ Petersburg Nuclear Physics Institute,\\
Gatchina, 188300, St.-Petersburg, Russia\\
$^2\,$ II. Institut $\,$ f\"{u}r$\,$ Theoretische Physik,\\
Universit\"{a}t Hamburg, Luruper Chausse 149,\\
22761, Hamburg, Germany }
\maketitle

\begin{abstract}
We consider scattering amplitudes in string models in the Regge limit of
high energies and fixed momentum transfers with the use of the unitarity in
direct channels. Intermediate states are taken in the multi-Regge
kinematics corresponding to the production of resonances with fixed
invariant masses and large relative rapidities. In QCD such kinematics leads
to the BFKL equation for the Pomeron wave function in the leading
logarithmic approximation. We derive a similar equation in the string theory
and discuss its properties. The purpose of this investigation is to find a
generalization of the BFKL approach to the region of  small momentum
transfers where
non-perturbative corrections to the gluon Regge trajectory and reggeon
couplings are essential. The BFKL equation in the string
theory contains additional contributions coming from a linear part of the
Regge trajectory and from the soft Pomeron singularity appearing already
in the tree
approximation. In higher dimensions in addition, a non-multi-Regge
kinematics corresponding to production of particles with large masses is
important. We solve the equation for the Pomeron wave function in the
string theory for $D=4$ and discuss integrability properties of
analogous equations for composite states of
several reggeised gluons in the multi-colour limit.
\end{abstract}

\titlepage

\newpage

\section{Introduction}

The derivation of the BFKL equation for QCD scattering amplitudes in the
Regge regime of high energies $E=\sqrt{s}$ and fixed momentum transfers $q=
\sqrt{-t}$ \cite{lip} is based on the fact that gluon is reggeized in
perturbation theory. In the leading logarithmic approximation (LLA) the
Pomeron singularity in the $j$-plane of the $t$-channel partial waves
appears as a composite state of two reggeized gluons. The gluon Regge
trajectory is known in two first orders of perturbation theory and the
integral kernel for the BFKL equation is calculated in the next-to-leading
approximation \cite{next}, which is important for the phenomenological
applications \cite{gamgam}.

It is reasonable to believe, that the gluon reggeization has a physical
meaning even beyond the QCD perturbation theory, although up to now its
Regge trajectory is calculated only at sufficiently large momentum transfers
$q$ where the effective coupling constant is small. For low momentum
transfers we should use non-perturbative methods. For example, one can
assume that the gluon trajectory in this region is approximately linear, as
it takes place for the hadron trajectories. The linearity of the Regge
trajectories was an important property in constructing the dual model by G.
Veneziano \cite{ven}. Later a string interpretation of the dual amplitudes
was developed \cite{gsw}. In the Born approximation the dual hadron models
include only particles lying on the secondary Regge trajectories. The
Pomeron-like singularity appearing in the open string scattering amplitudes
in one loop approximation was identified with a leading Regge trajectory for
the closed sector. Later it was found that four dimensional string theories
meet with difficulties, which were avoided in their superstring
generalizations to space-time dimensions $D=10$ \cite{gsw}. Now these
superstring models are considered as candidates for an unified theory of all
elementary particle interactions including the gravity. Moreover, all of
them are supposed to be various realizations of the same $M$-theory.

In the modern interpretation the closed string sector is
associated with the graviton Regge family rather than with the
Pomeron singularity\footnote{Note that the Regge asymptotics was
investigated also in the pure (super) gravity \cite{grav}.}, and
so the Pomeron does not directly present in the string theory. In
a line with the Maldacena proposal \cite{mald} for the $N=4 $
super-Yang-Mills model one might expect an appearance of a
colorless composite state becoming a graviton in the t'Hooft limit
$g^{2}N_{c}\rightarrow \infty $, where $g$ is a coupling constant
and $N_{c}$ is the number of colors. The Pomeron singularity seems
to be a candidate for such graviton state \cite{Pol}. However, in
this paper we treat the Pomeron similar to the case of
perturbative QCD where it is a composite state of two reggeized
gluons. Namely, this singularity should appear in the diagrams
where two open strings are exchanged in the $t$ channel. Such
Feynman graphs lead to the Mandelstam cut in the $j$-plane of the
$t$-channel partial wave $ \varphi _j(t)$. The sum of
contributions from the ladder-type diagrams in the string theory
corresponds to the BFKL-like equation. We hope that its string
modification is a reasonable model for non-perturbative effects in
the region of small momentum transfers. Indeed, the string models
in extra dimensions can lead to a dual description of gauge
theories including QCD \cite{mald}.

Note, that in the critical dimensions $D=10$ the multi-Regge kinematics for
intermediate particles in the $s$-channel is not unique even for small
coupling constants. Namely, one should take into account also the production
of resonances having large masses at high energies, which leads in
particular to the graviton contribution appearing in one loop. In the last
case the imaginary part of the corresponding non-planar diagram with the
graviton Regge pole in the $t$-channel appears from the production of two
resonances with masses $m \sim \sqrt{s}$. Below we does not discuss this
problem in details and consider mainly the $D=4$ case.

We use the superstring 4d model in the Ramond-Neveu-Schwarz version \cite
{gsw}, but the supersymmetry is involved only to remove the tachion from the
spectrum. It is known, that non-critical string models have difficulties
related to the absence of the $S$-matrix unitarity in higher loops. In
particular, to restore the unitarity in one loop approximation it is needed
to introduce an additional $2D$ gravity field \cite{gsw}. This field
provides the conformal symmetry on the tree and one-loop levels, and
restores the modular invariance of the one-loop closed string amplitudes
\cite{gtw}. Nevertheless, the modular invariance of higher loop closed
string amplitudes remains to be broken \cite{dan}. Thus, the higher loop
amplitudes for non-critical string models can not be constructed in a
self-consistent way.

The difficulties of non-critical string models are related mainly
to the closed string sector. In the perturbation theory with
respect to the closed string coupling constant $g_{cl}$ the
contributions from this sector grow with energy very rapidly $\sim
s^{2}g_{c}^{2n}\ln ^{n}(s)$. Such behavior in the case of
hadron-hadron interactions is not compatible with the $s$ -channel
unitarity. So one expects that once a relevant summation over $n$
being performed, the high asymptotics of the amplitude is reduced
to $A \sim s$. It is reasonable to omit initially the closed
string sector taking into account also, that $g_{cl}$ is quadratic
in the Yang-Mills coupling constant $g$ and for $N_{c}\rightarrow
\infty $ the open string terms in the amplitude are enhanced
comparing to closed string ones. Thus, we consider here the
contributions to the production amplitude only from the open
string states in crossing channels leading to the Mandelstam cuts
for the elastic $t $ -channel partial wave $\varphi _{j}(t)$ in
the angular momentum plane $ j=1+\omega $.

In the discussed model the gluon trajectory $\omega (t)$ is given by the
perturbative expansion
\begin{equation}
\omega (t)=\alpha ^{\prime }t+\omega _{1}(t)+...\,.  \label{omegas}
\end{equation}
where $\omega _{n}(t)\sim g^{2n}$ are radiative corrections, and the Regge
slope $\alpha ^{\prime }$ is a reversed square of an characteristic mass
scale. Below the correction $\omega _1(t)\sim g^2$ to the trajectory is also
taken into account. This correction is calculated from one-loop diagrams for
the scattering amplitude. One-loop non-planar diagrams contain also a
contribution destroying the unitarity, but the correction to the Regge
trajectory appears only from the planar graphs, where the problem with the
closed string sector does not exist. Providing that $-\alpha^{\prime}t
\ll 1$, the loop correction in the $D=4$ case has the infrared
divergency $\sim g^2N_c\ln(t/\lambda ^2)$ which is cancelled with the
contribution from the massless particle production. For
$-\alpha^{\prime}t\geq1$ the radiative correction to the Regge
trajectory has a complicated form.

An important difference between QCD and the string model is related to the
role of intermediate states with relatively large masses: $(\alpha ^{\prime
})^{-1}\ll M^{2}\ll s$ for produced resonances. These states are absent in
QCD. In the string theory the large mass states are responsible for the
appearance of the graviton contribution to the elastic scattering amplitude
in the one-loop approximation. Further, the impact factors for the
reggeon-particle scattering vanish for planar diagrams as a result of
integration over large masses. In particular, it leads to the absence of the
Mandelstam cuts in the color octet channel. In QCD the cancellation of these
cuts for the $t$-channel with gluon quantum numbers is provided by another
mechanism related to the so-called bootstrap relations for scattering
amplitudes \cite{klf}.

The large mass kinematics is responsible also for the additional term in the
kernel of the BFKL equation corresponding to the soft pomeron contribution.
It is important, that in the considered string model even in the tree
approximation there is a colorless state in the $t$-channel with vacuum
quantum numbers and a positive signature. In upper orders of the
perturbation theory its Regge trajectory is renormalized. At small $t$ this
state mixes with the Mandelstam cut constructed from two reggeized gluons.
The radiative corrections to its trajectory are calculated from ladder
diagrams in the $t$-channel. The $s$-channel imaginary part of scattering
amplitudes appears from the intermediate states in the above considered
kinematics with relatively large masses: $(\alpha ^{\prime })^{-1}\ll
M^{2}\ll s$ for produced resonances. Physically the $j$-plane singularity
with the vacuum quantum numbers in the tree approximation corresponds to the
soft Pomeron which can exist together with the hard BFKL Pomeron.

Similar to the perturbative QCD, we restrict ourselves to the
region $g^{2}N_{c}\ln (s/M^{2})\sim 1$. However, in the string
case, the region $ |\omega (t)|\ln (s/M^{2})\gg 1$ is possible also
because the Regge slope $ \alpha ^{\prime }=1/M^{2}$ has no
$g^{2}$ smallness. Some important properties of the BFKL equation
are related to this fact. In particular, we obtain that for
$-\alpha ^{\prime }t\gg g^{2}N_{c}$ its solution is concentrated
near the saddle point $q_{\perp }/2$ for the reggeon transverse
momenta $k_{\perp }$. For $D>4$ the fluctuations of this momentum
are small $ \Delta (k_{\perp }-q_{\perp }/2)^{2}\sim 1/(\alpha
^{\prime }\ln \alpha ^{\prime }s)$ and therefore the transverse
momenta of the emitted gluons are also small $|k_{\perp
}^{g}|^{2}\leq \alpha ^{\prime }(\ln \alpha ^{\prime }s)^{-1}$. In
the same time there are no similar restrictions on transverse
momenta of the virtual gluons entering in the loop corrections to
the gluon Regge trajectories. It means, that the contribution from
the multiple saddle points $k_{i\perp}\approx q/2$ is suppressed
by the reggeization effects.

The paper is organized as follows. In Sec. 2 the BFKL approach to the
perturbative QCD is briefly reviewed. In Sec. 3 the superstring model which
will be used later is introduced. In Sec. 4 the calculation of the
multi-Regge asymptotics of production amplitudes is presented. In Sec. 5 the
BFKL-like equation for the superstring model is derived. Also the vanishing
of the impact factors for planar diagrams is demonstrated. In more details
this problem is considered in Appendix D. In Sec. 6 the calculation of the
BFKL kernel is performed. In Sec. 7 the equation for the case $D=4$ is
discussed. Among other things, it is explained why in the space-time $D=10$
the non-Regge kinematics contributes to the Regge asymptotics of 
amplitudes. In Sec. 8 the solution of the BFKL equation at small values
of $\alpha 't$ is constructed. In Sec. 9 an algebraic approach to this problem
is developed and integrability properties of similar equations for composite 
states of several open strings in the multi-colour limit including a relation
with the Heisenberg spin model are discussed.
Appendices A, B and C contain some details of calculations.

\section{BFKL approach in the perturbation QCD}

As it was mentioned already, in the perturbative QCD the BFKL Pomeron
appears as a composite state of two reggeized gluons \cite{lip}. The gluon
is reggeized as a result of summing radiative corrections to the Born
amplitude $A_{Born}$ for the colored particle scattering $AB\rightarrow
A^{\prime }B^{\prime }$ in the Regge kinematics of large energies $\sqrt{s}$
and fixed momentum transfers $q=\sqrt{-t}$
\begin{equation}
A(s,t)=A_{Born}\,s^{\omega (t)}\,,  \label{reggeam}
\end{equation}
where $A_{Born}$ is given below
\begin{equation}
A_{Born}=2\,s\,g\,T_{A^{\prime }A}^{c}\,\delta _{\lambda _{A^{\prime
}}\lambda _{A}}\frac{1}{t}\,g\,T_{B^{\prime }B}^{c}\,\delta _{\lambda
_{B^{\prime }}\lambda _{B}}\,\,,\,\,\,[T^{c},T^{c^{\prime
}}]=i\,f_{cc^{\prime }d}\,T^{d}  \label{born}
\end{equation}
and $j=1+\omega (t)$ is the gluon Regge trajectory known in two first orders
of the perturbation theory
\begin{equation}
\omega (t)=\omega _{1}(t)+\omega _{2}(t)+...\,.  \label{omega}
\end{equation}

The trajectory contains logarithmic divergencies cancelled in the total
cross sections with the contributions from the production of soft gluons.
For example, in one loop approximation we have
\begin{equation}
\omega _{1}(-q^{2})=-\frac{g^{2}}{16\,\pi ^{3}}\,N_{c}\,\int d^{2}k\,\frac{
q^{2}+\lambda ^{2}}{(k^{2}+\lambda ^{2})((q-k)^{2}+\lambda ^{2})}\approx -
\frac{g^{2}}{8\,\pi ^{2}}\,N_{c}\,\ln \frac{q^{2}}{\lambda ^{2}}\,,
\label{traj}
\end{equation}
where $\lambda $ is a gluon mass introduced for the regularization of the
infraredly divergent integral. On the other hand, the amplitude for the
production of $n$ gluons with momenta $k_{r}$ in the multi-Regge kinematics
\begin{equation}
s\gg s_{r}=(k_{r-1}+k_{r})^{2}\gg q_{r}^{2}\,,
\end{equation}
has the factorized form
\begin{equation}
A=2\,s\,gT_{A^{\prime }A}^{c_{1}}\,\delta _{\lambda _{A^{\prime }}\lambda
_{A}}\frac{s_{1}^{\omega (t_{1})}}{t_{1}} g
\,T_{c_{2}c_{1}}^{d_{1}}C(q_{2},q_{1})\frac{s_{2}^{\omega (t_{2})}}{t_{2}}
\,g\,T_{c_{3}c_{2}}^{d_{2}}C(q_{3},q_{2})...gT_{B^{\prime }B}^{c}\,\delta
_{\lambda _{B^{\prime }}\lambda _{B}}\,\,,  \label{multreg}
\end{equation}
where the effective vertex $C(q_{2},q_{1})$ for an emission of the gluon
with a definite helicity is
\begin{equation}
C(q_{2},q_{1})=\frac{q_{1}\,q_{2}^{\ast }}{k_{1}}\,,\,\,k_{1}=q_{1}-q_{2}\,.
\,  \label{vertex}
\end{equation}

Here we introduced the complex coordinates
\begin{equation}
q_{r}=q_{r}^{x}+i\,q_{r}^{y}\,,\,\, k_{r}=k_{r}^{x}+i\,k_{r}^{y}
\end{equation}
for transverse components $q_{r}^{\perp },\,k_r^{\perp}$ of gluon momenta.
The contribution to the elastic scattering amplitude from the intermediate
state having a gluon with the momentum $k_{1}$ is proportional to the
expression
\begin{equation}
C(q_{2},q_{1})\,C^{\ast }(q_{2}^{\prime },\,q_{1}^{\prime })+C^{\ast
}(q_{2},q_{1})\,C(q_{2}^{\prime },\,q_{1}^{\prime })\,  \label{glucont}
\end{equation}
and contains the pole $1/|k_{1}|^{2}$. The integration over $k_1$ cancels
the infrared divergency in the gluon Regge trajectory appearing in the
virtual corrections to the production amplitudes.

It is convenient to present the elastic amplitude for the colorless particle
scattering in the form of the Mellin representation
\begin{equation}
A(s,t)=i\,s\int_{a-i\infty }^{a+i\infty }\frac{d\,\omega }{2\,\pi \,i}
\,s^{\omega }\,f_{\omega }(t)\,,  \label{mellin}
\end{equation}
where $f_{\omega }(t)$ is the $t$-channel partial wave
analytically continued to the complex values $j=1+\omega $ of the
angular momentum. The amplitude $A(s,t)$ contains only the
contribution from the $t$-channel state with vacuum quantum
numbers and the positive signature, corresponding to the BFKL
Pomeron. A positive value of the parameter $a$ in the above
representation is chosen from the condition, that all
singularities of $f_{\omega }(t)$ are situated to the left from
the integration contour.

The $t$-channel partial wave $f_{\omega }(t)$ can be expressed in terms of
the gluon-gluon scattering amplitude $f_{\omega }(q_{1},q_{2};q)$ integrated
with the impact-factors $\Phi (q_{i},q-q_{i})$
\begin{equation}
f_{\omega }(-q^{2})=\int \frac{d^{2}q_{1}}{(2\pi )^{2}}\frac{\Phi
(q_{1},q-q_{1})}{q_{1}^{2}\,(q-q_{1})^{2}}\int \frac{d^{2}q_{2}}{(2\pi )^{2}}
\,\frac{\Phi (q_{2},q-q_{2})}{q_{2}^{2}\,(q-q_{2})^{2}}\,f_{\omega
}(q_{1},q_{2};q)\,.  \label{partw}
\end{equation}
The impact-factors of colorless particles vanish at small gluon momenta

\begin{equation}
\Phi (0,q)=\Phi (q,0)=0\,,
\end{equation}
which leads to an infrared \ stability of $f_{\omega }(-q^{2})$. The partial
wave $f_{\omega }(q_{1},q_{2};q)$ satisfies the BFKL equation \cite{lip}
\begin{equation}
\omega \,\,f_{\omega }(q_{1},q_{2};q)\,=\omega \,\,f_{\omega
}^{0}(q_{1},q_{2};q)\,-\frac{g^{2}\,N_{c}}{8\pi ^{2}}\,H\,f_{\omega
}(q_{1},q_{2};q)\,.  \label{partsol}
\end{equation}
Here $f_{\omega}^{0}$ is a non-homogeneous term corresponding to the impact
factor. The hamiltonian $H$ is an integral operator, which can be defined by
its action on the Pomeron wave function $f(\vec{\rho}_{1},\vec{\rho}
_{1^{\prime }})$ in the coordinate representation \cite{integr}
\begin{equation}
H\,=\ln |\partial _{1}|^{2}+\ln |\partial _{2}|^{2}\,\,+\,\frac{1}{\partial
_{1}\partial _{2}^{\ast }}\ln \,|\rho _{12}|^{2}\,\,\partial _{1}\partial
_{2}^{\ast }\,+\frac{1}{\partial _{1}^{\ast }\partial _{2}}\ln \,|\rho
_{12}|^{2}\,\,\partial _{1}^{\ast }\partial _{2}-4\Psi (1)\,,  \label{hamil}
\end{equation}
where $\Psi (x)=(\ln \Gamma (x))^{\prime }$ and we introduced the complex
coordinates and momenta
\begin{equation}
\rho _{r}=x_{r}+iy_{r}\,,\,\,\partial _{r}=\frac{\,\partial }{\,\partial
\rho _{r}\,}\,,\,\,\,\rho _{12}=\rho _{1}-\rho _{2}\,.
\end{equation}

The hamiltonian has the property of the M\"{o}bius invariance, which
allows us to find its eigenfunctions \cite{conf}
\begin{equation}
E_{m,\widetilde{m}} (\vec{\rho}_1,\vec{\rho}_2; \vec{\rho}_0)=
\left(\frac{\rho _{12}}{\rho _{10}\,\rho _{20}}\right)^m
\left(\frac{\rho _{12}^*}{\rho _{10}^*\,\rho
_{20}^*}\right)^{\widetilde{m}}\,,
\label{polyasol}
\end{equation}
where
\begin{equation}
m=\frac{1}{2}+i\nu +\frac{n}{2}\,,\,\,
\widetilde{m}=\frac{1}{2}+i\nu -\frac{n}{2}
\end{equation}
are conformal weights.

The high energy asymptotics of the total cross-section is parametrized by
the Pomeron intercept $\Delta $
\begin{equation}
\sigma _{t}\sim s^{\Delta }\,  \label{totcrs}
\end{equation}
In the leading logarithmic approximation we have
\begin{equation}
\Delta =-\frac{g^{2}\,N_{c}}{8\pi ^{2}}\,E,  \label{delta}
\end{equation}
where $E=-8\ln 2$ is the ground state energy of the Hamiltonian
$H$. Therefore the cross-section $\sigma _t$ violates the
Froissart theorem $ \sigma _{t}<c\ln ^{2}(s)$. In the
next-to-leading approximation the cross-section grows also, but
not so rapidly (see \cite{gamgam}).

To verify the gluon reggeization one can use the $s$ and
$u$-channel unitarity constraints and dispersion relations to
calculate by iterations the scattering amplitude with the color
octet quantum numbers in the $t$-channel \cite{lip}. In LLA it is
enough to consider only the multi-Regge kinematics for 
intermediate particles in the direct channels. In this kinematics
the production amplitude has the multi-Regge form (\ref{multreg}
). The reggeization hypothesis should be in an agreement with the
$s$- and $ u $- channel unitarity. This requirement leads to the
so-called bootstrap relations. The simplest bootstrap relation
corresponds to the statement, that the scattering amplitude,
obtained from the solution of the Bethe-Salpeter equation for the
wave function of the composite state of two reggeized gluons in
the octet channel should coincide with the Regge pole anzatz for
the amplitude constructed in terms of the 
reggeized gluon exchange. In the
momentum space the equation for the $t$-chanel partial wave
$f_{\omega }^{G}( \overrightarrow{k},
\overrightarrow{q}-\overrightarrow{k})$ with the gluon quantum
numbers has the form \cite{lip}
\[
\omega \,f_{\omega }^{G}(\overrightarrow{k},\overrightarrow{q}-
\overrightarrow{k})=\frac{1}{\overrightarrow{q}^{2}+\lambda
^{2}}-\frac{ g^{2} }{8\pi ^{2}}\,N_{c}\,\int \frac{d^{2}k^{\prime
}}{2\pi }\frac{ \overrightarrow{q}^{2}+\lambda
^{2}}{\overrightarrow{k^{\prime }} ^{2}+\lambda
^{2}}\frac{f_{\omega }^{G}(\overrightarrow{k^{\prime }},
\overrightarrow{q}-\overrightarrow{k^{\prime
}})}{(\overrightarrow{q}- \overrightarrow{k^{\prime
}})^{2}+\lambda ^{2}}+
\]
\begin{equation}
\frac{g^{2}}{8\pi ^{2}}\,N_{c}\,\int \frac{d^{2}k^{\prime }}{2\pi }\left(
\frac{\overrightarrow{k}^{2}+\lambda ^{2}}{\overrightarrow{k^{\prime }}
^{2}+\lambda ^{2}}+\frac{(\overrightarrow{q}-\overrightarrow{k})^{2}+\lambda
^{2}}{(\overrightarrow{q}-\overrightarrow{k^{\prime }})^{2}+\lambda ^{2}}
\right) \,\frac{f_{\omega }^{G}(\overrightarrow{k^{\prime }},\overrightarrow{
q}-\overrightarrow{k^{\prime }})-f_{\omega }^{G}(\overrightarrow{k},
\overrightarrow{q}-\overrightarrow{k})}{(\overrightarrow{k}-\overrightarrow{
k^{\prime }})^{2}+\lambda ^{2}}\,,  \label{gluch}
\end{equation}
where the gluon mass $\lambda $ is introduced with the use of the Higgs
mechanism to regularize the infrared divergencies.

It is obvious, that in an accordance with the bootstrap requirement the
solution of the above equation corresponds to the Regge pole anzatz
\begin{equation}
f_{\omega }^{G}(\overrightarrow{k},\overrightarrow{q}-\overrightarrow{k})=
\frac{1}{\overrightarrow{q}^{2}+\lambda ^{2}}\,\frac{1}{\omega -\omega (-
\overrightarrow{q}^{2})}\,,  \label{glpolc}
\end{equation}
where $\omega (-\overrightarrow{q}^{2})$ is the gluon Regge trajectory.

\section{String model}

In the string and superstring models the scattering amplitude in
the tree approximation satisfies the duality requirement: namely,
the sum over the resonances in the $t$-channel related to its
Regge asymptotics in the $s$-channel is equal to the (analytically
continued) sum of resonances in the $ s $ and $u$-channels:
\begin{equation}
A(s,t,u)=A(s,t)+A(u,t)+A(s,u)\,,\,\,A(s,t)=\sum_{i}\frac{c_{i}(s)}{t-t_{i}}
=\sum_{i}\frac{c_{i}(t)}{s-s_{i}}\,.  \label{stram}
\end{equation}
The particles with squared masses equal to $t_{i}$ and integer
spins $ j=j_{i} $ lie on the linear Regge trajectories
\begin{equation}
j=j_{0}+\alpha ^{\prime }t \,,  \label{trajec}
\end{equation}
where $j_{0}$ and $\alpha ^{\prime }$ are their intercept and
slope, respectively. The slope $\alpha ^{\prime }$ \ is universal
for all excitations of the open string. For the closed strings it
is equal to $ \alpha ^{\prime }/2$. As for intercepts, in the
critical dimensions $D=26$ for the bosonic string and $D=10$ for
the superstrings, they are integer or half-integer numbers. In
particular, for the intercepts of the leading bosonic Regge
trajectories, corresponding to the massless vector ($V$) particle
- ''gluon'' and tensor ($T$) particle - ''graviton'' we have
respectively
\begin{equation}
j_{0}^{V}=1\,,\,\,j_{0}^{T}=2\,.
\end{equation}
We put $j_{0}^{V}=1$ also for the $D=4$ model to leave the gluon on the
trajectory. The ''graviton'' is absent in this case, instead one has a
non-physical cut in the $j$ -plane.

The Regge asymptotics of $A(s,t)$ in the dual models appears as a result of
summing over the poles in the $s$-channel. Really at large $s$ the
contributions $\sim s^{-k}$ with integer values of $k$ are cancelled and we
can substitute approximately the sum over $i$ by the dispersion integral
\[
A(s,t) \approx \frac{1}{\pi} \int _0 ^{\infty}\,\frac{ds^{\prime}}{
s-s^{\prime}}\,\Im A(s^{\prime},t)\,,\,\,s^{\prime}=s(i)\,,\,\, \Im
A(s^{\prime},t)=\pi c_i(t)\,.
\]
It agrees with the Regge asymptotics $A(s,t )\sim (-s)^{j(t)}$ providing
that $\Im A(s,t) \sim s^{j(t)}$.

In the Born approximation there are only stable particles in the
intermediate state, but with taking into account loop corrections
these particles acquire the widths due to their decay into lower
mass states. As a result, $\Im A(s,t)$ has the $\delta $-like
singularities only for a finite number of stable states and the
amplitude is a smooth function for large values of $s$. The
function $A(s,t)\sim s^{1+\alpha ^{\prime }t}$ can be expanded in
the series over the parameter $\alpha ^{\prime }t$ and one can
interpret the corresponding term of the expansion $\sim s\,(\ln
s)^{n}(\alpha ^{\prime }t)^{n}/n!$ as a contribution from the
production of $n$ particles in a multi-Regge kinematics. In QCD
such a non-perturbative contribution $ \omega (t)\sim \alpha
^{\prime }t$ to the Regge trajectory could appear from the
integration region $k^{2}\sim \Lambda _{QCD}^{2}$ in the loop
corrections of the type of (\ref{traj}). In this case the common
factor $t$ would lead to the linearity of the 
trajectory at small $t$.

To begin with, let us consider the Born amplitude for the tachyon-tachyon
scattering amplitude in the bosonic string theory
\begin{equation}
A(s,t)=g^{2}\,\frac{\Gamma (-\alpha (s))\,\Gamma (-\alpha (t))}{\Gamma
(-\alpha (t)-\alpha (s))}\,,\,\,\alpha (t)=1+\alpha ^{\prime }t\,,
\label{tach}
\end{equation}
where for simplicity we omitted the Chan-Paton factors. Asymptotically one
obtains
\begin{equation}
\lim_{s\rightarrow \infty }A(s,t)=-g^{2}\,\alpha ^{\prime }s\Gamma (-\alpha
(t))\,(-\alpha ^{\prime }s)^{\alpha ^{\prime }t}\,.  \label{tacas}
\end{equation}

This result corresponds to the Regge asymptotics described by the reggeized
gluon exchange in the $t$-channel. For other colliding particles there are
additional factors depending on their spins. They are related to different
residues for the corresponding Regge pole. Note, that the effective vertices
for reggeized gluon interactions in QCD were obtained also from the string
amplitudes in the limit $\alpha ^{\prime}\rightarrow 0$ \cite{BFstring}.

For the superstring models the multiplier $\Gamma (-\alpha (t))$ in (\ref
{tach}) is replaced by $\Gamma (-\alpha ^{\prime }t)$, which leads to the
absence of the tachyon pole at $\alpha ^{\prime }t=-1$. At small momentum
transfers both models give the same amplitude for the massless vector boson
scattering. Taking, however, into account that one should sum over other
intermediate $t$-channel states for the scattering amplitude with arbitrary
momentum transfers, it is natural to consider only the superstring model
where the tachyon disappears from the spectrum. The Regge limit of the
superstring scattering amplitude is given in the end of this section (see
\cite{gsw}).

As it was said in Introduction, we use the Ramond-Neveu-Schwarz
version of the open superstring model. In this model the
interaction vertices are calculated in terms of the scalar
superfield $X^{M}(z,\vartheta )$ where $z$ is a world-sheet
coordinate and $\vartheta $ is its superpartner. Here $M$ labels
the space-time coordinates, $M=0,\,1,\,\dots ,\,(D-1)$. The vertex
$V(z,\vartheta ;k,\xi )$ for the emission of a massless vector
boson with its momentum $k=\{k^{M}\}$ and polarization vector $\xi
=\{\xi ^{M}\}$ is given below \cite{gsw,fried}
\begin{equation}
V(z,\vartheta ;k,\xi )=\xi \,D\,X\,e^{-ikX}\,  \label{vert}
\end{equation}
where $kX\equiv k_{M}X^{M}(z,\vartheta )$ and $\xi \,D\,X\equiv
\xi _{M}\,D(z,\vartheta )\,X^{M}(z,\vartheta )$ are scalar
products of the corresponding $D$-dimensional vectors. As usually,
the relation $k\xi =0$ is valid for polarizations of external
vector bosons. In the contrast to the string tradition, in this
paper we use the ''mostly minus metrics'' $ ab=a_{0}b_{0}-
\overrightarrow{a}\overrightarrow{b}$. The covariant
super-derivative $D(z,\vartheta )$ appearing in (\ref{vert}) is given below
\begin{equation}
D(z,\vartheta )=\partial _{z}+\vartheta \, \partial_{\vartheta }\,,
\label{sder}
\end{equation}
where $\partial_{\vartheta }$ is the ''left'' derivative in $\vartheta $.
Note, that the gauge invariance $\xi \rightarrow \xi +ck $ of the amplitudes
is valid due to the relation
\begin{equation}
\int dz d\vartheta De^{-ikX}=0\,.
\end{equation}

The superfield vacuum correlator $\biggl<X^{M}(z,\vartheta )X^{N}(z^{\prime
},\vartheta ^{\prime })\biggl>$ in super-coordinates for $z>z^{\prime }$
equals
\begin{equation}
\biggl<X^{M}(z,\vartheta )X^{N}(z^{\prime },\vartheta ^{\prime
})\biggl> =2\alpha ^{\prime }\eta ^{MN}\ln (z-z^{\prime
}-\vartheta \vartheta ^{\prime })=2\alpha ^{\prime }\eta
^{MN}\biggl[\ln (z-z^{\prime })-\frac{\vartheta \vartheta ^{\prime
}}{z-z^{\prime }}\biggl]\,,  \label{scorr}
\end{equation}
where $\eta ^{MN}$ is the space-time metrics. The massless boson tree
amplitude is obtained by integrating the vacuum expectation of the product
of the vertices $V_{j}$ (\ref{vert}) over $(z_{j},\vartheta _{j})$. The
variables $(z_{j},\vartheta _{j})$ are assigned to the vertex for an
emission of the boson carrying the momentum $k_{j}$ and polarization $\xi
_{j}$. In the amplitude we do not integrate over three of coordinates $z_{j}$
using the integrand invariance under $SL(2,R)$-transformation. To conserve
this symmetry after fixing the variables $(z^{(1)},\,z^{(2)},\,z^{(3)})$ one
should include in the final expression the additional multiplier
\begin{equation}
r(z^{(1)},z^{(2)},z^{(3)})=
(z^{(1)}-z^{(2)})(z^{(1)}-z^{(3)})(z^{(2)}-z^{(3)}) \,,
\end{equation}
leading to an independence of the Born amplitude from the choice of these
variables.

Thus, the open string amplitude $A_{n}(\{k_{j},\xi _{j}\})$ for the
interaction of $n$ massless bosons in a tree approximation is given by
\begin{equation}
A_{n}(\{k_{j},\xi _{j}\})=\sum_{(r)}T_{(r)}A_{n}^{(r)}(\{k_{j},\xi _{j}\})
\,,  \label{opsam}
\end{equation}
where each a term corresponds to an ordering of the parameters
$z_{j}$ : $ \{(r):z_{j_{1}}>z_{j_{2}}>\dots >z_{j_{n}}\}$ and the
sum is taken over the configurations, which are non-equivalent
under the cyclic transmutations of indices $j_{r} $. The
coefficient $T_{(r)}$ is the Chan-Paton factor \cite {gsw} for the
given color group. Further, the expression $ A^{(r)}(\{k_{j},\xi
_{j}\})$ is the integral over $(z_{j},\vartheta _{j})$ from the
vacuum expectation of the product of interaction vertices
multiplied by the factor $r(z_{j_{1}},z_{j_{2}},z_{j_{n}})$:
\begin{eqnarray}
A_n^{(r)}(\{k_j,\xi_j\})=g^{n-2}
(z_{j_1}-z_{j_2})(z_{j_1}-z_{j_n})(z_{j_2}-z_{j_n})
\int\theta(z_{j_2}-z_{j_3})
\prod_{s=3}^{n-1}\theta(z_{j_s}-z_{j_{s+1}})dz_{j_s}  \nonumber \\
\times \biggl<d\vartheta_{j_1}V(z_{j_1},
\vartheta_{j_1};k_{j_1},\xi_{j_1})\dots d\vartheta_{j_n}V(z_{j_n},
\vartheta_{j_n};k_{j_n},\xi_{j_n})\biggl>\,,  \label{opsamt}
\end{eqnarray}
where $\theta (x)$ is the step function: $\theta (x)=1$ for $x>0$
and $ \theta (x)=0$ for $x<0$.

Since the correlator (\ref{scorr}) is singular at $z=z^{\prime }$, the
integral (\ref{opsamt}) is convergent only in a certain region of invariants
constructed from external particle momenta. Each of the terms $A_{n}^{(r)}$
in (\ref{opsam}) is calculated for such signs of the invariants where it is
convergent, and the result is analytically continued to their physical
values for the production kinematics. The integrand in (\ref{opsamt})
contains some contributions which do not contribute to the final result
because they are total derivatives in integration variables. One can
make their cancelation explicit using the fact, that the integrand in the
superstring case is invariant under the $SL(2,R)$-SUSY
transformation \cite{fried,volk}:
\begin{equation}
z=f(\tilde{z})\,,\,\,\, \vartheta = \sqrt{\frac{\partial f(\hat{z})}{
\partial \hat{z}}} \left(\hat{\vartheta} +\varepsilon (\hat{z})\right)
\left(1 -\frac{\beta \delta}{2} \right)\,,\,\,\, \tilde{z}=\hat{z}+\hat{
\vartheta}\varepsilon (\hat{z})\,,  \label{sfltr}
\end{equation}
where $\varepsilon (z)$ and $f(z)$ are given below
\begin{equation}
\varepsilon (z)=\beta z+\delta \,,\qquad
f(z)=\frac{az+b}{cz+d}\,\,, \,\,\,\,\, \sqrt{\frac{\partial
f(\hat{z})}{\partial \hat{z}}}= \frac{1}{cz+d }\,\,.  \label{spar}
\end{equation}
Here $a$, $b$, $c$, $d$ are bosonic parameters and $\beta $, $\delta $ are
their Grassmann partners. Note, that the superinterval is transformed in a
simpler way
\begin{equation}
z-z^{\prime}-\vartheta \vartheta ^{\prime}=Q^{-1}(\hat
z,\hat\vartheta) Q^{-1}(\hat z^{\prime},\hat\vartheta^{\prime})
\biggl(\hat{z}- \hat{ z^{\prime} }-\hat{\vartheta} \hat{\vartheta
^{\prime}}\biggl)\,, \label{superint}
\end{equation}
where
\begin{equation}
Q^{-1}(\hat z,\hat\vartheta)=D(\hat z,\hat\vartheta)\vartheta\,.
\label{supfa}
\end{equation}
Also, one can verify that
\begin{equation}
D(z,\vartheta)=Q(\hat z,\hat\vartheta)D(\hat z,\hat\vartheta).
\label{trsder}
\end{equation}

An appropriate transformation (\ref{sfltr}) of the integration
variables in (\ref{opsamt}) allows us to extract an explicit
dependence from two $ \vartheta _{j}$, which gives a possibility
to perform the integration over these variables. This symmetry is
non-splitted because it mixes the Grassmann variables to bosonic
ones. Note, that the step function factors in (\ref{opsamt}) lead
after the symmetry transformation to the $\delta $ -function type
terms which are multiplied by expressions vanishing in the
kinematical region where the integral is convergent. To avoid the
consideration of such terms, one can explicitly fix 5 variables
$(3|2)$ among all coordinates $(z_{j}|\vartheta _{j})$ using the
super-$SL(2,R)$ invariance. After that the integrand is multiplied
by a supersymmetric generalization of the above factor
$r(z_{j_{1}},z_{j_{2}},z_{j_{n}})$ \cite {danphl} (for details see
Appendix A). It is convenient to put $\vartheta _{j_{1}}=\vartheta
_{j_{2}}=0$. In this case the generalized factor $r$ is
$(z_{j_{1}}-z_{j_{n}})(z_{j_{2}}-z_{j_{n}})$. Thus, expression
(\ref{opsamt}) is replaced by
\begin{eqnarray}
A_{n}^{(r)}(\{k_{j},\xi _{j}\})
=g^{n-2}(z_{j_{1}}-z_{j_{n}})(z_{j_{2}}-z_{j_{n}})\int \theta
(z_{j_{2}}-z_{j_{3}})\prod_{s=3}^{n-1}\theta
(z_{j_{s}}-z_{j_{s+1}})dz_{j_{s}}  \nonumber \\ \times
\biggl<V(z_{j_{1}},0;k_{j_{1}},\xi
_{j_{1}})V(z_{j_{2}},0;k_{j_{2}},\xi _{j_{2}})d\vartheta
_{j_{3}}V(z_{j_{3}},\vartheta _{j_{3}};k_{j_{3}},\xi
_{j_{3}})\dots d\vartheta _{j_{n}}V(z_{j_{n}},\vartheta
_{j_{n}};k_{j_{n}},\xi _{j_{n}}) \biggl>\,.  \label{copsamt}
\end{eqnarray}

Using relation (\ref{scorr}) for the vacuum expectation of the product of
vertices (\ref{vert}), one finds finally
\begin{eqnarray}
A_{n}^{(r)}(\{k_{j},\xi _{j}\})
=g^{n-2}(z_{j_{1}}-z_{j_{n}})(z_{j_{2}}-z_{j_{n}})\int
\theta(z_{j_{2}}-z_{j_{3}})d\phi _{j_{1}}d\phi_{j_{2}}d\phi_{j_{n}}d
\vartheta_{j_{n}}\times  \nonumber \\
\times \Biggl(\prod_{s=3}^{n-1}\theta (z_{j_{s}}-z_{j_{s+1}})dz_{j_{s}}d\phi
_{j_{s}}d\vartheta _{j_{s}}\Biggl)  \nonumber \\
\times \exp \biggl[2\alpha ^{\prime} \sum_{m>n}[\xi _{j_{m}}\phi
_{j_{m}}D(z_{j_{m}},\vartheta _{j_{m}})-ik_{j_{m}}][\xi _{j_{n}}\phi
_{j_{n}}D(z_{j_{n}},\vartheta _{j_{n}})-ik_{j_{n}}] \ln
(z_{j_{m}}-z_{j_{n}}- \vartheta_{j_{m}}\vartheta _{j_{n}}) \biggl],
\label{term}
\end{eqnarray}
where $\vartheta _{j_{1}}=\vartheta _{j_{2}}=0$. The additional Grassmann
variables $\phi _{j_{s}}$ are introduced for each of the vertices
\begin{equation}
V(z,\vartheta ;k,\xi )=\int d\phi e^{(\phi \xi D-ik)X}\,.  \label{vertphi}
\end{equation}

So, the tree amplitude is presented by expression (\ref{opsam}),
where $ A_{n}^{(r)}(\{k_{j},\xi _{j}\})$ is given in eq.
(\ref{term}). Note, that under an anti-cyclic permutation the
amplitude $A_{n}^{(r)}(\{k_{j},\xi _{j}\})$ receives only the
factor $(-1)^{n}$. Provided that three variables are fixed as
$z_{j_{1}}=\infty $, $z_{j_{2}}=1$ and $z_{j_{n}}=0$, one can
verify this property with the use of transformation (\ref{sfltr})
for the integrand in (\ref{term}) choosing the functions
$f(\hat{z})=\hat{z} _{j_{n-1}}/\hat{z}$ and $\varepsilon
(\hat{z})= - \hat{\vartheta}_{j_{n}}-
\hat{z}(\hat{\vartheta}_{j_{n-1}}-\hat{\vartheta}
_{j_{n}})/\hat{z} _{j_{n-1}} $.

The Chan-Paton factor in (\ref{opsam}) is given by
\begin{equation}
T_{(r)}=trace[\lambda _{j_{1}}\dots \lambda _{j_{n}}]\,,  \label{chanp}
\end{equation}
where $\lambda _{s}$ is a color matrix for the corresponding group generator
in the fundamental representation. Below we discuss the oriented string, for
which $\lambda _{s}$ are $U(n)$-matrices in the fundamental representation.
In this case
\begin{equation}
trace(\lambda _{r}\lambda _{s})=\delta _{rs},\quad \sum_{j}(\lambda
_{j})_{ab}(\lambda _{j})_{cd}=\delta _{ad}\delta _{bc}\,.  \label{prmat}
\end{equation}
Hence
\begin{equation}
\lambda _{r}\lambda _{s}=\sum_{j}trace(\lambda _{r}\lambda _{s}\lambda
_{j})\lambda _{j}\,.  \label{pr1mat}
\end{equation}
We take $\lambda _{1}=I/\sqrt{n}$ as the $U(1)$-generator and the
matrices $ \lambda _{2},\dots ,\lambda _{n}$ as generators of the
$SU(n)$ group. They satisfy the following relations
\begin{equation}
\frac{1}{2}[\lambda _{r}\lambda _{s}-\lambda _{s}\lambda _{r}
]=\sum_{j}f_{rsj}\lambda _{j}\,,\quad \frac{1}{2}[\lambda
_{r}\lambda _{s}+\lambda _{s}\lambda _{r}]=\delta
_{rs}n^{-1}+\sum_{j}d_{rsj}\lambda _{j}\,,\quad
\sum_{j}d_{jjs}=0\,.  \label{pr2mat}
\end{equation}

Obviously, the tensor $d$ is symmetric in two first indices
$d_{rsj}=d_{srj}$ and the structure constants $f_{rsj}$ are
completely anti-symmetric. Furthermore, $f_{1sj}=0$,
$d_{11j}=d_{jr1}=0$, and, in addition, $d_{rsj}$ is symmetric in
all indices provided that both $s\neq 1$, $r\neq 1$ and $ j\neq
1$. Besides, $d_{s1j}=\sqrt{1/n}\delta _{sj}$ when $s\neq 1$ and $
j\neq 1$. We obtain also
\begin{equation}
\sum_{r,s}d_{rsj}d_{rsl}=\sum_{r,s}f_{rsj}f_{srl}=\frac{n}{2}[\delta
_{jl}-\delta _{j1}\delta _{l1}]\,.  \label{pr3mat}
\end{equation}

Below in the Regge kinematics ($s\gg -t\sim m^{2}$) we calculate
the amplitude $A_{(14)}^{(23)}$ describing the scattering
$a+b\rightarrow a^{\prime }+b^{\prime }$ of the vector massless
particles (gluons) with momenta $p_{i}$ ($p_{i}^{2}=0$). The 
corresponding kinematical invariants are $s=(p_{a}+p_{b})^{2}=(p_{a^{
\prime }}+p_{b^{\prime }})^{2}$ and $t=(p_{a}-p_{a^{\prime
}})^{2}=(p_{b}-p_{b^{\prime }})^{2}$. The gauge is chosen to be
$\xi_{i}^{0}=0$ for $i=a,a^{\prime },b,b^{\prime }$. In the Regge
limit the amplitude $A_{(ab)}^{(a^{\prime }b^{\prime })}$ is (cf.
(\ref {tacas}))
\begin{eqnarray}
\lim_{s\rightarrow\infty}A_{(ab)}^{(a^{\prime}b^{\prime})}=
-2g^2\alpha^{\prime}s \Gamma
(-\alpha^{\prime}t)(\alpha^{\prime}s)^{\alpha^{\prime}t}(\xi_a\xi_{a^{
\prime}})(\xi_b\xi_{b^{\prime}}) \Biggl\{\sum_s
f_{j_aj_{a^{\prime}}s}f_{sj_bj_{b^{\prime}}}(e^{-\pi i\alpha^{\prime}t}+1)
\nonumber \\
+(e^{-\pi i\alpha^{\prime}t}-1)\Biggl[\frac{\delta_{j_aj_{a^{\prime}}}
\delta_{j_bj_{b^{\prime}}}}{n} +\sum_sd_{j_aj_{a^{\prime}}s}
d_{sj_bj_{b^{\prime}}}\Biggl]\Biggl\}\,,  \label{scam}
\end{eqnarray}
where the color index $j_{i}$ refers to $U(n)$-quantum numbers of the
particle carrying the momentum $p_{i}$. The spin structure described by the
polarization vectors $\xi _{i}$ ($i=a,a^{\prime },b,b^{\prime }$)
corresponds to the conservation of helicities for each of colliding
particles. Various terms in (\ref{scam}) are associated with different
Regge contributions. Their quantum numbers are the $SU(n)$ singlet (with the
signature ''+'') and two adjoint $SU(n)$-representations (having the
dimension $n^{2}-1$ and the signatures ''+'' and ''-''). Note, that in QCD
the Regge asymptotics of the scattering amplitude in the Born approximation
contains only a contribution with the negative signature, corresponding to
an exchange of the reggeized gluon. The contribution from the positive
signature with octet quantum numbers appears only in upper orders of 
perturbation theory. Nevertheless, for large $N_{c}$ the Regge trajectories
with opposite signatures coincide each with another. The degeneracy of these
$t$-channel states is important for the duality symmetry between the
colorless composite states with different signatures \cite{dual}. As for the
Regge contribution with vacuum quantum numbers, it also takes place in
QCD only in upper orders of perturbation theory and corresponds to the BFKL
Pomeron. Its appearance in the superstring model already in a tree
approximation can be considered as a manifestation of the soft Pomeron
having a non-perturbative nature. When $n=N_{c}$ is large, one can neglect
this soft Pomeron contribution in expression (\ref{scam}).

To derive the above asymptotic behavior of the scattering amplitudes in the
superstring theory we used the relation
\begin{eqnarray}
\lambda _{r_{i}}\lambda _{r_{j}}e^{-i\pi (\alpha ^{\prime
}t_{ij}+1)}+\lambda _{r_{j}}\lambda _{r_{i}}=\frac{1}{n}
\delta_{r_{i}r_{j}}(e^{-i\pi (\alpha ^{\prime }t_{ij}+1)}+1)
+\sum_{s}d_{r_{i}r_{j}s}\lambda _{s}(e^{-i\pi (\alpha ^{\prime }t_{ij}+1)}+1)
\nonumber \\
+\sum_{s}f_{r_{i}r_{j}s}\lambda_{s}(e^{-i\pi (\alpha^{\prime
}t_{ij}+1)}-1)\,,  \label{wf1fac}
\end{eqnarray}
which follows from (\ref{pr2mat}). Note, that the soft pomeron
contribution appears also for the Chan-Paton factors corresponding
to the colour group $ O(n)$. Moreover, in the one-loop
approximation its Regge trajectory does not contain ultraviolet
divergencies for $n=32$ and $D=10$ \cite{gs}. In this model the
gluon Regge trajectory $\omega _1(t)$ is finite is
given below (see Appendix B)
\begin{equation}
\omega _1(t)=-8g^2n \int_{-1}^1\frac{d\lambda}{\lambda}\,\int_0^1 d \nu_2
\,(\sin\pi\nu _2)^2\,\left(\frac{L_2}{1+L_1} \right)^{-\alpha^{\prime}t}
\,(1+L_1)^{-1}\,,  \label{omega1}
\end{equation}
where $N=32$, and
\[
L_1= -2\sum_{n=1}^{\infty}\frac{\lambda ^n\,(1-\lambda ^n)^2}{ (1-2\lambda^n
\cos 2\pi \nu_2+\lambda ^{2n})^2}\,,
\]
\begin{equation}
L_2=\prod _{n=1}^{\infty} \frac{(1-\lambda ^n)^4}{ (1-2\lambda ^n \cos 2\pi
\nu_{2}+\lambda^{2n})^2}\,.  \label{definl}
\end{equation}

If we consider only a contribution of the planar diagram, the low limit of
integration over $\lambda$ in the above expression is zero and the gluon
Regge trajectory contains the logarithmic divergency at small $\lambda$,
which can be removed by a renormalization of the slope $\alpha ^{\prime}$
in the Born amplitude \cite{gs}. In a similar way for the $SU(n)$ group we
have
\begin{equation}
\omega _1(t)=-8g^2N_c \int_0^1\frac{d\lambda}{\lambda}\, \int_0^1 d \nu_2
(\sin\pi\nu_2)^2\,\biggl[\left(\frac{L_2}{1+L_1} \right)^{-\alpha^{\prime}t}
\,(1+L_1)^{-1}-1\biggl]\,,  \label{omegasu}
\end{equation}
where $n=N_c$ is the number of colors.

\section{Particle production in the multi-Regge kinematics}

Similar to the QCD case for string models the contribution of the ladder
diagrams Fig.1 is factorized in the multi-Regge kinematics. This
factorization was verified for the boson string theory \cite{weis} and it is
valid also for the superstring models. We are going to calculate the kernel
of the BFKL equation with the use of the $s$-channel unitarity by
integrating the square of inelastic amplitudes over the intermediate
particles in the multi-Regge kinematics.

In particular the diagram Fig.1b describes the production of one additional
resonance with a fixed mass and momentum $k-k^{\prime}$. In this diagram the
initial particles have non-vanishing color quantum numbers whereas usually
the solution of the BFKL equation for the gluon-gluon scattering should be
sandwiched between the impact factors for the colorless colliding objects to
avoid infrared divergencies. Note, however, that the integral kernel of the
BFKL equation for the Pomeron wave function does not depend on quantum
numbers of initial particles. To take into account a tower of the
intermediate string states for the middle line on Fig.1b we find in this
section the multi-Regge asymptotics of the amplitude for the tree diagram
Fig.2. Then we calculate the sum over residues in the poles over the
particle invariant mass $k^{2}$ and integrate over other kinematic variables
to obtain the BFKL kernel.

As far as a large number of colors $n$ is considered, only planar diagram
contributions are important and the kernel is proportional to $n=N_c$. In
the multi-Regge kinematics the momenta $k_{1},k_{2},k_{7},k_{8}$ on Fig.2
are almost collinear. Their space components are opposite in sign to the
corresponding components of the momenta $k_{4},k_{3},k_{6},k_{5}$.

To each particle with the momentum $k_{j}$, the string coordinates
$ z_{j},\vartheta _{j}$ and the color matrix $\lambda _{j}$ are
assigned. It is assumed that
$z_{8}<z_{7}<z_{6}<z_{5}<z_{4}<z_{3}<z_{2}<z_{1}$. We fix five
variables: $z_{1}=\infty ,\,z_{2}=1,\,z_{8}=0$ and $\vartheta
_{1}=\vartheta_{2}=0$. In amplitude (\ref{opsam}) one should sum
over the contributions of the diagrams which can not be obtained
from one configuration by cyclic or anti-cyclic transmutations of
gluon indices. We should take into account also the Chan-Paton
factors $T^{(+)}$
\begin{equation}
T^{(+)}=trace[\lambda _{r_{8}}\lambda _{r_{7}}\lambda _{r_{6}}\lambda
_{r_{5}}\lambda _{r_{4}}\lambda _{r_{3}}\lambda _{r_{2}}\lambda
_{r_{1}}+\lambda _{r_{5}}\lambda _{r_{6}}\lambda _{r_{7}}\lambda
_{r_{8}}\lambda _{r_{1}}\lambda _{r_{2}}\lambda _{r_{3}}\lambda_{r_{4}}]\,.
\label{cpfac}
\end{equation}

To calculate the kernel from Fig.2 only contributions having poles in
the invariant $k^{2}$ are essential. There are 16 diagrams of such type
corresponding to the configuration
\begin{equation}
(k_{1}=q_{1}^{\prime }\,,k_{2}=-p_{a^{\prime }})\,,\quad
(k_{3}=-p_{b^{\prime }}\,,k_{4}=q_{2}^{\prime })\,,\quad
(k_{5}=-q_{2}\,,k_{6}=p_{b})\,,\quad (k_{7}=p_{a}\,,k_{8}=-q_{1})
\label{cnfgr}
\end{equation}
and those obtained by the interchange $(k_{j}\rightleftharpoons
k_{l})$ inside each of the above brackets, which leads to the
signature factors. As it was pointed out already,
$p_{a}$,$\,p_{b}$ and $p_{a^{\prime }}$, $ p_{b^{\prime }}$ are
momenta of the initial and final particles, respectively. The
momenta $q_{1}$, $q_{2}$, $q_{1}^{\prime }$ and $ q_{2}^{\prime }$
correspond to intermediate particles. Obviously, for the
calculation of a discontinuity of the elastic amplitude the
relations $ q_{1}^{\prime }=-q_{1}$ and $q_{2}^{\prime }=-q_{2}$
are valid, but temporally we distinguish between $q_{i} $ and
$-q_{i}^{\prime }$ performing later an analytical continuation in
the invariants $(k+q_{1})^{2}$, $ (k+q_{2})^{2}$,
$(k-q_{1}^{\prime })^{2}$ and $(k-q_{2}^{\prime })^{2}$ to their
physical values. In the multi-Regge configuration the momentum $k$
on Fig.2 obeys some kinematical constraints. Namely, the
quantities $k^{2}, k_{\perp}^2 $ and $(k^{0})^{2}$ in the c.m.
system are assumed to be much smaller than the energy invariants
$s,\,s_{1}$ and $s_{2}$. Integral (\ref {term}) for each of 16
diagrams is calculated in the kinematics where it is convergent,
and subsequently
the result is analytically continued to the physical region of
the reaction.

In expression (\ref{term}) several polarization structures arise, but only
the term
\[
A_{s}\sim(\xi _{1}\xi _{2})(\xi _{3}\xi _{4}) (\xi _{5}\xi_{6})(\xi _{7}\xi
_{8})
\]
contributes to the multi-Regge asymptotics of the tree amplitude for Fig.2
\begin{equation}
A_{s}=g^{6}\widetilde{A}T^{(+)}(\xi _{1}\xi _{2})(\xi _{3}\xi _{4})(\xi
_{5}\xi _{6})(\xi _{7}\xi _{8})\,,  \label{spst}
\end{equation}
where the polarization vector $\xi _{j}$ is associated with the
momentum $ k_{j}$ and the Chan-Paton factor $T^{(+)}$ is given in
eq. (\ref{cpfac}). Fixing the parameters as follows: $z_{1}=\infty
$, $z_{2}=1$, $z_{8}=0$ and $ \vartheta _{1}=\vartheta _{2}=0$,
one obtains from eq. (\ref{term})
\begin{equation}
\widetilde{A}=\int \frac{\tilde{B}B\theta (1-z_{3})\theta (z_{7})d\vartheta
_{7}d\vartheta _{8}\,dz_{7}}{(z_{3}-z_{4}-\vartheta _{3}\vartheta
_{4})(z_{5}-z_{6}-\vartheta _{5}\vartheta _{6})(z_{7}-\vartheta
_{7}\vartheta _{8})}\prod_{s=3}^{6}\theta (z_{s}-z_{s+1})dz_{s}\,d\vartheta
_{s}\,.  \label{polar}
\end{equation}

Here the pre-factor $\tilde{B}$ is given below
\begin{eqnarray}
\tilde{B} &=&\biggl[1+\frac{2\alpha ^{\prime }(k_{3}k_{4})}{z_{3}-z_{4}}
\vartheta _{3}\vartheta _{4}+\frac{2\alpha ^{\prime }(k_{3}k_{5})}{
z_{3}-z_{5}}\vartheta _{3}\vartheta _{5}+\frac{2\alpha ^{\prime
}(k_{3}k_{6}) }{z_{3}-z_{6}}\vartheta _{3}\vartheta _{6}+\frac{2\alpha
^{\prime }(k_{3}k_{7})}{z_{3}-z_{7}}\vartheta _{3}\vartheta _{7}  \nonumber
\\
&&+\frac{2\alpha ^{\prime }(k_{3}k_{8})}{z_{3}-z_{8}}\vartheta _{3}\vartheta
_{8}\Biggl]\biggl[1+\frac{2\alpha ^{\prime }(k_{4}k_{5})}{z_{4}-z_{5}}
\vartheta _{4}\vartheta _{5}+\frac{2\alpha ^{\prime }(k_{4}k_{6})}{
z_{4}-z_{6}}\vartheta _{4}\vartheta _{6}+\frac{2\alpha ^{\prime
}(k_{4}k_{7}) }{z_{4}-z_{7}}\vartheta _{4}\vartheta _{7}  \nonumber \\
&&+\frac{2\alpha ^{\prime }(k_{4}k_{8})}{z_{4}-z_{8}}\vartheta _{4}\vartheta
_{8}\Biggl]\Biggl[1+2\frac{\alpha ^{\prime }(k_{5}k_{6})}{z_{5}-z_{6}}
\vartheta _{5}\vartheta _{6}+\frac{2\alpha ^{\prime }(k_{5}k_{7})}{
z_{5}-z_{7}}\vartheta _{5}\vartheta _{7}+\frac{2\alpha ^{\prime
}(k_{5}k_{8}) }{z_{5}-z_{8}}\vartheta _{5}\vartheta _{8}\Biggl]  \nonumber \\
&&\times \Biggl[1+\frac{2\alpha ^{\prime }(k_{6}k_{7})}{z_{6}-z_{7}}
\vartheta _{6}\vartheta _{7}+\frac{2\alpha ^{\prime }(k_{6}k_{8})}{
z_{6}-z_{8}}\vartheta _{6}\vartheta _{8}\Biggl]\Biggl[1+\frac{2\alpha
^{\prime }(k_{7}k_{8})}{z_{7}-z_{8}}\vartheta _{7}\vartheta _{8}\Biggl]
\label{snext}
\end{eqnarray}
and the expression $B$ coincides with the integrand for a multi-tachyon
scattering amplitude of the boson string theory:
\begin{equation}
B=\prod_{2\leq m<n\leq 8}(z_{m}-z_{n})^{-2\alpha ^{\prime }k_{m}k_{n}}\,.
\label{sstart}
\end{equation}

Similar to the case of bosonic strings \cite{detar} one concludes from
eq. (\ref{sstart}) that in the multi-Regge kinematics the essential values
of parameters are
\begin{equation}
z_{3}\rightarrow 0,\quad z_{3}=z_{4}+x\,,\quad z_{5}=z_{6}+y\,,\quad
x/z_{6}\rightarrow 0\,,\quad y/z_{6}\rightarrow 0\,,\quad
z_{7}/z_{6}\rightarrow 0\,.  \label{conf}
\end{equation}
In this configuration of variables the expression for $B$ is simplified as
follows
\begin{eqnarray}
B &\approx &x^{-2\alpha ^{\prime }k_{3}k_{4}}y^{-2\alpha ^{\prime
}k_{5}k_{6}}z_{7}^{-2\alpha ^{\prime }k_{7}k_{8}}z_{4}^{-2\alpha
^{\prime }(k_{3}+k_{4})(k_{7}+k_{8})}z_{6}^{-2\alpha ^{\prime
}(k_{5}+k_{6})(k_{7}+k_{8})}  \nonumber \\ &&\times
(z_{4}-z_{6})^{-2\alpha ^{\prime }(k_{3}+k_{4})(k_{5}+k_{6})}\exp
\biggl[2\alpha ^{\prime }k_{2}k_{3}x+2\alpha ^{\prime
}k_{2}k_{5}y+2\alpha ^{\prime }k_{2}(k_{3}+k_{4})z_{4}  \nonumber
\\ &&+2\alpha ^{\prime }k_{2}(k_{5}+k_{6})z_{6}-2\alpha ^{\prime
}k_{3}k_{7} \frac{z_{7}x}{z_{4}^{2}}+2\alpha ^{\prime
}k_{7}(k_{3}+k_{4})\frac{z_{7}}{ z_{4}}-2\alpha ^{\prime
}k_{5}k_{7}\frac{z_{7}y}{z_{6}^{2}}  \nonumber \\ &&+2\alpha
^{\prime }k_{7}(k_{5}+k_{6})\frac{z_{7}}{z_{6}}-2\alpha ^{\prime
}k_{3}(k_{7}+ke_{8})\frac{x}{z_{4}}-2\alpha ^{\prime
}k_{5}(k_{7}+k_{8}) \frac{y}{z_{6}}\biggl]\,.  \label{next}
\end{eqnarray}

In the multi-Regge limit we have
\begin{eqnarray}
k_{2}(k_{3}+k_{4}) &\rightarrow &-k_{2}(k_{5}+k_{6})\rightarrow
k_{1}k\,,\quad k_{7}(k_{3}+k_{4})\rightarrow -k_{7}(k_{5}+k_{6})\rightarrow
k_{8}k\,,  \nonumber \\
k_{3}(k_{7}+k_{8}) &\rightarrow &-k_{4}k\,,\quad
k_{5}(k_{7}+k_{8})\rightarrow -k_{6}k\,.  \label{limts}
\end{eqnarray}
The integral is convergent in the following kinematical region of invariants
\begin{eqnarray}
k_{2}k_{3} &<&0,\quad k_{2}k_{5}<0,\quad k_{3}k_{7}>0,\quad
k_{5}k_{7}>0,\quad k_{2}(k_{3}+k_{4})<0\,,\quad  \nonumber \\
k_{7}(k_{5}+k_{6}) &<&0\,,\,\,k_{3}(k_{7}+k_{8})>0\,,\qquad
k_{5}(k_{7}+k_{8})>0\,.  \label{neqs}
\end{eqnarray}

We redefine the variables as follows
\begin{eqnarray}
z_{4} &\rightarrow &\frac{z_{4}}{-2\alpha^{\prime }k_{2}(k_{3}+k_{4})}
\,,\quad z_{6}\rightarrow \frac{z_{6}}{2\alpha ^{\prime }k_{2}(k_{5}+k_{6})}
\,,\quad z_{7}\rightarrow \frac{z_{7}}{[2\alpha ^{\prime
}k_{2}(k_{3}+k_{4})][2\alpha ^{\prime }k_{7}(k_{5}+k_{6})]}\,,  \nonumber \\
x &\rightarrow &\frac{x}{-2\alpha ^{\prime }(k_{2}k_{3})},\quad y\rightarrow
\frac{y}{-2\alpha ^{\prime }(k_{2}k_{5})}\,.  \label{change}
\end{eqnarray}
The asymptotics of $\widetilde{A}$ in expression (\ref{spst}) can be written
as follows
\begin{equation}
\widetilde{A}=GA(t_{5678},\kappa
^{2},t_{3478},t_{34},t_{56},t_{12},t_{78})\,,  \label{tram}
\end{equation}
where the factor $G$ collects all large energy invariants, and
$ A(t_{5678},\kappa ^{2},t_{3478},t_{34},t_{56},t_{12},t_{78})$
depends only on fixed transverse momenta. We define the energy
invariants $s_{i}$ and $ s_{jk}$ as follows
\begin{equation}
s_{i}=((k_{1}+k)^{2}\approx 2(k_{1}k)\,,\quad s_{23}=-2(k_{2}k_{3})\,,\quad
s_{25}=-2(k_{2}k_{5})\,.  \label{lirs}
\end{equation}

Then the expression for $G$ in (\ref{tram}) has the factorized form
\begin{equation}
G=(-\alpha ^{\prime }s_{1})^{\alpha ^{\prime }t_{12}+1}(-\alpha ^{\prime
}s_{7})^{\alpha ^{\prime }t_{78}+1}(-\alpha ^{\prime }s_{4})^{\alpha
^{\prime }t_{34}+1}(-\alpha ^{\prime }s_{6})^{\alpha ^{\prime }t_{56}+1}\,,
\label{largf}
\end{equation}
where the fixed invariants are
\begin{equation}
t_{i}=(k_{1}+k_{j})^{2}\,,\quad t_{ijlm}=-(k_{i}+k_{j}+k_{l}+k_{m})^{2}\,.
\label{smirs}
\end{equation}
Note, that we have the kinematical constraint
\begin{equation}
t_{3456}+t_{3478}+t_{5678}=t_{12}+t_{34}+t_{56}+t_{78}\,.  \label{relt}
\end{equation}
The fixed invariant $\kappa ^{2}$ in (\ref{tram}) is given below
\begin{equation}
\kappa ^{2}=\alpha ^{\prime }s_{1}s_{4}/s_{23}=\alpha ^{\prime
}s_{1}s_{6}/s_{25}=\alpha ^{\prime }[(k^{0})^{2}-k_{\parallel }^{2})\,,
\label{kapa}
\end{equation}
where $k_{\parallel }$ is the longitudinal component of the momentum $k$. To
simplify the last factor in (\ref{tram}) one can use the following relations
valid in the multi-Regge kinematics due to eqs. (\ref{limts})
\begin{eqnarray}
(k_{5}k_{7})(k_{6}k_{8})-(k_{5}k_{8})(k_{6}k_{7})
&=&\frac{1}{2}\biggl[ ((k_{5}+k_{6})k_{7})((k_{6}-k_{5})k_{8}))
\nonumber \\ +((k_{5}-k_{6})k_{7})((k_{6}+k_{5})k_{8}))\biggl]
&=&\frac{1}{2} (k_{6}k_{8})[t_{5678}-t_{56}-t_{78}-\kappa ^{2}]\,
\label{k5k6d}
\end{eqnarray}
and
\begin{equation}
(k_{4}k_{7})(k_{3}k_{8})-(k_{4}k_{8})(k_{3}k_{7})=\frac{1}{2}
(k_{3}k_{8})[t_{4378}-t_{43}-t_{78}+\kappa ^{2}]\,.  \label{k4k3d}
\end{equation}

After redefinition (\ref{change}) of variables in expression (\ref{polar})
with the use of above simplifications one can perform the Grassmann
integrations. As a result, the last factor in (\ref{tram}) turns out to be
\begin{equation}
A(t_{5678},\kappa
^{2},t_{3478},t_{34},t_{56},t_{12},t_{78})=\int_{0}^{\infty
}dx\int_{0}^{\infty }dy\int_{0}^{\infty
}dz_{4}\int_{0}^{z_{4}}B_{s}V_{b}\,dz_{6}\,,  \label{mainin}
\end{equation}
where both $B_{s}$ and $V_{b}$ depend on integration parameters and
external variables. The expression for $V_{b}$ is the same as in the bosonic
string model, and the pre-factor $B_{s}$ arises due to the superstring
modifications. Explicitly,
\begin{eqnarray}
V_{b} &=&x^{-\alpha ^{\prime }t_{34}}y^{-\alpha ^{\prime
}t_{56}}z_{7}^{-\alpha ^{\prime }t_{78}}z_{4}^{-\alpha ^{\prime
}[t_{3478}-t_{34}-t_{78}]}z_{6}^{-\alpha ^{\prime
}[t_{5678}-t_{56}-t_{78}]}(z_{4}-z_{6})^{-\alpha ^{\prime
}[t_{3456}-t_{34}-t_{56}]}[xyz_{7}]^{-2}  \nonumber \\ &&\times
\exp \biggl[-\biggl(x+y+\frac{z_{7}x}{z_{4}^{2}}+\frac{z_{7}y}{
z_{6}^{2}}\biggl)-\biggl(z_{4}-z_{6}+\frac{z_{7}}{z_{6}}-\frac{z_{7}}{z_{4}}
\biggl)+\kappa
^{2}\biggl(\frac{x}{z_{4}}+\frac{y}{z_{6}}\biggl)\biggl]\,.
\label{form}
\end{eqnarray}
The integrals in (\ref{mainin}) are defined for $\kappa ^{2}<0$, and
\begin{eqnarray}
B_{s} &=&(\alpha ^{\prime }t_{34}+1)(\alpha ^{\prime
}t_{56}+1)(\alpha ^{\prime
}t_{78}+1)+\frac{xyz_{7}^{2}}{z_{4}^{2}z_{6}^{2}}[\alpha ^{\prime
}t_{35}+\alpha ^{\prime }t_{36}+\alpha ^{\prime }t_{45}+\alpha
^{\prime }t_{46}]  \nonumber \\ &&-y^{2}z_{7}^{2}(\alpha ^{\prime
}t_{34}+1)\Biggl[\frac{\alpha ^{\prime
}[t_{5678}-t_{56}-t_{78}]-\kappa
^{2}}{yz_{7}z_{6}^{2}}-\frac{1}{z_{6}^{4}} \Biggl]  \nonumber \\
&&-x^{2}z_{7}^{2}(\alpha ^{\prime }t_{56}+1)\Biggl[\frac{\alpha
^{\prime }[t_{3478}-t_{34}-t_{78}]+\kappa
^{2}}{xz_{7}z_{4}^{2}}-\frac{1}{z_{4}^{4}} \Biggl]\,.
\label{snex2t}
\end{eqnarray}

In expression (\ref{mainin}) one can perform easily the integration over the
variables $x$ and $y$
\begin{eqnarray}
A(t_{5678},\kappa ^{2},t_{3478},t_{34},t_{56},t_{12},t_{78}) &=&\Gamma
(-\alpha ^{\prime }t_{34})\Gamma (-\alpha ^{\prime }t_{56})  \nonumber \\
&&\times e^{\pi i(\alpha ^{\prime }t_{34}+\alpha ^{\prime
}t_{56})}I(t_{5678},\kappa ^{2},t_{3478},t_{34},t_{56},t_{12},t_{78})\,,
\label{samplitu}
\end{eqnarray}
where the factor $I(t_{5678},
\kappa^{2},t_{3478},t_{34},t_{56},t_{12},t_{78})$ is obtained from eq. (\ref
{mainin}). Its form can be essentially simplified as it is shown in Appendix
C. Below we present the final result using in addition the fact that the
calculated amplitude is symmetric under an interchange between the left and
right parts of the considered diagram (see Section 3). Thus, taking into
account the relation $t_{1256}=t_{3478}$ eq. (\ref{samplitu}) can be written
as follows
\begin{eqnarray}
A(t_{5678},\kappa ^{2},t_{3478},t_{34},t_{56},t_{12},t_{78}) &=&\Gamma
(-\alpha ^{\prime }t_{12})\Gamma (-\alpha ^{\prime }t_{78})e^{\pi i(\alpha
^{\prime }t_{12}+\alpha ^{\prime }t_{78})}  \nonumber \\
&&\times I(t_{5678},\kappa ^{2},t_{3478},t_{12},t_{78},t_{34},t_{56})\,,
\label{sampli1tu}
\end{eqnarray}
where
\begin{eqnarray}
I(t_{5678},\kappa
^{2},t_{3478},t_{12},t_{78},t_{34},t_{56})=(\kappa ^{2})^{-\alpha
^{\prime }t_{12}-\alpha ^{\prime }t_{78}-2}\int
dydzdfe^{-f-y}f^{-\alpha ^{\prime }t_{34}-1}y^{-\alpha ^{\prime
}t_{56}-1} \nonumber \\ \times z^{-\alpha ^{\prime
}t_{5678}-1}(1-z)^{-\alpha ^{\prime }t_{3456}+\alpha^{\prime
}t_{34}+\alpha ^{\prime }t_{56}} [f+yz-\kappa^{2}(1-z)]^{\alpha
^{\prime }t_{12}} [y+fz-\kappa ^{2}(1-z)]^{\alpha ^{\prime}t_{78}}
\nonumber \\ \times \Biggl[\alpha ^{\prime }t_{12}+\alpha ^{\prime
}t_{78}-\alpha ^{\prime }t_{3478}(1-z)+\frac{\alpha ^{\prime
}t_{12}y(1-z)}{f+yz-\kappa ^{2}(1-z)}  \nonumber \\ +\frac{\alpha
^{\prime }t_{78}f(1-z)}{y+fz-\kappa ^{2}(1-z)}+z-(f+y)\Biggl] \,.
\label{sit2u}
\end{eqnarray}
Here all integrations are performed from $0$ to $\infty $. The above
expression is convergent at $\kappa ^{2}<0$. For the factor in a front of
the integral we choose the condition $\arg \kappa ^{2}=-\pi $. Really the
phase arising in this case, is compensated by a similar phase in (\ref
{sampli1tu}) and $A$ is real for $\kappa <0$.

The final result is obtained by summing the contributions of 16 diagrams
listed in the beginning of this Section, every term being analytically
continued from the kinematical region where the corresponding integral (\ref
{term}) is convergent. Taking into account the spin structure (\ref{spst})
for each diagram and relations (\ref{tram}), (\ref{largf}) and (\ref
{sampli1tu}), we derive the following expression for $A^{(f)}$
\begin{eqnarray}
A^{(f)} &=&g^{6}s_{1}^{\alpha ^{\prime }t_{12}+\alpha ^{\prime
}t_{78}+2}s_{4}^{\alpha ^{\prime }t_{34}+\alpha ^{\prime
}t_{56}+2}(\xi _{1}\xi _{2})(\xi _{3}\xi _{4})(\xi _{5}\xi
_{6})(\xi _{7}\xi _{8})\Gamma (-\alpha ^{\prime }t_{12})\Gamma
(-\alpha ^{\prime }t_{78})e^{\pi i(\alpha ^{\prime }t_{12}+\alpha
^{\prime }t_{78})}  \nonumber \\ &&\times
\sum_{j_{1},j_{2},j_{3},j_{4}}\mathcal{F}
_{r_{1},r_{2},j_{1}}(t_{12})\mathcal{F}_{r_{3},r_{4},j_{2}}(t_{34})
\mathcal{ \ F}_{r_{5},r_{6},j_{3}}(t_{56})
\mathcal{F}_{r_{7},r_{8},j_{3}}(t_{78}) \nonumber \\ &&\times
T_{j_{1},j_{2},j_{3},j_{4}}^{(+)}I(t_{5678},\kappa
^{2},t_{3478},t_{12},t_{78},t_{34},t_{56})\,,  \label{whfac}
\end{eqnarray}
where
\begin{equation}
\mathcal{F}_{r_{s},r_{l},j}(t_{sl})=trace[(\lambda _{r_{s}}\lambda
_{r_{l}}e^{-\pi i(\alpha ^{\prime }t_{sl}+1)}+\lambda _{r_{s}}\lambda
_{r_{l}})\lambda _{j}]  \label{calf}
\end{equation}
and
\begin{equation}
T_{j_{1},j_{2},j_{3},j_{4}}^{(+)}=trace[\lambda _{j_{4}}\lambda
_{j_{3}}\lambda _{j_{2}}\lambda _{j_{1}}+\lambda _{j_{1}}\lambda
_{j_{2}}\lambda _{j_{3}}\lambda _{j_{4}}]\,.  \label{trpl}
\end{equation}

For the group $U(n)$ the index $r_{i}$ enumerates color states of
the particle carrying the momentum $k_{i}$ defined in
(\ref{cnfgr}). After an analytical continuation we put $s_{1}=\pm
s_{7}$ and $s_{4}=\pm s_{6}$. In a similar way, $I(t_{5678},\kappa
^{2},t_{3478},t_{12},t_{78},t_{34},t_{56})$ in (\ref{whfac} ) is
calculated using a similar continuation of expression
(\ref{sit2u}) to the region $\kappa ^{2}>0$. This procedure is
performed by the replacement $\kappa ^{2}\rightarrow \kappa
^{2}+i\delta $ with $\delta \rightarrow +0$. We have also the
condition $k^{(0)}>0$, and therefore due to (\ref{kapa}), our
prescription corresponds to the Feynman rule for going around the
singularity. After the analytic continuation the factor in front
of the integral turns out to be positive.

\section{BFKL equation in the string model}

Omitting the impact factors of colored particles in the left and
right hand sides of the contribution of the diagram Fig1.b one can
obtain expressions for higher order ladder diagrams by
iterating its interior part. To find the BFKL kernel in the
considered string model, one should calculate from expression
(\ref{whfac}) its contribution to the $t$-channel partial wave for
the scattering of massless particles. Also one-loop correction $
\omega _{1}(t)\sim g^2N_c$ to the trajectory (\ref{omegas}) should
be taken into account (see (\ref{omega1})). Thus, $\alpha ^{\prime
}t_{jl}$ in (\ref {whfac}) is replaced by the expression $\alpha
^{\prime }t_{jl}+\omega _{1}(t_{jl})$. Due to the presence of \
non-planar diagrams, the one-loop correction to the singlet
trajectory differs from that for the octet case. However, assuming
that the number of colors is large, below we neglect this
difference.

The contribution to the $t$-channel partial wave from the diagram
Fig1.b is given by the Mellin transformation in $\ln s$ applied
to the imaginary part of the amplitude. To calculate it one should
find in eq. (\ref{whfac}) the sum of residues for the poles in the
variable $\alpha ^{\prime }k^{2}$ and integrate the result over a
relevant phase volume. Initially we put $ q_{1}^{\prime }=-q_{1},$
$q_{2}^{\prime }=-q_{2}$, $r_{1}=r_{8}$, $ r_{4}=r_{5}$, $\xi
_{1}=\xi _{8} $, $\xi _{4}=\xi _{5}$ summing subsequently over
indices $r_{1}$, $r_{4}$ and polarization states $\xi _{1}$ and
$\xi _{4}$. The poles are situated at $\alpha ^{\prime }k^{2}=m$,
where $m$ is an integer number changing from $0 $ to $\infty $.

Below we denote by $l$, $l^{\prime }$ the transverse momenta of two
neighboring reggeons and by $q$ the total momentum transfer related to the
corresponding invariants as follows
\begin{equation}
l^{2}=-t_{12},\quad (l^{\prime })^{2}=-t_{34},\quad (q-l)^{2}=-t_{78},\quad
(q-l^{\prime })^{2}=-t_{56}\,.  \label{trmom}
\end{equation}
With these definitions, $\kappa ^{2}$ in (\ref{whfac}) is given
below (cf. (\ref{kapa}))
\begin{equation}
\kappa ^{2}=\alpha ^{\prime }[k^{2}+(l-l^{\prime })^{2}]\,\,,\qquad
k^{2}=t_{1234}=t_{5678}\,.  \label{newka}
\end{equation}

The multi-Regge kinematics implies, that the inequalities $s_{1}/k^{2}
\gg 1$ and $s_{4}/k^{2}\gg 1$ are fulfilled. The integration over this
region leads to the singularities of the $t$ -channel partial wave at
$\omega =j-1$. Here $j$ is the total angular momentum. The contribution
$F^{(b)}(\omega ;q^{2})$ to the $t$-channel partial wave from the
diagram Fig1.b including the correction to the trajectory $\omega
_{1}(t)$ (\ref{omegas}) is given below \begin{eqnarray} F^{(b)}(\omega
;q^{2}) =\sum_{r_{1},r_{2},r_{3},r_{4}}\int d^{D-2}l \widetilde{\Phi
}_{r_{a},r_{a^{\prime }},r_{1},r_{4}}(q;l)\Gamma (\alpha ^{\prime
}l^{2})\Gamma (\alpha ^{\prime }(q-l)^{2})e^{-\pi i(\alpha ^{\prime
}l^{2}+\alpha ^{\prime }(q-l)^{2})}  \nonumber \\
\times \frac{g^{2}}{4(2\pi )^{D-1}}T_{r_{1}r_{2}r_{3}r_{4}}^{(+)}
\sum_{m=0}^{\infty }\int_{s(m)}^{\infty }s^{-j-1}ds\int_{s(m)}^{\infty }
\frac{ds_{1}\,ds_{4}}{s^{2}}(\alpha ^{\prime }s_{1})^{-\beta (l^{2})-\beta
((q-l)^{2})+2}  \nonumber \\
\times \int d^{D-2}l^{\prime }\widetilde{I}_{m}(q;l,l^{\prime })
(\alpha^{\prime }s_{4})^{-\beta ((l^{\prime })^{2})-\beta ((q-l^{\prime
})^{2})+2}\delta (\alpha ^{\prime }s_{1}s_{4}/s-\alpha ^{\prime }k_{\perp
}^{2}-m)\widetilde{\Phi }_{r_{b},r_{b^{\prime }},r_{2},r_{3}}(q;l^{\prime
})\,.  \label{bcon}
\end{eqnarray}
Here
\begin{equation}
\beta (q^{2})=\alpha ^{\prime }q^{2}-\omega _{1}(-q^{2})\,,  \label{defco}
\end{equation}
and $D$ is the number of space-time dimensions. The quantity
$\widetilde{I} _{m}(q;l,l^{\prime })$ is the residue of the pole
at $\alpha ^{\prime }k^{2}=m$ in the integral $I(t_{5678},\kappa
^{2},t_{3478},t_{12},t_{78},t_{34},t_{56})$ appearing in
expression (\ref {whfac}) (see also definitions (\ref{trmom})).
Further, $s(m)$ is the low energy cut-off: $s(0)=s_{0}$ for $m=0$
and $s(m)=s_{0}\kappa _{m}^{2}$ for $ m\geq 1$. In this case
$\kappa _{m}^{2}=m+\alpha ^{\prime }k_{\perp }^{2} $. We impose
the condition $\alpha ^{\prime }s_{0}\gg 1$ because the production
amplitude is known only in the multi-Regge kinematics. The cut-off
is introduced to have a possibility to verify that the
non-multi-Regge kinematics is not essential. The factor $
T_{r_{1}r_{2}r_{3}r_{4}}^{(+)}$ is presented in eq. (\ref{trpl})
and
\begin{eqnarray}
\widetilde{\Phi }_{r_{a},r_{a^{\prime }},r_{1},r_{4}}(q;l) &=&(\xi _{a}\xi
_{a^{\prime }})\frac{2g^{2}}{(2\pi )^{D-1}}\sum_{r}\mathcal{F}_{r_{a^{\prime
}},r,r_{1}}(t_{12})\mathcal{F}_{r_{a},r,r_{4}}(t_{78})\,,  \nonumber \\
\widetilde{\Phi }_{r_{b},r_{b^{\prime }},r_{2},r_{3}}(q;l^{\prime }) &=&(\xi
_{b}\xi _{b^{\prime }})\frac{2g^{2}}{(2\pi )^{D-1}}\sum_{r}\mathcal{F}
_{r_{b^{\prime }},r,r_{2}}(t_{34})\mathcal{F}_{r_{b},r,r_{3}}(t_{56})\,,
\label{tilph}
\end{eqnarray}
where the function $\mathcal{F}$ is defined in (\ref{calf}).
Expressions (\ref{tilph}) are massless state contributions to the impact 
factors. 
The total impact factor for the
planar diagram Fig1b being the sum of a tower of string states is
equal to zero, which can be verified with the use of the
quasi-elastic asymptotics of the production amplitude (for more
details see Appendix D). Its vanishing ensures the cancellation of
the Amati-Fubini-Stangelini cuts in the $j$-plane for the planar
diagrams.

Once the integration over $s$, $s_{1}$ and $s_{4}$ being performed, eq.(\ref
{bcon}) is represented as follows
\begin{eqnarray}
F^{(b)}(\omega ;q^{2}) =\sum_{r_{1},r_{2},r_{3},r_{4}}\int d^{D-2}l
\widetilde{\Phi }_{r_{a},r_{a^{\prime }},r_{1},r_{4}}(q;l) \Gamma
(\alpha^{\prime }l^{2})\Gamma(\alpha^{\prime }(q-l)^{2}) e^{-\pi i(\alpha
^{\prime}l^{2}+\alpha ^{\prime }(q-l)^{2})}  \nonumber \\
\times \frac{g^{2}(\alpha ^{\prime })^{\omega }}{4(2\pi )^{D-1}}\int
d^{D-2}l^{\prime }T_{r_{1}r_{2}r_{3}r_{4}}^{(+)} \frac{(\alpha^{
\prime}s_{0})^{-\omega -\alpha ^{\prime }l^{2}- \alpha ^{\prime
}(q-l)^{2}}(\alpha^{\prime }s_{0})^{-\omega - \alpha^{\prime }(l^{\prime
})^{2}-\alpha^{\prime } (q-l^{\prime })^{2}}}{[ \omega +\beta
(l^{2})+\beta((q-l)^{2})][\omega + \beta ((l^{\prime })^{2})-\beta
((q-l^{\prime })^{2})]}  \nonumber \\
\times \widehat{I}(q ;l,l^{\prime }) \widetilde{\Phi }_{r_{b},r_{b^{\prime
}},r_{2},r_{3}}(q;l^{\prime })\,.  \label{bconf}
\end{eqnarray}
Here
\begin{equation}
\widehat{I}(q;l,l^{\prime })=S_{\omega }^{(0)}(q;l,l^{\prime })+S_{\omega
}(q;l,l^{\prime })  \label{diisc}
\end{equation}
and, in turn,
\begin{eqnarray}
S_{\omega }(q;l,l^{\prime })= \sum_{m=1}^{\infty }(\kappa_{m}^{2})^{-\omega
-\alpha^{\prime }l^{2}-\alpha ^{\prime }(q-l)^{2}-\alpha ^{\prime
}(l^{\prime })^{2}-\alpha ^{\prime }(q-l^{\prime })^{2}+2} \widetilde{I}
_{m}(q:l,l^{\prime })\,,  \nonumber \\
S_{\omega }^{(0)}(q;l,l^{\prime })=\alpha ^{\prime }(\kappa
_{0}^{2})^{\omega +2}\widetilde{I}_{0}(q;l,l^{\prime })\,.  \label{iker}
\end{eqnarray}

In this expression the factor $\widetilde{I}_{m}(q;l,l^{\prime })$
is the same as in eq. (\ref{bcon}). We remind, that the quantity $
\kappa_{m}^{2}=m+\alpha ^{\prime } k_{\perp}^{2}$ coincides with
$\kappa^{2}$ on the mass shell $\alpha^{\prime }k^{2}=m$ ($m$ is
an integer number). Expression (\ref{bconf}) is correct only in
the domain where it does not depend on the cut-off $s_{0}$, which
means, that the power of $s_{0}$ in this expression should be much
smaller than unity. In an accordance with eq. (\ref{bconf}), it is
convenient to present the contribution $F(\omega ;q^{2}) $ to the
partial wave as a sum of contributions of the diagrams Fig.1
starting from Fig.1b written in the form
\begin{eqnarray}
F(\omega ;q^{2}) &=&\frac{g^{2}}{4(2\pi )^{D-1}}
\sum_{r_{1},r_{2},r_{3},r_{4}}\int d^{D-2}l\,\int d^{D-2}l^{\prime }
\widetilde{\Phi }_{r_{a},r_{a^{\prime }},r_{1},r_{2}}(q;l)\Gamma (\alpha
^{\prime }l^{2})\Gamma (\alpha ^{\prime }(q-l)^{2})  \nonumber \\
&&\times e^{-\pi i(\alpha ^{\prime }l^{2}+\alpha ^{\prime }(q-l)^{2})}\frac{
R_{r_{1}r_{2}r_{3}r_{4}}(\omega ;q;l,l^{\prime })}{\omega +\beta ((l^{\prime
})^{2})+\beta ((q-l^{\prime })^{2})}\widetilde{\Phi }_{r_{b},r_{b^{\prime
}},r_{3},r_{4}}(q;l^{\prime })\,.  \label{pwave}
\end{eqnarray}

The particle-particle-reggeon vertices $\widetilde{\Phi}$
contained in eq.(\ref{pwave}) can be extracted from
eq.(\ref{scam}). Omitting these vertices in eq. (\ref{pwave}), one
can verify that the amplitude $ R_{r_{1}r_{2}r_{3}r_{4}}(
\omega;q;l,l_{1})$ for Fig.1 obeys the BFKL-like equation
\begin{eqnarray}
(\omega +\beta(l^{2})+\beta (q-l)^{2})R_{r_{1},r_{2},r_{3},r_{4}}(\omega
;q;l,l_{1}) =\widehat{I} (q:l,l_{1})T_{r_{1},r_{4},r_{3},r_{2}}^{(+)}
\nonumber \\
+\frac{\alpha ^{\prime }g^{2}}{4(2\pi )^{D-1}}\int d^{D-2}l^{\prime }
\widehat{I}(q ;l,l^{\prime })\sum_{r,r^{\prime }}T_{r_{1}r^{\prime
}rr_{2}}^{(+)}R_{rr^{\prime }r_{3}r_{4}}(\omega ;q;l^{\prime },l_{1})
\nonumber \\
\times \Biggl[\sum_{\sigma ,\sigma ^{\prime }}[e^{\pi i(\alpha ^{\prime
}(l^{\prime })^{2}+1)}+(-1)^{\sigma }][e^{i\pi (\alpha ^{\prime
}(q-l^{\prime })^{2}+1)}+(-1)^{\sigma ^{\prime }}]e^{-\pi i(\alpha ^{\prime
}(l^{\prime })^{2}+\alpha ^{\prime }(q-l^{\prime })^{2}}\Biggl]\,,
\label{prleq}
\end{eqnarray}
where the summation over $(\sigma ,\sigma ^{\prime })$ is associated with
the signatures for the corresponding trajectories having their color group
quantum numbers denoted, respectively, by $(r,r^{\prime })$. So, $\sigma
,\sigma ^{\prime }$ are $0$ for a positive signature and $1$ for the
negative one.

The number of colors is considered to be large and therefore one
can neglect the color-singlet reggeons in (\ref{pwave}). In this
case $ r_{i},r,r^{\prime} $ coincide with color indices of the
corresponding adjoint representations. Because $\sigma $ and
$\sigma ^{\prime }$ take values $0$ and $1$, the expression inside
the large square brackets in eq. (\ref{prleq}) is equal to $4$.
Using eqs. (\ref{pr2mat}) and (\ref{trpl}) one finds that
\begin{equation}
T_{r_{1}r^{\prime }rr_{2}}^{(+)}=\tilde{T}_{r_{1}r^{\prime
}rr_{2}}^{(+)}+2\delta _{r_{1}r_{2}}\delta _{rr^{\prime }}/n\,\,,\qquad
\sum_{r}\tilde{T}_{r_{1}rrr_{2}}^{(+)}=0\,.  \label{trsin}
\end{equation}
So, $\tilde{T}_{r_{1},r^{\prime },r,r_{2}}^{(+)}$ annihilates the singlet
state. Furthermore,
\begin{equation}
R_{r_{1}r_{2}r_{3}r_{4}}(\omega;q;l,l_{1})=
2f_{\omega}^{(0)}(q;l,l_{1})\delta _{r_{1}r_{2}} \delta _{rr^{\prime
}}/n+f_{\omega }^{(1)}(q;l,l_{1})\tilde{T}_{r_{1}r^{\prime }rr_{2}}^{(+)}\,,
\label{strr}
\end{equation}
where $f_{\omega }^{(0)}(q;l,l_{1})$ and $f_{\omega }^{(1)}(q;l,l_{1})$ are
partial waves for the vacuum channel and for the state belonging to the
adjoint representation of the $SU(N_{c})$ group, respectively.

Using expression (\ref{prleq}) one can derive, that the partial
waves $ f_{\omega }^{(s)}(q;l,l_{1})$ with $s=0,1$ obey the
BFKL-like equation
\begin{eqnarray}
&&[\omega +\alpha^{\prime }l^{2} +\alpha ^{\prime
}(q-l)^{2}-\omega_{1}(-l^{2}) -\omega _{1}(-(q-l)^{2})]f_{\omega
}^{(s)}(\omega ;q;l,l_{1})  \nonumber \\
&=&\widehat{I}(q;l,l_{1})+\frac{g^{2}N_{c}c_{s}}{(2\pi )^{D-1}} \int
\widehat{ I}(q;l,l^{\prime}) f_{\omega }^{(s)}(q;l^{\prime
},l_{1})d^{D-2}l^{\prime }\,,  \label{equ}
\end{eqnarray}
where $c_{0}=1$ and $c_{1}=1/2$. The integral kernel $\widehat{I}(q;l,l_{1})$
is calculated in the next Section.

\section{Integral kernel}

With the use of (\ref{iker}) one can verify, that the massless state
contribution to $I$ (\ref{sit2u}) is given by
\begin{eqnarray}
S_{\omega }^{(0)}(q;l,l^{\prime })/\alpha ^{\prime }=[\alpha ^{\prime
}(l-l^{\prime })^{2}]^{\omega +\alpha ^{\prime }l^{2}+\alpha ^{\prime
}(q-l)^{2}}\int dfdye^{-(y+f)}f^{\alpha ^{\prime }(l^{\prime
})^{2}-1}y^{\alpha ^{\prime }(q-l^{\prime })^{2}-1}  \nonumber \\
\times \lbrack f-\alpha ^{\prime }(l-l^{\prime })^{2}-i\epsilon ]^{-\alpha
^{\prime }l^{2}}[y-\alpha ^{\prime }(l-l^{\prime })^{2}-i\epsilon ]^{-\alpha
^{\prime }(q-l)^{2}}\Biggl[-\alpha ^{\prime }q^{2}+\alpha ^{\prime
}(l^{\prime })^{2}+\alpha ^{\prime }(q-l^{\prime })^{2}  \nonumber \\
-\frac{\alpha ^{\prime }l^{2}y}{[f-\alpha ^{\prime }(l-l^{\prime
})^{2}-i\epsilon ]}-\frac{\alpha ^{\prime }(q-l)^{2}f}{[y-\alpha ^{\prime
}(l-l^{\prime })^{2}-i\epsilon ]}-(f+y)\Biggl]\,.  \label{nsker}
\end{eqnarray}

The integral in the above expression can be written in terms of
the Whittaker function $ W_{\lambda ,\mu }(-\kappa ^{2}-i\epsilon
)$ defined as follows
\begin{equation}
J(a,b;z)\equiv \int_{0}^{\infty
}e^{-t}t^{a}(z+t)^{b}dt=z^{(a+b)/2}e^{z/2}\Gamma
(b+1)W_{(b-a)/2,(a-b+1)/2}(z)\,  \label{witt}
\end{equation}
which has the following representation
\begin{equation}
J(a,b;z)=z^{a+b+1}\Gamma (a+1)\frac{\Gamma (-a-b-1)}{\Gamma (-b)}\Phi
(a+1,a+b+2,z)+\Gamma (a+b+1)\Phi (-b,-a-b,z)  \label{inhyp}
\end{equation}
as a linear combination of the confluent hypergeometric function
$\Phi (a,b,z)$
\begin{equation}
\Phi (a,b,z)=1+\frac{a}{b}z+\frac{a(a+1)}{2b(b+1)}z^{2}+\dots \,.
\label{hgeom}
\end{equation}
Indeed, we obtain for $\widetilde{I}_0$ (\ref{iker})
\begin{eqnarray}
(\kappa _{0}^{2})^{\omega +2}\widetilde{I}_{0}(q ;l,l^{\prime })=[\alpha
^{\prime }(l-l^{\prime })^{2}]^{\omega +\alpha ^{\prime }l^{2}+\alpha
^{\prime }(q-l)^{2}}\Biggl[[-\alpha ^{\prime }q^{2}+\alpha ^{\prime
}(l^{\prime })^{2}+\alpha ^{\prime }(q-l^{\prime })^{2}]  \nonumber \\
\times J(\alpha ^{\prime }(l^{\prime })^{2}-1,-\alpha ^{\prime
}l^{2};-\alpha ^{\prime }(l-l^{\prime })^{2})J(\alpha ^{\prime }(q-l^{\prime
})^{2}-1,-\alpha ^{\prime }(q-l)^{2};-\alpha ^{\prime }(l-l^{\prime })^{2})
\nonumber \\
-\alpha ^{\prime }l^{2}J(\alpha ^{\prime }(l^{\prime })^{2}-1,-\alpha
^{\prime }l^{2}-1;-\alpha ^{\prime }(l-l^{\prime })^{2})J(\alpha ^{\prime
}(q-l^{\prime })^{2},-\alpha ^{\prime }(q-l)^{2};-\alpha ^{\prime
}(l-l^{\prime })^{2})  \nonumber \\
-\alpha ^{\prime }(q-l)^{2}J(\alpha ^{\prime }(l^{\prime })^{2},-\alpha
^{\prime }l^{2};-\alpha ^{\prime }(l-l^{\prime })^{2})J(\alpha ^{\prime
}(q-l^{\prime })^{2}-1,-\alpha ^{\prime }(q-l)^{2}-1;-\alpha ^{\prime
}(l-l^{\prime })^{2})  \nonumber \\
-J(\alpha ^{\prime }(l^{\prime })^{2},-\alpha ^{\prime }l^{2};-\alpha
^{\prime }(l-l^{\prime })^{2})J(\alpha ^{\prime }(q-l^{\prime
})^{2}-1,-\alpha ^{\prime }(q-l)^{2};-\alpha ^{\prime }(l-l^{\prime })^{2})
\nonumber \\
-J(\alpha ^{\prime }(l^{\prime })^{2}-1,-\alpha ^{\prime }l^{2};-\alpha
^{\prime }(l-l^{\prime })^{2})J(\alpha ^{\prime }(q-l^{\prime })^{2},-\alpha
^{\prime }(q-l)^{2};-\alpha ^{\prime }(l-l^{\prime })^{2})\Biggl]\,.
\label{nske1r}
\end{eqnarray}

To calculate the massive state contribution $S(q;l,l^{\prime })$
to the kernel (\ref{diisc}) it is convenient to change the
integration variables $f\to\kappa^{2}f,\, y\to\kappa^{2}y$ in
expression (\ref{sit2u}). As a result, the factor being a power of
$\kappa_{m}^{2}$ is extracted from the integral
\begin{eqnarray}
&&(\kappa ^{2})^{-\omega -\alpha ^{\prime }l^{2}+\alpha ^{\prime
}(q-l)^{2}-\alpha ^{\prime }(l^{\prime })^{2}-\alpha ^{\prime }(q-l^{\prime
})^{2}+2}I(t_{5678},\kappa
^{2},t_{3478},t_{12},t_{78},t_{34},t_{56})\,\alpha ^{\prime }  \nonumber \\
&=&\frac{1}{\Gamma (\omega +\alpha ^{\prime }l^{2}+\alpha ^{\prime
}(q-l)^{2})}\int dfdydvv^{\omega +\alpha ^{\prime }l^{2}+\alpha ^{\prime
}(q-l)^{2}-1}e^{-\kappa ^{2}(y+f+v)}z^{-\alpha ^{\prime }k^{2}-1}  \nonumber
\\
&&\times \Biggl[V_{1}(z,f,y;q,l,l^{\prime })+\frac{1}{v}[\omega
+\alpha ^{\prime }l^{2}+\alpha ^{\prime
}(q-l)^{2}-1]V_{2}(z,f,y;q,l,l^{\prime }) \Biggl]\,,
\label{sdisc}
\end{eqnarray}
where $V_{i}(z,f,y;q,l,l^{\prime })$ does not depend on $\kappa ^{2}$
\begin{eqnarray}
V_{1}(z,f,y;q,l,l^{\prime })=f^{\alpha ^{\prime }(l^{\prime
})^{2}-1}y^{\alpha ^{\prime }(q-l^{\prime })^{2}-1}(1-z)^{\alpha ^{\prime
}q^{2}-\alpha ^{\prime }(l^{\prime })^{2}-\alpha ^{\prime }(q-l^{\prime
})^{2}}q_{1}^{-\alpha ^{\prime }l^{2}}q_{2}^{-\alpha ^{\prime }(q-l)^{2}}
\nonumber \\
\times \Biggl[[-\alpha ^{\prime }q^{2}+\alpha ^{\prime }(l^{\prime
})^{2}+\alpha ^{\prime }(q-l^{\prime })^{2}+\alpha ^{\prime }l^{2}+\alpha
^{\prime }(q-l)^{2}-\alpha ^{\prime }(l-l^{\prime })^{2}](1-z)-\alpha
^{\prime }l^{2}  \nonumber \\
-\alpha ^{\prime }(q-l)^{2}-\alpha ^{\prime }l^{2}\frac{y}{q_{1}}
(1-z)-\alpha ^{\prime }(q-l)^{2}\frac{f}{q_{2}}(1-z)+z\Biggl]\,,  \label{sv}
\\
V_{2}(z,f,y;q,l,l^{\prime }) =f^{\alpha ^{\prime }(l^{\prime
})^{2}-1}y^{\alpha ^{\prime }(q-l^{\prime })^{2}-1}(1-z)^{\alpha ^{\prime
}q^{2}-\alpha ^{\prime }(l^{\prime })^{2}-\alpha ^{\prime }(q-l^{\prime
})^{2}}q_{1}^{-\alpha ^{\prime }l^{2}}q_{2}^{-\alpha ^{\prime }(q-l)^{2}}
\nonumber \\
\times \lbrack (1-z)-(f+y)]\,.  \label{s1v}
\end{eqnarray}
In these expressions we denoted
\begin{equation}
q_{1}=f+yz-(1-z)-i\epsilon \,,\quad q_{2}=y+fz-(1-z)-i\epsilon \,,\quad
\epsilon \rightarrow 0\,,  \label{qq2q}
\end{equation}
where $\epsilon \rightarrow +0$. In integral (\ref{sdisc}) the
residue in the pole at $k^{2}=m$ depends on $m$ only through the
exponent $\exp [-m(f+y+v)]$ multiplied by the derivative
$\partial_{z}^{m-1} V_{(i)}(z,f,y;q,l,l^{\prime })/(m-1)!$
calculated at $z=0$. Therefore after summing the residues over $m$
we obtain
\begin{eqnarray}
\sum_{m=1}^{\infty}e^{-m(f+y+v)}
\partial_{z}^{m-1}V_{(i)}(z,f,y;q,l,l^{\prime })/(m-1)!\biggl|_{z=0}
&=&V_{i}(e^{-(f+y+v)},f,y;q,l,l^{\prime })  \nonumber \\
&&-V_{i}(0,f,y;q,l,l^{\prime }) \,.  \label{suvi}
\end{eqnarray}
Thus, the quantity $S_{\omega }(q;l,l^{\prime })$ can be written as follows
\begin{eqnarray}
S_{\omega }(q;l,l^{\prime })/\alpha ^{\prime }=\frac{1}{\Gamma
(\omega +\alpha ^{\prime }l^{2}+\alpha ^{\prime }(q-l)^{2})}\int
dfdydvv^{\omega +\alpha ^{\prime }l^{2}+\alpha ^{\prime
}(q-l)^{2}-1}\times  \nonumber \\ \times e^{-\alpha ^{\prime
}(l-l^{\prime })^{2}(y+f+v)}\Biggl[
[V_{1}(e^{-(f+y+v)},f,y;q,l,l^{\prime })-V_{1}(0,f,y;q,l,l^{\prime
})] \nonumber \\ +\frac{1}{v}[\omega +\alpha ^{\prime
}l^{2}+\alpha ^{\prime }(q-l)^{2}-1][
V_{2}(e^{-(f+y+v)},f,y;q,l,l^{\prime })-V_{2}(0,f,y;q,l,l^{\prime
})]\Biggl]. \label{sdis1c}
\end{eqnarray}

Integrating the last term in (\ref{sdis1c}) over $v$ by parts we obtain
\begin{eqnarray}
S_{\omega}(q;l,l^{\prime })/\alpha^{\prime } =\frac{1}{\Gamma (\omega
+\alpha ^{\prime }l^{2}+ \alpha^{\prime }(q-l)^{2})}\int dfdydvv^{\omega
+\alpha^{\prime }l^{2}+\alpha ^{\prime }(q-l)^{2}-1} e^{-\alpha ^{\prime
}(l-l^{\prime })^{2}(y+f+v)}  \nonumber \\
\times \Biggl[f^{\alpha^{\prime}(l^{\prime })^{2}-1}y^{\alpha^{\prime
}(q-l^{\prime })^{2}-1}(1-e^{-f-y-v})^{\alpha ^{\prime }q^{2}
-\alpha^{\prime }(l^{\prime })^{2} -\alpha ^{\prime }(q-l^{\prime })^{2}}
\nonumber \\
\times [f+ye^{-f-y-v}-(1-e^{-f-y-v})-i\epsilon ]^{-\alpha
^{\prime}l^{2}}[y+fe^{-f-y-v}-(1-e^{-f-y-v})-i\epsilon ]^{-\alpha^{\prime
}(q-l)^{2}}  \nonumber \\
\times \mathcal{B}(e^{-f-y-v},f,y;q,l,l^{\prime })-V_{1}(0,f,y;q,l,l^{\prime
})-\alpha ^{\prime }(l-l^{\prime })^{2}V_{2}(0,f,y;q,l,l^{\prime})\Biggl]\,,
\label{s2vp}
\end{eqnarray}
where
\begin{eqnarray}
\mathcal{B}(z,f,y;q,l,l^{\prime }) =-\alpha^{\prime }(l-l^{\prime
})^{2}(f+y)+\alpha ^{\prime }l^{2}y+\alpha^{\prime }(q-l)^{2}f
+[\alpha^{\prime }(q-l^{\prime })^{2}  \nonumber \\
+\alpha ^{\prime }(l^{\prime })^{2}-\alpha ^{\prime }q^{2}] \biggl[1-\frac{
z(f+y)}{1-z}\biggl]- \frac{\alpha^{\prime}l^{2}fy(1-z)}{f+yz-(1-z)-i\epsilon
}- \frac{\alpha^{\prime }(q-l)^{2}fy(1-z)}{y+fz-(1-z)-i\epsilon }\,.
\label{calbp}
\end{eqnarray}

Really the leading contribution to (\ref{s2vp}) arises from the
region of small integration variables. In particular, it results
in a pole at $ \omega =\alpha ^{\prime }q^{2}$, as well as in a
Mandelstam cut term. To find the main part of (\ref{s2vp}) we cut
from below the integration variables in eq. (\ref{s2vp}) by a
parameter $\lambda \ll 1$. Then from eq. (\ref{calbp}), one can
obtain, that the leading contribution to $S_{\omega }$ is given by
the expression
\begin{eqnarray}
S_{\omega }(q;l,l^{\prime })/\alpha ^{\prime } \rightarrow
-e^{\alpha ^{\prime }l^{2}+\alpha ^{\prime
}(q-l)^{2}}\frac{[\alpha ^{\prime }q^{2}-\alpha ^{\prime
}(l^{\prime })^{2}-\alpha ^{\prime }(q-l^{\prime })^{2}]}{\Gamma
(\omega +\alpha ^{\prime }l^{2}+\alpha ^{\prime
}(q-l)^{2})}\Biggl[\int_{0}^{\lambda }dfdydvv^{\omega }f^{\alpha
^{\prime }(l^{\prime })^{2}-1}  \nonumber \\ \times y^{\alpha
^{\prime }(q-l^{\prime })^{2}-1}(v+f+y)^{\alpha ^{\prime
}q^{2}-\alpha ^{\prime }(l^{\prime })^{2}-\alpha ^{\prime
}(q-l^{\prime })^{2}-1}-\frac{1}{\alpha ^{\prime }(l^{\prime
})^{2}\alpha ^{\prime }(q-l^{\prime })^{2}(\omega +\alpha ^{\prime
}l^{2}+\alpha ^{\prime }(q-l)^{2})}\Biggl]\,,  \label{spol1e}
\end{eqnarray}
where the pole term arises from two last terms in (\ref{calbp}).

To calculate the integral (\ref{spol1e}) the integration region is
divided into 6 domains: $v>f>y$, $v>y>f$, $f>v>y$, $f>y>v$,
$y>v>f$ and $y>f>v$. In the first domain we replace initially
$y\rightarrow fy$ and then $ f\rightarrow vf$. As a result, the
$v$-dependence of the integrand turns out to be $v^{\omega +\alpha
^{\prime }q^{2}}$. Integrating it over $v$ we observe the pole at
$\omega =-\alpha ^{\prime }q^{2}$. The similar procedure is
performed in each of the rest domains. As far as, in addition, the
expression  $\omega +\alpha ^{\prime }l^{2}+\alpha ^{\prime
}(q-l)^{2}$ is implied to be small $\sim 1/\ln s$, the factor
$\exp [\alpha ^{\prime }l^{2}+\alpha ^{\prime }(q-l)^{2}]$ in
(\ref{spol1e}) should be replaced by unity. For the same reason
$\Gamma (\omega +\alpha ^{\prime }l^{2}+\alpha ^{\prime
}(q-l)^{2})=\Gamma (1+\omega +\alpha ^{\prime }l^{2}+\alpha
^{\prime }(q-l)^{2})/(\omega +\alpha ^{\prime }l^{2}+\alpha
^{\prime }(q-l)^{2})\approx 1/(\omega +\alpha ^{\prime
}l^{2}+\alpha ^{\prime }(q-l)^{2})$. Using these simplifications
expression (\ref{spol1e}) is given below
\begin{eqnarray}
S_{\omega }(q;l,l^{\prime })/\alpha ^{\prime } \approx \frac{1}{(\omega
+\alpha ^{\prime }q^{2})}[\alpha ^{\prime }q^{2}-\alpha ^{\prime
}l^{2}-\alpha ^{\prime }(q-l)^{2}][\alpha ^{\prime }q^{2}-\alpha ^{\prime
}(l^{\prime })^{2}-\alpha ^{\prime }(q-l^{\prime })^{2}]  \nonumber \\
\times \Biggl[\tilde{F}(\alpha ^{\prime }(l^{\prime })^{2},\alpha ^{\prime
}(q-l^{\prime })^{2};\alpha ^{\prime }q^{2}-\alpha ^{\prime }(l^{\prime
})^{2}-\alpha ^{\prime }(q-l^{\prime })^{2})  \nonumber \\
+\tilde{F}(1-\alpha ^{\prime }q^{2},\alpha ^{\prime }(q-l^{\prime
})^{2};\alpha ^{\prime }q^{2}-\alpha ^{\prime }(l^{\prime })^{2}-\alpha
^{\prime }(q-l^{\prime })^{2})  \nonumber \\
+\tilde{F}(1-\alpha ^{\prime }q^{2},\alpha ^{\prime }(l^{\prime
})^{2};\alpha ^{\prime }q^{2}-\alpha ^{\prime }(l^{\prime })^{2}-\alpha
^{\prime }(q-l^{\prime })^{2})\Biggl]\,,  \label{s1pole}
\end{eqnarray}
where
\begin{eqnarray}
\tilde{F}(a,b,c)
=\int_{0}^{1}df\int_{0}^{1}dyf^{a+b-1}(1+f+fy)^{c-1}[y^{b-1}+y^{a-1}]=
\nonumber \\
=\sum_{n,m=0}^{\infty }\frac{\Gamma (c)}{\Gamma (c-m-n)\Gamma (m+1)\Gamma
(n+1)(a+m)(b+n)}\,.  \label{tfpol}
\end{eqnarray}
One can verify that at small momenta $\alpha ^{\prime }q^{2}\sim \alpha
^{\prime }l^{2}\sim \alpha ^{\prime }l^{\prime }\ll 1$ the first term
in the large square brackets of eq. (\ref{s1pole}) gives the main
contribution \begin{equation} S_{\omega }^{sing}(q;l,l^{\prime
})=\frac{[\alpha ^{\prime }q^{2}-\alpha ^{\prime }l^{2}-\alpha ^{\prime
}(q-l)^{2}][\alpha ^{\prime }q^{2}-\alpha ^{\prime }(l^{\prime
})^{2}-\alpha ^{\prime }(q-l^{\prime })^{2}]}{\alpha ^{\prime }(\omega
+\alpha ^{\prime }q^{2})(l^{\prime })^{2}(q-l^{\prime })^{2}}\,.
\label{s3pole} \end{equation}
Expressions (\ref{s1pole}) and (\ref{s3pole}) are correct in a
neighbourhood of the pole and of zeros of the numerator with the
deviations being $\sim 1/\ln s\sim N_{c}g^{2}$. As far as the
numerator does not vanish at $\omega +\alpha ^{\prime
}l^{2}+\alpha ^{\prime }(q-l)^{2}=\omega +\alpha ^{\prime
}l^{\prime 2}+\alpha ^{\prime }(q-l^{\prime })^{2}=0$, it
contributes to both the Mandelstam cuts and the pole at $\omega
=-\alpha ^{\prime }q^{2}$.

The pole at $\omega =-\alpha ^{\prime}q^2$ corresponds to the soft
Pomeron which exists already in the Born expression (\ref{scam})
for the elastic amplitude. Relatively large masses
$1\ll \alpha^{\prime}M^2\ll \alpha^{ \prime}s$ of produced resonances
contribute to this pole. Therefore in the box diagram Fig.1a we
expect a pole of the second order from the integration over large
masses of two intermediate $s$-channel resonances. This second
order pole appears as a result of the perturbative expansion of
the Pomeron Regge pole over the one-loop correction
$\omega_1(t)\sim g^2$. In the two-loop approximation,
corresponding to Fig.1b, we should have the third order pole with
the residue proportional to $\omega _1^2(t)$. In this diagram,
apart from the pole (\ref{s3pole}) there is a product of two pole
singularities $1/(\omega +\alpha ^{\prime}q^2)$ from the
integration over the large masses of resonances produced in the
fragmentation regions of initial particles. In the multi-Regge
kinematics one obtains also the poles $ 1/(\omega +\alpha
^{\prime}l^2+\alpha ^{\prime}(q-l)^2)$ and $1/(\omega +\alpha
^{\prime}l^{\prime 2}+\alpha ^{\prime}(q-l^{\prime})^2)$ leading
after the integration over $l$ and $l^{\prime}$ to the Mandelstam
cuts (we put here $l=k_{\perp}$ and
$l^{\prime}=k^{\prime}_{\perp}$). Because the residue of the pole
(\ref{s3pole}) in the BFKL kernel is small due to the smallness of
the expressions in the square brackets, it cancels approximately
the neighboring poles depending on $l$ and $l^{\prime}$ and
therefore one can attempt to extract from the contribution for
Fig.1b the third order pole being the second order term in the
expansion of the soft Pomeron pole in $\omega _1(t)$.

Indeed, let us present the numerator of the pole in eq. (\ref{s3pole}) in
the form
\[
[ \alpha^{\prime}q^2- \alpha ^{\prime }l^{2}-\alpha ^{\prime }(q-l)^{2} ]
[\alpha ^{\prime}q^{2}-\alpha ^{\prime}(l^{\prime })^{2} -\alpha
^{\prime}(q-l^{\prime})^{2}]= [\omega +\alpha ^{\prime }l^{2}+\alpha
^{\prime }(q-l)^{2}] [\omega +\alpha ^{\prime}(l^{\prime })^{2} +\alpha
^{\prime}(q-l^{\prime})^{2}]
\]
\begin{equation}
- [\omega +\alpha ^{\prime }q^{2}] \left[\omega -\alpha ^{\prime }q^{2}+
\alpha ^{\prime }l^{2}+\alpha ^{\prime }(q-l)^{2}+ \alpha ^{\prime
}l^{\prime 2}+\alpha ^{\prime }(q-l^{\prime})^{2} \right]\,.  \label{expnum}
\end{equation}
Then the second term in the right hand side of this equality,
killing the pole $1/(\omega +\alpha ^{\prime}q^2)$ in
(\ref{s3pole}), contributes only to the Mandelstam cuts. As for
the first term in (\ref{expnum}), it corresponds to the second
term of expansion for the soft Pomeron pole. Indeed, its numerator
cancels the neighboring propagators for Mandelstam cuts. Therefore
the corresponding integrals over the relative rapidities $ \ln
s_{12}$ and $\ln s_{23}$ are convergent for the large invariants
$s_{12}$ and $s_{23}$. So, we should calculate these integrals
exactly without simplifications corresponding to the multi-Regge
kinematics. It is plausible, that as a result of such calculation
the pole in expression (\ref {s3pole}) together with additional
poles $1/(\omega +\alpha ^{\prime}q^2)$ from two impact factors
would reproduce the total one-loop correction $\sim \omega _1^2$
from Fig.1b in the second order expansion of the Pomeron pole.

The first term in (\ref{expnum}) is important also for a
cancellation of the singularities in (\ref{s3pole}) at
$(l,l^{\prime})\to0$ and $ (l,l^{\prime})\to q$ leading to a
convergence of the corresponding integrals over the multi-Regge
region. In addition, it has a non-trivial funcional dependence
containing both poles and cuts in $\omega$. So, for the
investigation of the BFKL equation in the $D=4$ model we use the 
whole expression
(\ref{s3pole}) without neglecting the soft Pomeron pole.
Simultaneously, we add a piece from
the non-multi-Regge kinematics.

For a general case of the ladder Fig.1 one can perform a decompositions
similar to (\ref{expnum}) for each kernel. The contributions 
appearing from the first
terms in the right hand sides of (\ref{expnum}) correspond to the particles
produced in a non-multi-Regge kinematics. The form of production
amplitudes in this region can not be extracted from our above results. Probably
this contribution corresponds to a geometric progression appearing from an
expansion of the soft Pomeron pole in $\omega _1(t)$.

Presumably one can represent the partial wave as follows $f_\omega(-q^2)$ as
\begin{equation}
f_\omega(-q^2)=f_\omega^{(p)}(-q^2)+f_\omega^{mr}(-q^2) \,,  \label{regam}
\end{equation}
where the first term corresponds to the soft Pomeron contribution in the
form of the geometrical progression and the term $f_\omega^{mr}(-q^2)$
results from the multi-Regge kinematics. In principle there can be a more
complicated situation with an interference between the Regge pole and cut.

\section{BFKL equation in the $D=4$ string model}

It follows from the above discussion that the singularities of the
$t$=channel partial waves arise from the region where
$\omega+\alpha^{
\prime}l^{\prime \,2}+\alpha^{\prime}(q-l^{\prime})^2 \sim 1/\ln s$.
For $D>4$ after the integration over the region
$\alpha ^{\prime}l^{\prime \,2}$ the corresponding
contribution is suppressed by powers of logarithms $\sim (\ln
s)^{-(D-4)/2}$ for each produced particle, which leads to a
possibility to find the solution of the BFKL equation as a series in this
small parameter. In principle, it is not excluded that for very
large energies the number of produced particles grows so rapidly,
that the averaged pair energies $ s_{k,k+1}$ for these particles
are not so large to justify the saddle-point method of
calculations of the integrals. In this case the BFKL equation which sums
contributions from the multi-Regge kinematics could have
non-trivial solutions even for $D>4$. Here, however, we restrict
ourselves to the $D=4$ case hoping to return to the discussion of
other values of $D$ in future publications. Moreover, only the
amplitude with vacuum quantum numbers in the crossing channel is
considered.

At $D=4$ the BFKL equation has a non-trivial solution in
terms of the function
$f_{\omega}^{(0)}(q;l)$ defined by the relation
\begin{equation}
f_{\omega}^{(0)}(q;l)=\int f_{\omega}^{(0)}(q;l,l_{1}) \Phi(q;l_1)d^2l_1\,,
\label{deff}
\end{equation}
where $\Phi(q;l_1)$ is an impact factor. Generally the solution contains
contributions from non-planar diagrams.

One loop
correction $ \omega_1(t)$ to the gluon trajectory for $D=4$
(\ref{omegas}) has the form
\begin{equation}
\omega_1(t)=-\frac{g^2N_c}{8\pi^2}\ln(q^2/\lambda^2)+\omega_1^{(m)}(q^2)
\label{tra1j}
\end{equation}
where the first contribution corresponds to massless states in the
$t$ -channel and the second term non-singular at $q^2=0$ appears
from the massive string excitations (cf. expression (\ref{traj})
in QCD).

To begin with, let us discuss the region of small $t$, where
$\alpha^{\prime }q^{2}\sim g^{2}N_{c}$. In this case for $D=4$
the small gluon virtualities $\alpha^{\prime }l^{2}\sim
g^{2}N_{c}\sim \alpha ^{\prime }(l^{\prime })^{2}\sim g^{2}N_{c}$
are important. For such momenta $l$ and $l^{\prime }$ the pole
contribution (\ref{s3pole}) dominates in $S_{\omega}(q:l,l^{\prime
})$ and the singularities of the $t$-channel partial wave are
situated for small $g^2$ at $\omega \sim g^{2}$. Because the
infra-red divergencies in the integral kernel are cancelled
between the contribution from the real particle emission and
one-loop correction to the Regge trajectories, the factor
$[(l-l^{\prime })^{2}]^{\omega
+\alpha^{\prime}l^2+\alpha^{\prime}(q-l)^2}$ in the right hand
side of eq. (\ref{nske1r}) can be omitted. Hence, from expression
(\ref{nsker}) we obtain the following contribution to the kernel
(\ref{sit2u}) corresponding to the massless state production

\begin{equation}
S_{\omega }^{(0)}(q:l,l^{\prime })=-\frac{q^{2}}{(l^{\prime
})^{2}(q-l^{\prime })^{2}}+\frac{l^{2}}{(l-l^{\prime })^{2}(l^{\prime })^{2}}
+\frac{(q-l)^{2}}{(l-l^{\prime })^{2}(q-l^{\prime })^{2}}\,.  \label{li1m}
\end{equation}

Expression (\ref{li1m}) coincides with the corresponding result
\cite{klf} in QCD. The massive state term in  (\ref{tra1j}) is
expected to vanish at $t\rightarrow 0$. So, the radiative
correction to the gluon trajectory for small momentum transfers
$l$ and $q-l$ also can be approximated by the QCD expression
(\ref{traj}). As a result, the BFKL equation (\ref{equ}) for the
vacuum channel at $D=4$ and $\alpha 'q^2 \sim g^2N_c$ is
drastically simplified
\begin{eqnarray}
[\omega +\alpha^{\prime }l^{2}+\alpha^{\prime }(q-l)^{2}]
f_{\omega}^{(0)}(q;l)=\Phi(q;l)+ \frac{g^{2}N_{c}}{8\pi^{3}}\int
\Biggl\{ S_{\omega}^{(0)}(q:l,l^{\prime }) f_{\omega
}^{(0)}(q;,l^{\prime }) \nonumber \\ -\frac{1}{(l-l^{\prime
})^{2}}\Biggl[\frac{l^{2}}{[(l-l^{\prime })^{2}+(l^{\prime
})^{2}]}+\frac{(q-l)^{2}}{[(l-l^{\prime })^{2}+(q-l^{\prime
})^{2}]} \Biggl]f_{\omega }^{(0)}(q;,l)\Biggl\} d^{2}l^{\prime}
\nonumber \\ +\frac{g^{2}N_{c}}{8\pi^{3}}\int
\frac{[\alpha^{\prime}q^2 -\alpha^{\prime }l^{2}-\alpha ^{\prime
}(q-l)^{2}][q^{2}-(l^{\prime })^{2}-(q-l^{\prime })^{2}]}{(\omega
+\alpha ^{\prime }q^{2})(l^{\prime })^{2}(q-l^{\prime })^{2}
}f_{\omega}^{(0)}(q;,l^{\prime})d^{2}l^{\prime}\,,  \label{equsm}
\end{eqnarray}
where the contribution from Fig.1a is also taken into account.

In eq. (\ref{equsm}) we performed a relevant subtraction of the
Regge trajectory contribution to obtain the integral
kernel in the BFKL
form (cf. \cite{lip}), and the expression for $S_{\omega
}^{(0)}(q:l,l^{\prime })$ is given in (\ref{li1m}). Equation
(\ref{equsm}) differs from the BFKL equation in QCD only by 
terms linear in squared gluon momenta at its left hand side and by
an additional pole term $\sim 1/(\omega+\alpha^{\prime }q^{2})$ in
the kernel. The terms $\sim l^2$ and $\sim (q-l)^2$ improve the
properties of its kernel at $ l\rightarrow\infty $. As a result,
unlike the QCD case in LLA, eq.(\ref{equsm}) is expected to have a
discrete spectrum at nonzero values of $q^2$.

Comparing the large-$l$ behaviour of the left and right hand sides
of eq.(\ref{equsm}) we conclude, that the linear terms in the
gluon trajectories in eq. (\ref{equsm}) lead to a constant
behaviour of $f_{\omega}^{(0)}(q;l)$ at $l\to\infty$. As a result,
the integral
\begin{equation}
h_\omega(q)= \int \frac{[q^{2}-(l^{\prime })^{2}-(q-l^{\prime })^{2}]}{
(l^{\prime })^{2}(q-l^{\prime })^{2} }f_{\omega}^{(0)}(q;,l^{
\prime})d^{2}l^{\prime}  \label{homeg}
\end{equation}
in the last term on its right hand side is divergent. Taking into
account, that this term plays role of an additional inhomogenious
contribution to eq.(\ref{equsm}) we present
$f_{\omega}^{(0)}(q;l)$ in the form
\begin{equation}
f_{\omega}^{(0)}(q;l)= \frac{g^{2}N_{c}[\alpha^{\prime}q^2
-\alpha^{\prime }l^{2}-\alpha ^{\prime }(q-l)^{2}]}{8\pi^{3}
(\omega+\alpha^{\prime }q^{2}) [\omega +\alpha^{\prime
}l^{2}+\alpha^{\prime}(q-l)^{2}]} h_\omega(q)+ \frac{ g^{2}N_{c}
h_\omega(q)}{8\pi^{3}(\omega+\alpha^{\prime }q^{2})} \hat
f_{\omega}^{(0)}(q;l)+\tilde f_{\omega}^{(0)}(q;l)\,,
\label{solsm}
\end{equation}
where $h_\omega (q)$ is given by (\ref{homeg}), while $\hat
f_{\omega}^{(0)}(q;l)$ and $\tilde f_{\omega}^{(0)}(q;l)$ are determined
from the equation (below $F_{\omega}^{(0)}(q;l)$ is denoted either by $\hat
f_{\omega}^{(0)}(q;l)$ or $\tilde f_{\omega}^{(0)}(q;l)$)
\begin{eqnarray}
[\omega +\alpha^{\prime }l^{2}+\alpha^{\prime }(q-l)^{2}]
F_{\omega}^{(0)}(q;l)=\tilde\Phi(q;l)+
\frac{g^{2}N_{c}}{8\pi^{3}}\int
\Biggl\{S_{\omega}^{(0)}(q:l,l^{\prime}) F_{\omega
}^{(0)}(q;,l^{\prime}) \nonumber \\ -\frac{1}{(l-l^{\prime
})^{2}}\Biggl[\frac{l^{2}}{[(l-l^{\prime })^{2}+(l^{\prime
})^{2}]}+\frac{(q-l)^{2}}{[(l-l^{\prime })^{2}+(q-l^{\prime
})^{2}]} \Biggl]F_{\omega }^{(0)}(q;,l)\Biggl\} d^{2}l^{\prime}\,.
\label{soltil}
\end{eqnarray}
Here for $F_{\omega}^{(0)}(q;l)=\hat f_{\omega}^{(0)}(q;l)$ we have
\begin{eqnarray}
\tilde\Phi(q;l)= \int S_{\omega}^{(0)}(q:l,l^{\prime }) \frac{
[\alpha^{\prime}q^2 -\alpha^{\prime }(l^{\prime})^{2}-\alpha
^{\prime }(q-l^{\prime})^{2}]}{\omega
+\alpha^{\prime}(l^{\prime})^{2}+\alpha^{\prime}(q-l^{\prime})^{2}}
d^2l'- \nonumber \\ \int \frac{d^2l' }{(l-l^{\prime })^{2}}
\Biggl[\frac{l^{2}}{[(l-l^{\prime })^{2}+(l^{\prime })^{2}]}+
\frac{ (q-l)^{2}}{[(l-l^{\prime })^{2}+(q-l^{\prime })^{2}]}
\Biggl] \frac{ [\alpha^{\prime}q^2 -\alpha^{\prime } l^{2}-\alpha
^{\prime }(q-l)^{2}]}{ \omega +\alpha^{\prime }l^{2}+
\alpha^{\prime}(q-l)^{2}}  \,, \label{psoltil}
\end{eqnarray}
and for $F_{\omega}^{(0)}(q;l)=\tilde f_{\omega}^{(0)}(q;l)$,
\begin{equation}
\tilde\Phi(q;l)=\Phi(q;l) \,.  \label{psolil}
\end{equation}

Using (\ref{solsm}) and (\ref{homeg}) one obtains $h_\omega(q)$ as
the solution of a linear equation
\begin{equation}
h_\omega(q)=\frac{
\omega+\alpha^{\prime}q^2}{\omega+\alpha^{\prime}q^2
-\tilde\beta(\omega,q^2)}\,\int
\frac{[q^2-l^2-(q-l)^2]}{l^2(q-l)^2} \tilde
f_{\omega}^{(0)}(q;,l)d^2l \,,  \label{home1g}
\end{equation}
where
\begin{eqnarray}
\tilde\beta(\omega,q^2)=
\frac{g^2N_c}{8\pi^3}\Biggl[\tilde\beta_0+ \biggl[
\int_{l^2<\Lambda^2} \frac{\alpha^{\prime}[q^2-l^2-(q-l)^2]^2}{
[\omega +\alpha^{\prime}l^2+\alpha^{\prime}(q-l)^2]l^2
(q-l)^2}d^2l-\pi\ln\Lambda^2 \biggl]_{\Lambda^2\to\infty}
\nonumber \\ +\int \frac{[q^2-l^2-(q-l)^2]}{l^2(q-l)^2} \hat
f_{\omega}^{(0)}(q;,l)d^2l \Biggl]\,.  \label{home2g}
\end{eqnarray}

We subtracted the logarithmic divergency from the second term in
the brackets assuming that subtraction term is added to the
quantity $\tilde\beta_0$ determined by the integration region $
\alpha^{\prime}k^2\sim1$. So, $\beta_0$ depends also on the
non-multi-Regge configurations, leading to the renormalisation
$\omega _1$ of the soft Pomeron Regge trajectory. This conclusion
follows from expression (\ref{bconf}) for the production cross-section,
where the kernel dependence from the cut-off $s_0$ is essential,
and from our discussion of eq. (\ref{expnum}). It is natural to
expect that $\tilde\beta_0\sim1$. So, the solution of eq. (\ref{equsm})
depends on the additional parameter $\tilde\beta_0$. The equation
$\omega+\alpha^{\prime}q^2=\tilde\beta(\omega,q^2)$ allows to find
the Regge trajectories . In addition, one can conclude
from (\ref{home2g}) that $\tilde\beta(\omega,q^2)$ contains
the Mandelstam cuts in the $\omega$-plane.

In the region $g^{2}N_{c}\ll \alpha ^{\prime }q^{2}\ll 1$ the
asymptotic behaviour of the scattering amplitude is related to 
singularities of the integral $\int f_{\omega
}^{(0)}(q;,l^{\prime},l_{1})d^{2}l^{\prime }$ near
$\omega\approx-q^{2}/2$. They appear from the kinematics, in which 
the
solution of eq. (\ref{equsm}) is concentrated at $l=q/2$. Let us
introduce the new momenta $v$ and $v^{\prime }$ according to the
definition
\begin{equation}
l=q/2+v\,,\qquad l^{\prime}=q/2+v^{\prime }\,, \qquad
v^{2}\ll q^{2}\,,\quad (v^{\prime })^{2}\ll q^{2}\,.  \label{sadkin}
\end{equation}
Leaving only leading terms, eq. (\ref{nsker}) is simplified as follows
\begin{equation}
S_{\omega }^{(0)}(q:l,l^{\prime})=2\, \frac{[\alpha ^{\prime }(l-l^{\prime
})^{2}]^{\omega +\alpha^{\prime }q^{2}/2}}{(l-l^{\prime})^{2}}\,,
\label{li2m}
\end{equation}
where $l-l^{\prime }=v-v^{\prime }$. The numerator in
(\ref{li2m}) is different from unity only in the region $\alpha
^{\prime}r^2=\alpha ^{\prime}(l-l^{\prime })^{2}\leq s_0/s$. Due
to eq. (\ref{kapa}) for a massless intermediate state the value of
$r^{2}$ in the multi-Regge kinematics $s_{1},s_{2}>s_{0}\gg 1/\alpha
^{\prime }$ is restricted by the condition $r^{2}\geq
s_{0}^{2}/s$.

However, according to the generalized Gribov theorem
the gluon production amplitude for
the momenta $r^{2}\ll 1/\alpha ^{\prime } $
is also large in the quasi-elastic regions $s_{1}\leq 1/\alpha
^{\prime } $ and $s_{2}\leq 1/\alpha ^{\prime }$ and equals to the
elastic amplitude multiplied by a bremstrahlung factor (see for
example \cite{brems}). Therefore the integral over $r^{2}$ is not
bounded from below by $s_{0}^{2}/s$ being infraredly divergent.
As usually, this divergency is cancelled with the contribution
from the virtual corrections proportional to the gluon Regge
trajectories. Thus, we substitute by unity the numerator in
(\ref{li2m}) and represent the massless contribution to the gluon
trajectory correction as follows
\begin{eqnarray}
g^2N_c\ln (l^2/\lambda^2)= \int\frac{d^2l^{\prime }}{(l-l^{\prime
})^{2}}\Biggl[ \frac{l^{2}}{[(l-l^{\prime })^{2}+(l^{\prime
})^{2}]}+ \frac{(q-l)^{2}}{ [(l-l^{\prime })^{2}+ (q-l^{\prime
})^{2}]}-\frac{4v^{2}}{(v^{2}+(v^{\prime })^{2})}\Biggl] \nonumber
\\ +\int\frac{4v^2d^2l^{\prime }}{(v^{2}+(v^{\prime
})^{2})(v-v^{\prime })^{2}}\,. \label{krnlsq}
\end{eqnarray}
Here $l$ and $v$ are related according to eqs. (\ref{sadkin}).
Performing the expansion in $v$ in the right hand side of eq.
(\ref{equ}) one can write it as follows
\begin{eqnarray}
[\omega+\alpha^{\prime}q^{2}/2+2\alpha^{\prime}
v^{2}+\frac{g^{2}N_{c} }{ 4\pi ^{2}}\ln (q^{2}/64v^{2})]
f_{\omega}^{(0)}(q;v,v_{1})=\Phi(q;q/2) \nonumber \\
+\frac{g^{2}N_{c}}{8\pi ^{3}}\int \frac{2}{(v-v^{\prime
})^{2}}\Biggl[ f_{\omega }^{(0)}(q;,v^{\prime
},v_{1})-\frac{2v^{2}}{(v^{2}+(v^{\prime })^{2})}f_{\omega
}^{(0)}(q;,v,v_{1})\Biggl]d^{2}l^{\prime }\,. \label{equsms}
\end{eqnarray}
The impact factor in (\ref{equsms}) is taken at $l=q/2$ because it
is expected to be a smooth function of $l$ near $l=q/2$. For
$\alpha^{\prime }q^{2}\geq 1$ the radiative correction
(\ref{traj}) to the gluon trajectory at a small momentum transfer
should be replaced by $\omega_{1}(t)$ taken at $ t=-q^{2}/4$.
Thus, the final equation valid for both restrictions $
g^{2}N_{c}\ll \alpha^{\prime }q^{2}\ll 1$ and $\alpha^{\prime
}q^{2}\geq 1$ is obtained from (\ref{equsms}) by the substitution
$\omega\rightarrow \omega^{\prime}(q^{2})$ where
\begin{equation}
\omega^{\prime}(q^{2})=\omega+2\biggl[-
\omega_{1}^{(m)}(-q^{2}/4)+ \frac{ g^{2}N_{c}}{8\pi^{2}} \ln
(q^{2}/4 \lambda^{2}) \biggl]\,.  \label{toml}
\end{equation}
The corresponding quantities are defined in eqs. (\ref{omegas})
and (\ref {tra1j}). Note, that the infra-red divergency at
$\lambda \rightarrow 0$ in the last term is cancelled with a
similar divergency in the right hand side of eq. (\ref{krnlsq}) at
$v'\rightarrow v$.

\section{Solution of the equation at small momentum transfers}

At $q\neq 0$ the integral kernel of the BFKL equation
for the string theory at $D=4$ is non-singular at small
momenta. In this
case one can expect that for the $t$-channel partial wave the cut
at $ \omega =\omega_{0}$ disappears, and instead of a fixed
singularity of $ f_{\omega}(q^{2})$ in the $\omega$-plane there
are only Regge poles. Here we demonstrate this phenomenon in the
case of small values of $\alpha '\vec{q}^2$, where there exists an
analytic solution of the equation in the $D=4$ string theory.
The pole contribution to the kernel corresponding to the soft
Pomeron will be neglected.

In the domain of  relatively small $\vec{q}^{\,2}$
\begin{equation}
 \alpha '\vec{q}^{\,2}\ll g^2\,N_c\,.
\end{equation}
one can divide the region of possible values of $\vec{\rho}^{\,2}\sim
\vec{k}^{\,-2}$ into two subregions $ \vec{\rho}^2\sim \vec{q}^{-2}
$ and $\vec{\rho}^{\,2} \sim \alpha ' (g^2\,N_c)^{-1}$, where $\rho
=\rho _{12}$. In the first
subregion one can use the conformal (M\"{o}bius) invariance and
the eigenfunction in the mixed representation coincides with the Fourrie
transformation
in the c.m. coordinate $\vec{\rho}_0$ from the function
$E_{m,\widetilde{m}}(\vec{\rho}_1,\vec{\rho}_2;\vec{\rho}_0)$
(\ref{polyasol}). Its asymptotics at small $\vec{\rho}^2$ has the
form \cite{conf}
 \begin{equation}
 E_{m,\widetilde{m}}(\vec{q},\vec{\rho})\sim
  \rho ^m\,(\rho ^*)^{\widetilde{m}}
+e^{i\,\delta
_{m,\widetilde{m}}(\vec{q})}\,
 \rho ^{1-m}\,(\rho ^*)^{1-\widetilde{m}}\,,
 \label{smrho}
\end{equation}
where
\begin{equation}
  e^{i\,\delta
_{m,\widetilde{m}}(\vec{q})}=(-1)^n\,
\left(\frac{|q|}{4}\right)^{-4i\nu}\,
\left(\frac{q}{q^*}\right)^n\,
\frac{
\Gamma (m+\frac12)\,\Gamma (\widetilde{m}+\frac12)}{
\Gamma (-m+\frac32)\,\Gamma (-\widetilde{m}+\frac32)}\,.
\label{smrho1}
 \end{equation}

For simplicity we consider the case $n=0$, where
\begin{equation}
  e^{i\,\delta
_{m,m}(\vec{q})}= \left(\frac{|q|}{4}\right)^{-4i\nu}\, \frac{
\Gamma ^2(1+i \nu )} {\Gamma ^2(1-i \nu )} \label{phaseq}
 \end{equation}
 and the wave function for small $|\rho|$
 \begin{equation}
E_{m,m}(\vec{q},\vec{\rho})\sim |\vec{\rho}|^{1+2i\nu}+
e^{i\,\delta _{m,m}(\vec{q})}\,|\vec{\rho}|^{1-2i\nu}\,.
\label{confass}
 \end{equation}
After the Fourrie transformation to the momentum space we obtain
\begin{equation}
\Psi (\vec{q},\vec{k})=\int d^2\rho
\,e^{i\vec{\rho}\vec{k}}\,E_{m,m}(\vec{q},\vec{\rho})\sim
|k/q|^{-3-2i\nu}+
e^{i\,\delta (\nu )}\,|k/q|^{-3+2i\nu}\,,
\label{waveasmom}
\end{equation}
where
\begin{equation}
e^{i\,\delta (\nu )}=2^{4i\nu}\,
\frac{
\Gamma ^2(1+i \nu )\,\Gamma
(-\frac{1}{2}-i\nu )\,\Gamma
(\frac{3}{2}-i\nu )} {\Gamma ^2(1-i \nu )\,\Gamma
(-\frac{1}{2}+i\nu )\,\Gamma
(\frac{3}{2}+i\nu )}\,.
\label{phasemom}
\end{equation}

On the other hand, in the region
$\vec{\rho}^2 \sim \vec{k}^{-2} \alpha ' (g^2\,N_c)^{-1}$ one can
put $\vec{q}=0$ and after the redefinition of the wave function and its
argument
\begin{equation}
\Psi (\vec{q},\vec{k})=|k|^{-3}\,\phi (z)\,,\,\,z=\ln (\alpha '\vec{k}^2)
\end{equation}
the BFKL homogeneous equation in the string model can be written as the
Schr\"{o}dinger equation
\begin{equation}
E\phi =H\phi \,,\,\,\omega
=-\frac{g^2N_c}{4\pi
^2}\,E\,,\,\,H=H_{BFKL}(i\partial /\partial z)+\lambda e^z \,,
\end{equation}
where
\begin{equation}
H_{BFKL}(\nu)=\psi (i \nu +1/2)+\psi (-i \nu +1/2)-2\psi
(1)\,,\,\,\lambda = \frac{4\pi ^2}{g^2N_c}\,. \label{Hzeron}
\end{equation}

The analogy with the Schr\"{o}dinger equation is especially fruitful in
the diffusion approximation, where
\begin{equation}
H_{BFKL}(i\partial /\partial z)= -4\ln 2 -14 \,\zeta (3)
\left(\partial /\partial z\right)^2\,
\end{equation}
has the form of the non-relativistic kinetic energy.
The potential energy $\lambda e^z$ grows rapidly at large positive $z$
and therefore the wave function $\phi$ should tend to zero in this
region
\begin{equation}
\lim _{z\rightarrow \infty} \phi (z)\sim \exp
\left(-2\sqrt{\frac{\lambda}{14\,\zeta (3)} }\,e^{z/2} \right).
\label{asymz}
\end{equation}
For
$z\rightarrow -\infty$ the potential energy vanishes, which agrees
with a
possibility to neglect the string effects at small $\vec{k}^2$.
In the momentum representation
\begin{equation}
\phi (p)=\int _{-\infty}^{\infty}\,e^{ipz}\,\phi (z)\,dz,
\end{equation}
where $p=i\partial /\partial z$, the BFKL
equation is reduced to the equation in finite differences
\begin{equation}
(E-H_{BFKL}(p)) \phi (p)=  \lambda \phi (p-i)\,.
\end{equation}
The function $\phi (p)$ can have the singularities (poles) only in the
upper semi-plane. It is analytic in the lower semi-plane to provide a
rapidly decreasing behaviour of $\phi
(z)$ at $z \rightarrow +\infty$. The positions of the poles is given
below
\begin{equation} p_r=p_0+ir\,,\,\, (r=0,1,2,...)\,,
\end{equation}
where the possible
values of $p_0$ satisfy the equation \begin{equation}
H_{BFKL}(p_0)=E\,.
\end{equation}
For example, in the diffusion approximation, where
\begin{equation}
E-H_{BFKL}(p)=E+4\,\ln 2-14\,\zeta (3)\,p^2\,,
\end{equation}
the solution of
the above recurrent relation is
\begin{equation}
\phi (p)= \phi _0(p)\, \left(7\,\zeta (3)\,\frac{g^2\,N_c}{2\pi
^2}\right)^{ip}\, \Gamma \left(i\,p-i\sqrt{\frac{E+4\ln
2}{14\,\zeta (3)}}\right)\, \Gamma \left(i\,p+i\sqrt{\frac{E+4\ln
2}{14\,\zeta (3)}}\right) \label{soldif} \end{equation} up to a
periodic function satisfying the relation $\phi _0(p)=\phi
_0(p+i)$. We should substitute this function  by a constant
\begin{equation}
\phi _0(p)=const\,,
\end{equation}
because in an opposite case  for $p
\rightarrow \pm i\infty$ the wave function does not decrease
sufficiently rapidly due to the additional factors
$\sim \exp (\mp 2\pi i p)$.
Indeed, for $\phi _0(p)=1$ the normalization integral \begin{equation}
\int _{-\infty }^{\infty}dp\, |\phi (p)|^2= \int
_{-\infty}^{\infty}dp\, \frac{\pi ^2/\left(p^2- \frac{E+4\ln
2}{14\,\zeta (3)} \right)}{ \sinh \left(\pi\,p-\pi \,\sqrt{\frac{E+4\ln
2}{14\,\zeta (3)}}\right)\, \sinh \left(\pi\,p+\pi \,\sqrt{\frac{E+4\ln
2}{14\,\zeta (3)}}\right)}
\label{normali}
\end{equation}
is convergent at $p
\rightarrow \pm \infty$.  Moreover, for $\phi _0 (p)=1$ the wave
functions $\phi (p)$ with different $E$ are orthogonal. Note, that
the integrand (\ref{normali}) contains the poles. After their
appropriate regularization it leads to the $\delta$-function $\sim
\delta (E-E')$ in the orthonormality conditions.

Let us go to the $z$
representation
\begin{equation}
\phi (z)=\int _{-\infty -i0}^{\infty
-i0}\frac{dp}{2\pi }\, e^{-ip\,z}   \, \phi (p)\,.  \label{phiz}
\end{equation}
For large positive $z$ the contour of the
integration over $p$ should be shifted in the lower semi-plane up to
the saddle point situated at
\[
z=\psi \left(i\,p-i\sqrt{\frac{E+4\ln
2}{14\,\zeta (3)}}\right) +\psi \, \left(i\,p+i\sqrt{\frac{E+4\ln
2}{14\,\zeta (3)}}\right)+\ln \left(7\,\zeta (3)\,\frac{g^2\,N_c}{2\pi
^2}\right)
\]
\begin{equation}
\approx \ln \left((ip)^2\,7\,\zeta
(3)\,\frac{g^2\,N_c}{2\pi ^2}\right)\,.
\end{equation}
We can
estimate $\phi (z)$ by the value of the integrand  in (\ref{phiz}) at
this point
\begin{equation}
\phi (z) \approx e^{-ip\,z}\,\phi
(p)\approx  \exp \left(-2\,\sqrt{\frac{2\pi ^2}{7\,\zeta
(3)\,g^2N_c}}\,e^{z/2}\right)
\end{equation}
in an accordance with eq. (\ref{asymz}).

At small $E+4\ln 2$, where the diffusion
approximation is valid, the solution near the poles at small values of
$p$ is \begin{equation} \phi (p) \sim
\frac{\left(7\zeta (3)\,g^2\,N_c/(2\pi ^2)\right)^{ip}}{E+4\ln 2 -14\,
\zeta (3) \,p^2}\,.  \end{equation} Thus, $\phi (z)$ at  $z =\ln
(\alpha '\vec{k}^2)\rightarrow -\infty$ behaves as follows
\begin{equation} \phi (z) \sim \left(\frac{7\zeta (3)\,g^2\,N_c}{2\pi
^2\alpha '\vec{k}^2}\right)^{i \sqrt{\frac{E+4\ln 2}{14\,\zeta (3)}}}-
\left(\frac{7\zeta (3)\,g^2\,N_c}{2\pi ^2\alpha '\vec{k}^2}\right)^{-i
\sqrt{\frac{E+4\ln 2}{14\,\zeta (3)}}}\,.
\end{equation}

By comparing this result with expressions (\ref{waveasmom}) and
(\ref{phasemom}) for small $\nu$ in the intermediate region $ \alpha
'/(g^2N_c)\ll
\vec{\rho}^2\ll 1/\vec{q}^2$ we obtain the quantization of the
Regge trajectories
\begin{equation}
2\,\sqrt{\frac{E_r+4\ln 2}{14\,\zeta (3)}}\ln \left(\frac{7\zeta
(3)\,g^2\,N_c}{2\pi ^2\alpha ' \vec{q}^2}\right)=2\pi
(r+1/2)\,,\,\,r=0,1,2,...    \end{equation} for $n=0$ and small
\begin{equation}
E+4\ln 2=4\,\frac{\omega _0-\omega }{g^2N_c}\ll 1\,.
\end{equation}

For comparatively large energies $E$ in the diffusion
approximation one can use the semiclassical approximation near the
turning point $z=z_0$, where
\begin{equation}
\lambda e^{z_0}=E+4\ln2\,,\,\,\lambda = \frac{4\pi ^2}{g^2N_c}\,,
\end{equation}
corresponding to the following simplification of the solution
(\ref{soldif}) at $p\ll \sqrt{E+4\ln 2}$
\begin{equation}
\phi (p)\sim  \left((E+4\ln 2)\,\frac{g^2\,N_c}{4\pi
^2}\right)^{ip}\, \exp \left( -i\,\frac{14 \zeta (3)}{E+4\ln 2}\,
\,\frac{p^3}{3}\right)\,. \label{return}
\end{equation}
The Fourrie transformation to the $z$-representation can be
performed with the use of the saddle-point method
\begin{equation}
\phi (z) \sim \exp \left(i\,(-\Delta
z)^{3/2}\frac{2}{3}\sqrt{\frac{E+4\ln 2}{14 \zeta
(3)}}-i\frac{\pi}{4}\right)+\exp \left(-i\,(-\Delta
z)^{3/2}\frac{2}{3}\sqrt{\frac{E+4\ln 2}{14 \zeta
(3)}}+i\frac{\pi}{4}\right)\,,
\end{equation}
where $\Delta z=z-z_0$. Therefore in the diffusion approximation
of small $\nu$ the wave function at $z \rightarrow -\infty$ equals
\begin{equation}
\phi (z) \sim e^{-i\pi /4}\,\left(\frac{7\zeta (3)\,g^2\,N_c}{2\pi
^2\alpha '\vec{k}^2}\right)^{i \sqrt{\frac{E+4\ln 2}{14\,\zeta
(3)}}}+e^{i\pi /4}\, \left(\frac{7\zeta (3)\,g^2\,N_c}{2\pi
^2\alpha '\vec{k}^2}\right)^{-i \sqrt{\frac{E+4\ln 2}{14\,\zeta
(3)}}}
\end{equation}
and the quantization condition for energies is
\begin{equation}
2\,\sqrt{\frac{E_r+4\ln 2}{14\,\zeta (3)}}\ln \left(\frac{7\zeta
(3)\,g^2\,N_c}{2\pi ^2\alpha ' \vec{q}^2}\right)=2\pi (r+1/4)
\end{equation}
for large integer $r$.

We investigate below a general case of arbitrary $\nu$ for small
$\alpha ' t$ without using the diffusion approximation. To begin
with, one can verify, that here in the semiclassical approach
expression (\ref{return}) for the wave function is also valid near
the returning point $z=z_0$, where $p=0$. The only difference with
the diffusion approximation is an additional $\nu$-dependence of the
phase $\delta (\nu)$ in (\ref{phasemom}), which leads to the
modified quantization condition
\begin{equation}
2\,|\nu _r|\ln \left(\frac{7\zeta (3)\,g^2\,N_c}{2\pi ^2\alpha '
\vec{q}^2}\right)=\delta (\nu _r)+2\pi (r+1/4)\,,\,\,r=0,1,2,...
\end{equation}
and the corresponding quantized energies can be obtained from the
relation $E=H_{BFKL}(\nu)$ (see (\ref{Hzeron})).

To derive an exact solution of the BFKL equation for small $\alpha
' \vec{q}^2$ let us introduce the new variables
\begin{equation}
x=2\alpha^{\prime }l^{2}\,,\qquad x^{\prime }= 2\alpha^{\prime
}(l^{\prime})^{2}\,, \qquad x_{1}=2\alpha ^{\prime }l_{1}^{2}\,.
\label{redef}
\end{equation}
In these variables the inhomogeneous BFKL equation has the form
\begin{eqnarray}
[\omega+x]f(x)=\hat\Phi(x)+ c\int _0^{\infty}\Biggl[
\frac{f(x^{\prime})}{|x-x^{\prime}|} -
\biggl[\frac{1}{|x-x^{\prime}|} -\frac{1}{\sqrt{x^2+
4{x^{\prime}}^2}} \biggl]\frac{x}{x^{\prime}}f(x)\Biggl]
dx^{\prime}\,,\quad c=\frac{g^{2}N_{c} }{4\pi^{2}}\,. \label{eqsm}
\end{eqnarray}
Here $f(x)\sim \phi (x)/\sqrt{x}$ is
$F_{\omega}^{(0)}(0;l)/x$ averaged over the angle
$\varphi$ between $l$ and $l^{\prime }$, and $\hat\Phi(x)$ is
$\tilde\Phi(0;l)/x$ averaged over $\varphi$. We expect that
$\tilde\Phi(0;l)\to0$ at $x\to0$ and $\hat\Phi(x)$ is finite at
$x=0$.

The above BFKL equation differs from that in QCD \cite{klf} by the
presence of an additional term proportional to $x$ in its left
hand side. As in the QCD case, we search the solution in the form
of the Mellin transformation
\begin{equation}
f(x)=\int_{-i\infty }^{i\infty }(x)^{\sigma -1/2} C(\sigma
)\frac{d\sigma }{2\pi i}\,, \,\,\sigma =i\nu \,.
\label{mod1eq}
\end{equation}
Similarly, the inhomogeneous term is presented as follows
\begin{equation} \hat\Phi(x)=\int_{-i\infty
}^{i\infty }(x)^{\sigma -1/2} \hat\Phi_1(\sigma)
\frac{ d\sigma }{2\pi i
}\,.  \label{mod2eq} \end{equation}

To obtain an equation for $C(\sigma)$ one collects the terms
proportional to $ x^\sigma$. For the contribution $xf(x)$ the
integration contour should be moved to the line $ \Re \sigma =-1$
and therefore the function $C(\sigma )$ can not have any
singularities inside the strip $-1< \Re \,\sigma <0$. If this
condition is fulfilled, we have
\begin{equation}
C(\sigma -1)=\Phi_1(\sigma)-[c\, b(\sigma )+\omega ]C(\sigma )\,,
\label{solmod}
\end{equation}
where
\begin{equation}
b(\sigma)=\psi (\sigma +1/2)+\psi (-\sigma +1/2)-2\psi (1)  \label{defb}
\end{equation}
and $\psi (x)=d\ln \Gamma (x)/dx$ is the derivative of the logarithm of
the gamma-function.

It is convenient to introduce the new variable $\xi$
according to the definition
\begin{equation}
\sigma =-\frac{\ln \xi }{2\pi i}\,.  \label{solrep}
\end{equation}
In the new variables eq. (\ref{solmod}) can be written as follows
\begin{equation}
\tilde{C}(\xi e^{-2\pi i})=\widehat\Phi_1(\sigma)-[c \,b(\sigma )+\omega
] \tilde{C} (\xi ) \,,  \label{solmo1d}
\end{equation}
where $\tilde{C}(\xi )\equiv C(\sigma (\xi ))$. The calculation of
$C(\xi )$ is reduced to the known mathematical problem of finding
a function satisfying the requirement, that its discontinuity is
proportional to the same function. Let us define an
auxiliary function $\phi (\sigma ,\widehat\sigma_{1})$ being a solution
of the homogeneous
equation
\begin{equation} \phi (\sigma
-1,\widehat\sigma_{1})=[c\,b(\sigma )+\omega ]\phi (\sigma
,\widehat\sigma_{1})\,.  \label{hsolmod} \end{equation}
with
$\widehat\sigma _{1}$ being an arbitrary
subtraction
point, where $\Phi =1$. Note, that the sign in the right hand side
of eq. (\ref{hsolmod}) is opposite in comparison with the sign in front
of the corresponding term in eq. (\ref{solmod}).

The explicit expression for such function is given below
\begin{equation}
\phi (\sigma ,\widehat\sigma_{1})=\exp
\Biggl[\int_{-i\infty }^{i\infty }
\frac{\sin \pi (\widehat\sigma _{1}-\sigma )\ln [c\,b(\sigma ^{\prime
})+\omega ]}{ \sin \pi (\sigma ^{\prime }-\widehat\sigma _{1})\sin \pi
(\sigma ^{\prime }-\sigma )}\frac{d\sigma ^{\prime }}{2i} \Biggl]
\label{hsol} \end{equation}
where it is implied that $\Re \sigma  <0$ and $\Re \widehat\sigma_1<0$.
At $\Re \sigma >0$ the result is obtained by an analytic
continuation  of (\ref{hsol}) from the region $\Re \sigma <0$.
Furthermore, it is implied, that the solution for $\omega <\omega
_0=(g^2N_c\ln 2)/\pi^2$ can be derived also by an analytic continuation
from the region $\omega >\omega _0$, where the argument of the logarithm
has two zeros situated on the imaginary axes and pinching the integration
contour at $\omega \rightarrow \omega _0$.

The
integral over
$\sigma^{\prime}$ is convergent at large $\sigma^{\prime}$ since from
(\ref{defb}) one obtains
\begin{equation}
\ln b(\sigma )\rightarrow \ln\ln|\Im\,\sigma |
\label{solas}
\end{equation}
at $\Im \,\sigma
\rightarrow \pm \infty $.

Let us show, that indeed expression (\ref{hsol}) is  a solution
of eq.  (\ref{hsolmod}). The pole at $\sigma '=\sigma$ is situated to the
left of the integration contour and can pinch only the right
singularity of the logarithm  situated at the zero of its
argument. The pole
at $\sigma '=\sigma -1$ being to the right from the contour pinches with
the left singularity of the logarithm.  It means, that
the function $ \phi (\sigma ,\widehat\sigma _{1})$ has no singularities
in the strip $-1\leq \Re\,\sigma \leq 0$. To verify that solution
(\ref{solas}) satisfies eq.
(\ref{hsolmod}) it is enough to note that after the shift $\sigma
\rightarrow \sigma -1$ the pole at $\sigma ^{\prime }=\sigma +1$ of the
integrand moves to the point $\sigma ^{\prime }=\sigma $ which was
earlier to
the left from the integration contour. The initial and final expressions
differ
each from another by an additional term in the exponent. This term is
obtained by taking the residue in the pole at $\sigma^{\prime}=\sigma$.
As a result, relation (\ref{hsolmod}) is fulfilled.

It is useful to investigate the positions of zeroes and poles of $
\phi (\sigma ,\widehat\sigma _{1})$. Both of them are obtained due
to pinching the poles $1/\sin \pi (\sigma '-\sigma)$ with the
singularities of the logarithm situated at zeros and poles of its
argument $\left[c\,b(\sigma')+\omega\right]$. The poles are
situated at $\sigma '=\pm (n+1)$, where $n=0,1,2,...$. The zeros
are situated between these poles. We denote their position by
$\sigma _m^{(+)}$ for $\Re \sigma _m>0$ and $\sigma _m^{(-)}$ for
$\Re \sigma _m<0$. It is obvious, that $|\Re \sigma_m ^{(\pm)
}|<|\Re \sigma_n ^{(-) }|$ for $m<n$. The function $\phi (\sigma
,\widehat\sigma _{1})$ has zeroes at $\sigma =\sigma _m^{(-)}-r$,
where $r$ is an integer or zero for $m=1,2,...$ and $r\ne 0$ for
$m=0$. Indeed, due to the above discussion $\sigma =\sigma
_0^{(-)}$ is not a singularity of the exponent. Furthermore, $\phi
(\sigma ,\widehat\sigma _{1})$ has zeros in the right half-plane
at $\sigma = n+1/2$, where $n$ is an integer or zero. The poles
are situated in the right half-plane at $\sigma =\sigma
_m^{(+)}+n$ and in the left half-plane at $\sigma =-(n+3/2)$ for
$n=0,1,2,...$. Similar to the case $\sigma _0^{(-)}$ the point
$\sigma =-1/2$ does not corresponds to a singularity of the
exponent. In the above discussion we used the relation
\begin{equation}
\frac{\sin \pi (\widehat{\sigma} _1-\sigma)}{\sin \pi (\widehat{\sigma}
_1-\sigma')\,\sin \pi (\sigma -\sigma')}=\cot \pi (\sigma
'-\widehat{\sigma}_1)-\cot \pi (\sigma
'-\sigma)\,.
\end{equation}

Using expression (\ref{hsol}) one can find for large $\sigma =r+iy$
\begin{equation} \phi (\sigma ,\widehat\sigma_{1})
\rightarrow \exp \Biggl[-(\ln\ln|y|)[iy+r+1/2]\Biggl]   \label{sol3}
\end{equation}
up to a phase independent from $\sigma$. Here in the essential
region of integration $\sigma ^{\prime }\sim \sigma $ we replaced
the logarithmic function $ c\,b(\sigma^{\prime})+\omega$ by its
asymptotic value at $\sigma^{\prime }=\sigma $.

In a similar way one can check that for $\omega >\omega_0 $ the
solution $C(\sigma )$ of the inhomogeneous equation is given by
\begin{equation}
C(\sigma )=\int_{-i\infty }^{i\infty }\frac{\phi (\sigma',
\widehat\sigma_{1}) \hat\Phi_1(\sigma ') d\sigma ^{\prime }}{2i\phi
(\sigma^{\prime},\widehat\sigma_{1}) [c\,b(\sigma ^{\prime })+ \omega
]\sin \pi (\sigma^{\prime }-\sigma )}\, \label{soli}\,. \end{equation}
In an agreement with general arguments the continuation of the partial
wave in the complex plane from the integer points is performed from the
region $\omega >\omega _0$. Similar to (\ref{hsol}) in (\ref{soli}) the
conditions $\Re \sigma <0$ and $\widehat{\sigma}_1<0$ are assumed to be
fulfilled and the expression in the region $\Re \sigma >0$ are obtained by
an analytic continuation. It can be written in the
equivalent form
\begin{equation} C(\sigma )=\int_{-i\infty }^{i\infty
}\frac{\widehat\Phi_1(\sigma')\phi (\sigma
^{\prime },\sigma )
d \sigma^{\prime}}{2i[c\,b(\sigma ^{\prime })+\omega ]\sin \pi
(\sigma ^{\prime }-\sigma)}\,  \label{sol} \end{equation} with
the same conventions concerning the signs of $\Re\,\sigma$ and $\omega$.

As in the case of QCD  \cite{klf}, the leading singularity in the
$\omega $-plane is situated at $\omega =\omega_{0}=(g^{2}N_{c}\ln
2)/\pi^{2}$. It is obtained from the region $\sigma ^{\prime }\sim
\sigma \rightarrow 0$ in eq. (\ref{sol}). In this limit the
corresponding denominator is approximated by the diffusion
expression $\omega -\omega _{0}-a(\sigma^{\prime })^{2}$. Calculating
the integral at $ \sigma \rightarrow0$, one obtains
\begin{equation}
C(\sigma )\sim 1/(\sigma -\sqrt{\omega-\omega_{0}})\,,  \label{aprc}
\end{equation}
where the omitted factor has no singularity at small $\sigma $. Thus, at
$x \rightarrow 0$ the solution is $\sim x^{\sqrt{\omega -\omega _0}-1/2}$.
In QCD there are singularities in both points $\sigma =\pm \sqrt{\omega
-\omega _{0}}$, but in the string model only the singularity at
$\sigma =\sqrt{\omega-\omega_{0}}$ survives. Another singularity is
absent because at large momenta the kernel of the equation is
non-singular due to the linear term in the trajectory on the left
hand side of eq. (\ref{eqsm}).

In the important case of the leading singularity, where the diffusion
approximation
\begin{equation} c\,b(\sigma)+\omega\approx
a(\sigma_0^2-\sigma^2)\,,\,\, a=-7\zeta (3)\,\frac{g^2}{2\pi^2}\,,
\,\,\, \sigma_0^2=-4\ln 2-\omega/[g^2\,7\zeta (3)/2\pi ^2]
\label{difap}
\end{equation}
is valid,  the function $\phi(\sigma)$  is given by
\begin{equation}
\phi(\sigma)=a^{-\sigma}\Gamma(\sigma_0-\sigma)
[\Gamma(\sigma_0+\sigma+1)]^{-1} \,.
\label{psdif}
\end{equation}
It is related to the solution  $\tilde\phi(\sigma)$ of the
homogenious equation as
follows
\begin{equation}
\tilde\phi(\sigma)=-\pi\phi(\sigma)/\sin\pi(\sigma+\sigma_0)\,.
\label{eigf}
\end{equation}
At $\omega<0$, as  it was discussed above, $\tilde{\phi} $
describes the wave
function of the particle with an energy equal to
$-c\omega$, which is rejected from the potential barrier $e^z$.
In this case $z=\ln x$, and $-i\sigma$ is the momentum of
the colliding particle.

According to (\ref{sol}), the function $C(\sigma)$   in (\ref{mod1eq})
determining the solution of eq. (\ref{eqsm}) is given below
\begin{equation}
C(\sigma)=\frac{a^{-\sigma}\Gamma(\sigma_0-\sigma)}{
\Gamma(\sigma_0+\sigma+1)}\int_{-i\infty}^{+i\infty}
\frac{a^{\sigma'-1}\hat\Phi(\sigma')
\Gamma(\sigma'+\sigma_0)\,d\sigma'}{2i\,
\Gamma(\sigma_0-\sigma'+1)\sin\pi(\sigma'-\sigma)}\,.
\label{difsol}
\end{equation}
In (\ref{difsol}) it is implied that $\Re\,\sigma<0$, and so that
the pole at $\sigma'=\sigma$ is twisted with the right side. At
$\Re\,\sigma>0$ the result is obtained by an analytic continuation
in $\sigma$.  Furthermore, in (\ref{mod1eq}) the pole at
$\sigma=\sigma_0$ is on the right hand side from the integration
contour. It is solely the solution at $\sigma_0>0$ because in this
case (\ref{eigf}) determines a function of $x$ increasing at
$x\to\infty$. At $\sigma_0<0$ the solution is not unique because
(\ref{eigf}) might be added to (\ref{difsol}). In an agreement
with general arguments one should chose the solution which is an
analytical continuation of the solution (\ref{difsol}) to the
region $\sigma_0<0$.  The result is presented by eq.
(\ref{difsol}) where the integration contour is defined in an
accordance with these arguments.

\section{Heisenberg spin model and integrability}
To investigate the region of $\alpha '\vec{q}^2 \sim g^2\,N_c$ it
is convenient to use the conformal invariance of the BFKL kernel
in QCD (see \cite{conf}). In the coordinate representation for the
wave function describing the composite state of two reggeized
gluons with the impact parameters $\vec{\rho}_1$ and
$\vec{\rho}_2$ we have the expression \cite{conf} (see (\ref{polyasol}))
\begin{equation} E_{m,
\widetilde{m}}(\vec{\rho}_1,\vec{\rho}_2;\vec{\rho}_0) =
\left(\frac{\rho _{12}}{\rho _{10}\rho _{20}}\right)^m
\left(\frac{\rho ^*_{12}}{\rho ^*_{10}\rho ^*_{20}}
\right)^{\widetilde{m}}\,,\,\,\rho _{12}=\rho _1-\rho _2 \,,
 \label{polya}
\end{equation}
where $\vec{\rho}_0$ is the coordinate of the Pomeron, $\rho
_r=x_r+iy_r$ and $\rho ^*_r=x_r-iy_r$ are respectively the
holomorphic and anti-holomorphic variables, $\rho _{rs}=\rho _r
-\rho _s$ and
\begin{equation}
m=\frac12+i\nu+\frac{n}{2}\,,\,\,\widetilde{m}=\frac12+i\nu-\frac{n}{2}
\label{mm}
\end{equation}
 are conformal weights related to the
eigenvalues of the Casimir operators of the M\"{o}bius group
\begin{equation}
\vec{M}^2E_{m,\widetilde{m}}=m(m-1)E_{m,\widetilde{m}}\,,\,\,\,
\vec{M}^{*\, 2}E_{m,\widetilde{m}}=
\widetilde{m}(\widetilde{m}-1)E_{m,\widetilde{m}}\,,\,\,
\vec{M}^2=-\rho _{12}^2\frac{\partial}{\partial \rho
_1}\frac{\partial}{\partial \rho _2}\,.
\end{equation}
Note, that in (\ref{mm}) the conformal spin $n$ is integer
$n=0,\pm 1,\pm 2,...$ and the parameter $\nu$ is a real number for
the principal series of the unitary representations.

The operator $\vec{M}^2$ is related to the generators of the
M\"{o}bius group $\vec{M}$ \begin{equation}
\vec{M}=\vec{M}_1+\vec{M}_2\,,\,\,M_r^z= \rho
_r\frac{\partial}{\partial \rho _r} \,,\,\,M_r^-=
\frac{\partial}{\partial \rho _r}\,,\,\, M_r^+=-\rho
^2_r\frac{\partial}{\partial \rho _r}\,. \label{Moebgen}
\end{equation}

The generators satisfy the following commutation relations
\begin{equation}
[M^z,M^{\pm}]=\pm M^{\pm}\,,\,\,[M^+,M^-]=2M^z\,,\,\,
[M^{*\,z},M^{*\,\pm}]=\pm
M^{*\,\pm}\,,\,\,[M^{*\,+},M^{*\,-}]=2M^{*\,z}\,.  \end{equation}
For the solution of the BFKL equation in the string theory it is
convenient to introduce also the generators
\begin{equation}
\vec{N}=\vec{M}_1-\vec{M}_2\,.
\end{equation}
Together with the operators $\vec{M}$ they produce the Lie algebra
for the Lorentz group
\begin{equation}
[M^z,N^{\pm}]=\pm N^{\pm}\,,\,\,[M^+,N^-]=2N^z\,,\,\,
[M^{\pm},N^{z}]=\mp N^{\pm}\,,\,\,[M^-,N^+]=-2N^z\,,
\end{equation}
\begin{equation}
[N^z,N^{\pm}]=\pm M^{\pm}\,,\,\,[N^+,N^-]=2M^z\,.
\end{equation}
We can find the representation of this algebra in the space of the
functions $E_{m}$
\begin{equation}
E_m(\rho _1,\rho _2;\rho _0)= \left(\frac{\rho _{12}}{\rho
_{10}\rho _{20}}\right)^m
\end{equation}
as follows
\[
M^zE_m=(-\rho _0\partial _0-m)E_m \,,\,\,\,M^+E_m=(\rho
_0^2\partial _0+2m\rho _0)E_m \,,\,\,\, M^-E_m=-\partial _0E_m \,,
\]
\[
N^-E_m=\frac{m(m-1)}{2m-1}\left(E_{m+1}+\frac{\partial
_0^2}{(m-1)^2}E_{m-1}\right)\,,\,\,\, (N^z-\rho
_0N^-)E_m=\frac{m}{m-1}\partial _0E_{m-1}\,,
\]
\begin{equation}
(N^++2\rho _0N^z-\rho _0^2N^-)E_m=-2mE_{m-1}
\end{equation}
and analogously for the representation of $\overrightarrow{M^*}$
and $\overrightarrow{N^*}$ on functions $E^*_{\widetilde{m}}$.

The BFKL integral operator $K_{BFKL}$ is diagonal in the
$(m,\widetilde{m})$-representation and its eigenvalue has the
property of the holomorphic separability
\begin{equation}
\omega _{BFKL}=- \frac{g^2}{8\pi ^2}\,N_c\,\epsilon
_{m,\widetilde{m}} \,,\,\, \epsilon_{m,\widetilde{m}}=\epsilon
_m+\epsilon _{\widetilde{m}}\,, \label{omegBFKL}
\end{equation}
where the holomorphic energies are the following functions of the
conformal
  weights $m$ and $\widetilde{m}$ \begin{equation} \epsilon_{m}=\psi
(m)+\psi (1-m)-2\psi (1)\,,\,\, \epsilon_{\widetilde{m}}=\psi
(\widetilde{m})+ \psi (1-\widetilde{m})-2\psi (1)\,.
\end{equation}

In the case of the string theory in the eigenvalue equation for
the Pomeron wave function $f$ in the dimension $D=4$ we have the
additional contribution $\Delta K_{BFKL}$ (neglecting the pole
term from the soft Pomeron)
\begin{equation}
\omega \,f=K\,f\,,\,\,K=K_{BFKL}+\Delta K\,,\,\, \Delta K=-\alpha
'\vec{p}^2_1-\alpha '\vec{p}_2^2\,.
\end{equation}
It is convenient to use the mixed representation
($\vec{q}=\vec{p}_1+\vec{p}_2,\,\vec{\rho} =\vec{\rho} _{12}$),
where the additional string contribution to $K$ has the form
\begin{equation}
\Delta K =-\alpha' \left(\frac{\vec{q}^2}{2}-2\, \frac{\partial
^2}{(\partial  \rho_{\mu})^2}\right)\,,\,\, \frac{\partial
^2}{(\partial  \rho_{\mu})^2}= N^-N^{*-}\,,\,\,N^-=\partial
_1-\partial _2\,,\,\,\rho =\rho _1-\rho _2 \,.
\end{equation}
In this representation the Pomeron wave function in QCD can be
obtained by the Fourrie transformation
\begin{equation}
E_{m,\widetilde{m}}(\vec{q},\vec{\rho}) =\int d^2R\, e^
{i\,\vec{q}\vec{R}}\, \left(\frac{\rho}{(R+\frac{\rho
}{2})(R-\frac{\rho }{2})}\right)^m \left(\frac{\rho
^*}{(R^*+\frac{\rho ^*}{2})(R^*-\frac{\rho ^*}{2}}
\right)^{\widetilde{m}}\,,\,\,R=\frac{\rho _1+\rho _2}{2}\,.
\label{mixed}
\end{equation}

Let us present the solution of the BFKL homogenious equation in
the string theory as a superposition of the above functions
\begin{equation}
f(\vec{q},\vec{\rho}) =\int _{-\infty}^{\infty}d \nu \sum
_{n=-\infty}^{\infty}
   C_{m,\widetilde{m}}(\vec{q})\,
\Gamma (m)\,\Gamma (\widetilde{m})\,
E_{m,\widetilde{m}}(\vec{q},\vec{\rho})\,. \label{reprsol}
\end{equation}
Here we extracted the factor $\Gamma (m)\,\Gamma (\widetilde{m})$
from coefficients $C_{m,\widetilde{m}}(\vec{q})$ to simplify the
relations between them. The operators $N^-$ and $N^{*-}$ act on
the functions $ E_{m,\widetilde{m}}(\vec{q},\vec{\rho})$ as
follows
\begin{equation}
N^-E_{m,\widetilde{m}}=
\frac{m(m-1)}{2m-1}\left(E_{m+1,\widetilde{m}}-\frac{q^{*\,2}}
{4(m-1)^2}E_{m-1,\widetilde{m}}\right)\,,
\end{equation}
\begin{equation}
N^{*\,-}E_{m,\widetilde{m}}=
\frac{\widetilde{m}(\widetilde{m}-1)}{2\widetilde{m}-1}
\left(E_{m,\widetilde{m}+1}- \frac{q^{2}} {4(\widetilde{m}-1)^2}
E_{m,\widetilde{m}-1}\right)\,.
\end{equation}

Therefore the function $ f(\vec{q},\vec{\rho})$ is a solution of
the homogeneous BFKL equation in the string theory if the
coefficients $ C_{m,\widetilde{m}}(\vec{q}) $ in (\ref{reprsol})
satisfy the following recurrent relation
\[
\left(\omega + \frac{g^2}{8\pi ^2}\,N_c\,(\epsilon _{m}+\epsilon
_{\widetilde{m}}) +\alpha'\frac{\vec{q}^2}{2}\right)\,
  C_{m,\widetilde{m}}(\vec{q})=
\]
\[
2\alpha' \left(\frac{m-2}{2m-3}\,
\frac{\widetilde{m}-2}{2\widetilde{m}-3}\,
C_{m-1,\widetilde{m}-1}(\vec{q})- \frac{m+1}{2m+1}\,
\frac{\widetilde{m}-2}{2\widetilde{m}-3}\, \frac{q^{*\,2}} {4}
C_{m+1,\widetilde{m}-1}(\vec{q})\right.
\]
\begin{equation}
-\left. \frac{m-2}{2m-3}\,
\frac{\widetilde{m}+1}{2\widetilde{m}+1}\, \frac{q^{2}} {4}
C_{m-1,\widetilde{m}+1}(\vec{q})+ \frac{m+1}{2m+1}\,
\frac{\widetilde{m}+1}{2\widetilde{m}+1}\, \frac{q^{*\,2}} {4}
\frac{q^{2}} {4} C_{m+1,\widetilde{m}+1}(\vec{q})\right)\,.
\end{equation}
By introducing the new function
\begin{equation}
\phi
_{m,\widetilde{m}}(\vec{q})=(2m-1)^{-1}(2\widetilde{m}-1)^{-1}\,
(q/2)^{\widetilde{m}}(q^*/2)^{m}\,C_{m,\widetilde{m}}(\vec{q})
\end{equation}
one can write this recurrent relation in a simpler form
\[
\left(\omega + \frac{g^2}{8\pi ^2}\,N_c\,(\epsilon _{m}+\epsilon
_{\widetilde{m}}) +\alpha'\frac{\vec{q}^2}{2}\right)\,
(2m-1)(2\widetilde{m}-1)
  \phi _{m,\widetilde{m}}(\vec{q})=
\]
\[
\alpha'\frac{\vec{q}^2}{2} \left((m-2)\, (\widetilde{m}-2)\, \phi
_{m-1,\widetilde{m}-1}(\vec{q})- (m+1)\, (\widetilde{m}-2)\, \phi
_{m+1,\widetilde{m}-1}(\vec{q})\right.
\]
\begin{equation}
-\left. (m-2)\, (\widetilde{m}+1)\, \phi
_{m-1,\widetilde{m}+1}(\vec{q})+ (m+1)\, (\widetilde{m}+1)\, \phi
_{m+1,\widetilde{m}+1}(\vec{q})\right)\,. \label{threeq}
\end{equation}
 One
should add to this recurrent relation the information about the
asymptotic behavior of the coefficients
$C_{m,\widetilde{m}}(\vec{q})$ at large $m$ and $\widetilde{m}$
corresponding to $|k|\gg |q|$ investigated above. Note, that contrary to
the case of small $\alpha '\vec{q}^2$, considered in the previous section,
now the eigenfunctions contain a mixture of states with different 
conformal spins. Expanding $\phi _{m,\widetilde{m}}$ in the basis of
the functions $x^mx^{*\,\widetilde{m}}$ one can reduce the 
recurrent relation (\ref{threeq})  in the diffusion
approximation to a differential equation, which can be solved, for
example, by the semi-classical methods similar to those used
in the previous section.

In the case of the colourless state constructed from several reggeized
gluons \cite{BKP} the homogeneous equation for its wave function in the
string theory is given in the multi-colour limit
$N_c\rightarrow \infty$ below (cf. \cite{integr})
\begin{equation}
E\,
\phi (\vec{\rho} _1,\vec{\rho}_2,...,\vec{\rho}_n)=H\,
\phi (\vec{\rho} _1,\vec{\rho}_2,...,\vec{\rho}_n)\,,\,\,\omega =-
\frac{g^2N_c}{8\pi ^2}\,E\,,
\end{equation}
where
\begin{equation}
H=H_{BFKL}^{(n)} +
  l^2\,
\sum _{r=1}^n (\vec{p}_r)^2\,,\,\,
  l^2=\frac{\alpha '\,8\pi ^2}{g^2N_c}\,,\,\,
  p_r^{\mu}=i\,\frac{\partial}{\partial \rho _r^{\mu}}\,.
\label{strHn}
\end{equation}
Here $H_{BFKL}^{(n)}$ has the property of the holomorphic separability
\begin{equation}
H_{BFKL}^{(n)} =
h_{BFKL}^{(n)} +
h_{BFKL}^{(n)\,*}\,,\,\, h_{BFKL}^{(n)}=\sum
_{r=1}^nh_{BFKL}^{(r,r+1)}\,,
\end{equation}
\begin{equation}
h_{BFKL}^{(r,r+1)}=\psi (\hat{m}_{r,r+1})+\psi
(1-\hat{m}_{r,r+1})-2\psi (1)
\,,\,\,\hat{m}_{r,r+1}(\hat{m}_{r,r+1}-1)=-\rho _{r,r+1}^2\partial
_r\partial _{r+1}
\end{equation} and $h_{BFKL}^n$ is the local
hamiltonian for the integrable XXX model \cite{Heis} with the
spins coinciding with the generators of the M\"{o}bius group
(\ref{Moebgen}). Really we have two independent spin chains for
holomorphic and anti-holomorphic subspaces. The term $\sim l^2$ in
eq. (\ref{strHn}) describes an additional interaction between
these two spin chains because according to (\ref{Moebgen})
\begin{equation}
(\vec{p}_r)^2= -4\,M_r^{-}M^{-\,*}_r\,.
\end{equation}
This term violates the M\"{o}bius symmetry for $H$ and leaves only
its invariance under translations and rotations. Therefore the
eigenvalues of $H$ can depend on $\vec{q}^2$, which leads to the
Regge trajectories for composite states of reggeized
gluons. We do not know, if the corresponding Heisenberg spin model
is integrable or not. But in the region $\alpha '\vec{q}^2 \ll
g^2N_c$ it is possible to apply the integrability of the QCD hamiltonian 
for calculating the Regge trajectories. Indeed, as in the previous 
section, one can divide the essential momenta in two regions
$\vec{k}^2_r\sim \vec{q}^2_r$ and $\vec{k}_r^2\sim 1/\alpha '$. In
the first region we can use the integrability of the BFKL
hamiltonian to obtain the wave function of the composite state.
For the leading singularity the integrals of motion are quantised
and depend only on the conformal weights $m,\widetilde{m}$
\cite{integr,Heis,sol}. Therefore the corresponding energy
$E_{BFKL}$ for this leading singularity is a function of these
variables
\begin{equation}
E_{BFKL}=E(m,\widetilde{m})\,.
\end{equation}
It means, that for the solution of the equation in the second
region $\vec{k}^2\sim 1/\alpha '$ we can use the same methods
which were used in the previous section for the calculation of the
Pomeron trajectory. We hope to return to the problem of finding
the Regge trajectories for the Odderon and other gluon composite
states in our future publications.

\vspace*{2cm}

\centerline{\bf\Large Acknowledgments}
\vspace*{0.5cm}
The work is partially supported by the Russian State Grant Scientific School
RSGSS--1124.2003.2, the RFBR Grant 04-02-17094 and the Marie Curie Grant. 
We thank J. Bartels, V. Fadin, V. Kudryavtsev, A. Sabio Vera and other participants 
of the PNPI Winter School for helpful discussions. 

\appendix

\setcounter{equation}{0}

\setcounter{equation}{0}

\section{Conformal factor for  SL(2)-SUSY transformations}

If the fixed variables are $(z_{1}^{(0)}|\vartheta_{1}^{(0)})$, $
(z_{2}^{(0)}|\vartheta _{2}^{(0)})$ and $z_{3}^{(0)}$ while the
superpartner $\vartheta $ of $z_{3}^{(0)}$ is not fixed, then the
discussed factor $ H(z_{1}^{(0)},z_{2}^{(0)},z_{3}^{(0)},\vartheta
_{1}^{(0)},\vartheta _{2}^{(0)},\vartheta )$ turns out to be \cite{danphl}
\begin{equation}
H(z_{1}^{(0)},z_{2}^{(0)},z_{3}^{(0)},\vartheta _{1}^{(0)},\vartheta
_{2}^{(0)},\vartheta )=(z_{1}^{(0)}-z_{3}^{(0)})(z_{2}^{(0)} -z_{3}^{(0)})
\left[ 1-\frac{\vartheta _{1}^{(0)}\vartheta }{2(z_{1}^{(0)}-z_{3}^{(0)})}-
\frac{\vartheta _{2}^{(0)}\vartheta }{2(z_{2}^{(0)}-z_{3}^{(0)})}\right] \,.
\label{factorg}
\end{equation}
When $\vartheta _{1}^{(0)}=\vartheta _{2}^{(0)}=0$ this factor is reduced to
the expression given in Sec.3 of the paper. Refixing the above variables to
the new values $(\hat{z}_{1}^{(0)}, \hat{z}_{2}^{(0)},\hat{z}_{3}^{(0)},
\hat{\vartheta}_{1}^{(0)}, \hat{\vartheta}_{2}^{(0)})$ can be achieved by
the following transformations.

Firstly, both $\vartheta _{1}^{(0)}$ and $\vartheta _{2}^{(0)}$
are pushed to vanishing values. The supersymmetric $SL(2)$
transformation (\ref{sfltr}), which preserves the variables
$z_{1}$, $z_{2}$ and $z_{3}$ but adjustes to $\vartheta _{1}$ and
$\vartheta _{2}$ the zero values, is given by
\begin{equation}
f(\hat{z})=\hat{z}-\frac{(\hat{z}-z_{1})(\hat{z}-z_{2})}{
(z_{3}-z_{1})(z_{3}-z_{2})}\hat{\vartheta}_{3}\varepsilon_{0}(z_{3})\,,
\quad \varepsilon (\hat{z})=\frac{\vartheta _{1}(\hat{z} -z_{2}) }{
(z_{1}-z_{2})\sqrt{f^{\prime }(z_{1})}} -\frac{\vartheta _{2}(\hat{z }
-z_{1}) }{(z_{1}-z_{2})\sqrt{f^{\prime }(z_{2})}}\,,  \label{trnsf}
\end{equation}
where $\varepsilon _{0}(z)=[\vartheta _{1}(z-z_{2})
-\vartheta_{2}(z-z_{1})]/(z_{1}-z_{2})$. Evidently, we have
$f^{\prime }(z_{1})f^{\prime }(z_{2})=1$. Secondly, by the usual
$L(2)$ transformation one changes
$(z_{1}^{(0)},z_{2}^{(0)},z_{3}^{(0)})$ to new values $(\hat{z}
_{1}^{(0)},\hat{ z }_{2}^{(0)},\hat{z}_{3}^{(0)})$. Finally, using
the change of variables inversed to transformation (\ref{trnsf})
with preserving the values
$(\hat{z}_{1}^{(0)},\hat{z}_{2}^{(0)},\hat{z} _{3}^{(0)})$ one can
give the new values $(\hat{z} _{1}^{(0)},\hat{z}_{2}^{(0)})$ to
the vanishing superpartners of the bosonic coordinates
$(\hat{\vartheta} _{1}^{(0)},\hat{ \vartheta}_{2}^{(0)})$. To
verify that with the factor (\ref {factorg}) the amplitude is
independent of the values $ (z_{1}^{(0)},z_{2}^{(0)},z_{3}^{(0)},
\vartheta_{1}^{(0)},\vartheta_{2}^{(0)})$ of the fixed world-sheet
variables, one should take into account that under the $\Gamma $
-transformation (\ref{sfltr}) the integrand being $SL(2)$
covariant, receives the factor $Q_{\Gamma
}(\hat{z},\hat{\vartheta})$ for each world sheet variable
$(z|\vartheta )$, see eqs. (\ref{supfa}) and (\ref{trsder}). The
above factor is cancelled by the factor $1/Q_{\Gamma
}(\hat{z},\hat{ \vartheta})$ from the corresponding transformation
jacobian for all variables $(z|\vartheta )$ except the fixed ones
together with the superpartner $\vartheta $ of $z_{3}^{(0)}$,
because the last jacobian is different from $1/Q_{\Gamma
}(\hat{z}_{3}^{(0s)},\hat{\vartheta})$. One can verify, that these
additional extra-factors are just compensated by the corresponding
change of factor (\ref{factorg}). One can also check that the
amplitude is not changed when another set of variables is fixed.

\section{One-loop Regge trajectory for the critical superstring}

\setcounter{equation}{0}

The integral for the one-loop amplitude, corresponding to the sum of the
planar and non-oriented diagrams for the gluon-gluon sccattering
\begin{eqnarray}
A_{pl,no}=8K\int_{-1}^1\frac{d\lambda}{\lambda}\,\int _0^1 \left(\prod
_{I=1}^3\theta (\nu _{I+1}-\nu _I)d\nu _I\right) \,R\,,  \label{planbox}
\end{eqnarray}
is convergent at $\lambda =0$ \cite{gs}. In the above expression the
integrand is
\begin{eqnarray}
R= \left(\frac{B(\nu _1-\nu _2, \lambda ) B(\nu _3-1, \lambda )}{B(\nu
_1-\nu _3, \lambda ) B(\nu _2-1, \lambda )}\right) ^{-\alpha^{\prime}s}
\left(\frac{B(\nu _1-1, \lambda ) B(\nu _2-\nu _3, \lambda )}{B(\nu
_1-\nu_3, \lambda ) B(\nu _2-1, \lambda )}\right)^{-\alpha ^{\prime}t}
\label{integrand}
\end{eqnarray}
and the function $B$ is given below
\begin{equation}
B(\nu ,\lambda)=\sin \pi \nu \,\prod _{n=1}^{\infty} \frac{1-2\lambda ^n
\cos 2\pi \nu +\lambda ^{2n}}{(1-\lambda ^n)^2}.
\end{equation}
The factor $K$ includes the colour matrices $T$ and the products
of polarization vectors. In the Regge limit $-s \gg -t$ it equals
(cf.(\ref{scam}))
\begin{equation}
K= \pi^3g^4N\,T\,(\alpha^{\prime}s)^2
(\xi_a\xi_{a^{\prime}})(\xi_b\xi_{b^{\prime}})  \label{Kfact}
\end{equation}
where $N=32$ is the dimension of the $SO(32)$ group. In the same limit the
region $\nu_{32}=\nu_3-\nu_2 \sim 1/(\alpha^{\prime}s) \ll 1$ is essential
and we have the following simplifications
\begin{equation}
\frac{B(\nu _{12}, \lambda ) B(\nu _3-1, \lambda )}{B(\nu _{13}, \lambda )
B(\nu _2-1, \lambda )} \approx 1-\frac{ \sin \pi \nu _1 \, \sin \pi \nu
_{32} }{\sin \pi \nu _2 \,\sin \pi \nu _{31}}-4\pi \nu _{32}\,l_1\,,
\end{equation}
\begin{equation}
\frac{B(\nu_1-1, \lambda ) B(\nu_{23}-1, \lambda )}{B(\nu _{13}, \lambda )
B(\nu _2-1, \lambda )} \approx \frac{ \sin \pi \nu _1 \,
\sin \pi \nu _{32} }{
\sin \pi \nu _2 \,\sin \pi \nu _{31}}\,l_2\,,
\end{equation}
where
\begin{equation}
l_1=\sum _{n=1}^{\infty}\left( \frac{\lambda ^n \sin 2\pi \nu _{21}}{
1-2\lambda ^n \cos 2\pi \nu _{21}+\lambda ^{2n}} -\frac{\lambda ^n \sin 2\pi
\nu _{2}}{1-2\lambda ^n \cos 2\pi \nu _{2}+\lambda ^{2n}} \right)\,,
\end{equation}
\begin{equation}
l_2=\prod _{n=1}^{\infty} \frac{(1-2\lambda ^n \cos 2\pi \nu _1+\lambda
^{2n})(1-\lambda ^n)^2}{ (1-2\lambda ^n \cos 2\pi \nu _{31}+\lambda ^{2n})
(1-2\lambda ^n \cos 2\pi \nu _{2}+\lambda ^{2n})}\,.
\end{equation}
Instead of $\nu _1$ it is convenient to introduce the new integration
variable
\begin{equation}
y= \frac{ \sin \pi \nu _1 \,\sin \pi \nu _{32} }{\sin \pi \nu _2 \,\sin \pi
\nu _{31}} \,,\,\,x=1-y= \frac{ \sin \pi \nu _3 \,\sin \pi \nu _{21} }{\sin
\pi \nu _2 \,\sin \pi \nu _{31}}
\end{equation}
with the inverse transformation
\begin{equation}
\tan \pi \nu _1=\frac{(1-x)\, \sin \pi \nu_2 \sin \pi \nu _3 }{ \cos \pi \nu
_2 \sin \pi \nu _3 -x \cos \pi \nu _3 \sin \pi \nu _2}\,.
\end{equation}
Then the integral can be written as follows
\begin{eqnarray}
A=8K\int_{-1}^1\frac{d\lambda}{\lambda}\,\int _0^1 \frac{dx}{\pi} \int_{\nu
_1}^1d\nu _2 \int _{\nu _2}^1d\nu _3\, \sin \pi \nu _2 \,\sin \pi \nu _3
\,\sin \pi \nu _{23}  \nonumber \\
\times \frac{((1-x)l_2)^{-\alpha ^{\prime}t}\, (x -4\pi \nu
_{32}l_1)^{-\alpha ^{\prime}s}}{(\sin\pi \nu _3-x\sin \pi \nu _2)^2+
4x\,\sin \pi \nu_2 \,\sin \pi \nu _3 \,\sin^2 \frac{\pi \nu_{32}}{2}}\,.
\label{soam}
\end{eqnarray}
In the Regge limit the essential region of integration over $\nu_{32}$ is
\begin{equation}
1-x=y \sim (\alpha ^{\prime}s)^{-1}\ll \nu _{32}\ll 1\,,\,\, \nu _1 \sim
\frac{y}{\nu _{32}} \ll 1  \label{essen}
\end{equation}
where the integral is simplified as follows
\begin{eqnarray}
A=8K\int_{-1}^1\frac{d\lambda}{\lambda}\, \frac{\Gamma (1-\alpha^{\prime}t)
}{\pi ^2} \,\int _0^1 d \nu _2 \frac{(\sin \pi \nu_2)^2}{\pi}\,\ln
(-1/\alpha ^{\prime}s) (-\alpha ^{\prime}s)^{-1+\alpha^{\prime}t}  \nonumber
\\
\times \left(\frac{L_2}{1+L_1} \right)^{-\alpha^{\prime}t}\,(1+L_1)^{-1}.
\label{soamcon}
\end{eqnarray}
Here both
\[
L_1=4 (\sin \pi \nu_2)^2\frac{\partial l_1}{\partial (\pi\nu _1)}_{|\nu
_1=\nu _{32}=0}
\]
and $L_2$ are given explicitly by (\ref{definl}). As the result we obtain
for the Regge trajectory eq.(\ref{omega1}) in the text.

\section{ Multi-Regge production amplitudes}

\setcounter{equation}{0}

In integral (\ref{mainin}) we redefine $z_{7}\rightarrow
z_{7}z_{4}z_{6}/(z_{4}-z_{6})$ and introduce $f=z_{4}-z_{6}$
instead of $ z_{4}$. In addition, we replace $z_{6}\rightarrow
f\,z_{6}$. Then the integral $I(t_{5678},\kappa
^{2},t_{3478},t_{34},t_{56},t_{12},t_{78})$ in ( \ref{samplitu})
is given by expression
\begin{eqnarray}
I(t_{5678},\kappa ^{2},t_{3478},t_{34},t_{56},t_{12},t_{78}) =(\kappa
^{2})^{-\alpha ^{\prime }t_{34}-\alpha ^{\prime }t_{56}-2}\int
dz_{7}\,dz_{6}\,df\,e^{-f-z_{7}}f^{-\alpha ^{\prime }t_{12}-2}  \nonumber \\
\times z_{7}^{-\alpha ^{\prime }t_{78}}z_{6}^{-\alpha ^{\prime
}t_{5678}}(1+z_{6})^{-\alpha ^{\prime }t_{3478}}\hat{q}_{1}^{\alpha ^{\prime
}t_{34}}\hat{q}_{2}^{\alpha ^{\prime }t_{56}}  \nonumber \\
\times \Biggl[\frac{(\alpha ^{\prime }t_{78}+1)\hat{q}_{1}\hat{q}_{2}}{
z_{6}^{2}(1+z_{6})^{2}z_{7}^{2}}+\frac{(\alpha ^{\prime }t_{5678}-\alpha
^{\prime }t_{56}-\alpha ^{\prime }t_{78}-\kappa ^{2})\hat{q}_{1}}{
(1+z_{6})z_{6}^{2}z_{7}}+\frac{\alpha ^{\prime }t_{56}\hat{q}_{1}}{z_{6}^{2}
\hat{q}_{2}}  \nonumber \\
+\frac{(\alpha ^{\prime }t_{3478}-\alpha ^{\prime }t_{34}-\alpha ^{\prime
}t_{78}+\kappa ^{2})\hat{q}_{2}}{(1+z_{6})^{2}z_{6}z_{7}}+\frac{\alpha
^{\prime }t_{34}\hat{q}_{2}}{(1+z_{6})^{2}\hat{q}_{1}}+\frac{\alpha ^{\prime
}t_{35}+\alpha ^{\prime }t_{36}+\alpha ^{\prime }t_{45}+\alpha ^{\prime
}t_{46}}{z_{6}(1+z_{6})}\Biggl],  \label{sint}
\end{eqnarray}
where
\begin{equation}
\hat{q}_{1}=f(1+z_{6})+z_{7}z_{6}-\kappa ^{2}-i\epsilon \,,\quad \hat{q}
_{2}=fz_{6}+z_{7}(1+z_{6})-\kappa ^{2}-i\epsilon \,,\quad \epsilon
\rightarrow 0\,.  \label{qqq}
\end{equation}
Integrating it by parts, one obtain the following result
\begin{eqnarray}
I(t_{5678},\kappa
^{2},t_{3478},t_{34},t_{56},t_{12},t_{78})=(\kappa ^{2})^{-\alpha
^{\prime }t_{34}-\alpha ^{\prime }t_{56}-2}\int
dz_{7}dz_{6}dfe^{-f-z_{7}}f^{-\alpha ^{\prime }t_{12}-1}
\nonumber \\ \times z_{7}^{-\alpha ^{\prime
}t_{78}-1}(1+z_{6})^{-\alpha ^{\prime }t_{3478}}z_{6}^{-\alpha
^{\prime }t_{5678}}\hat{q}_{1}^{\alpha ^{\prime
}t_{34}}\hat{q}_{2}^{\alpha ^{\prime }t_{56}}\Biggl[\frac{(\alpha
^{\prime }t_{5678}+1)}{z_{6}^{2}}+\frac{\alpha ^{\prime
}t_{3478}+1}{(1+z_{6})^{2}} \nonumber \\ -\frac{\kappa
^{2}}{z_{6}(1+z_{6})}\Biggl(\frac{\alpha ^{\prime }t_{34}}{
\hat{q}_{1}}+\frac{\alpha ^{\prime
}t_{56}}{\hat{q}_{2}}\Biggl)-\frac{ f+z_{7}
}{z_{6}(1+z_{6})}\Biggl]\,.  \label{sintu}
\end{eqnarray}
To derive eq. (\ref{sintu}), one integrates the first term in the brackets
in eq. (\ref{sint}) over $z_{7}$ by parts. As a result, we obtain the
expression similar to eq. (\ref{sintu}) but the terms inside the brackets
turn out to be
\begin{eqnarray}
-\frac{\hat{q}_{1}\hat{q}_{2}}{z_{6}^{2}(1+z_{6})^{2}z_{7}}+\frac{(\alpha
^{\prime }t_{5678}+1-\alpha ^{\prime }t_{78}-\kappa ^{2})\hat{q}_{1}}{
(1+z_{6})z_{6}^{2}z_{7}}+\frac{\alpha ^{\prime }t_{56}\hat{q}_{1}}{z_{6}^{2}
\hat{q}_{2}}  \nonumber \\
+\frac{(\alpha ^{\prime }t_{3478}+1-\alpha ^{\prime }t_{78}+\kappa ^{2})
\hat{q}_{2}}{(1+z_{6})^{2}z_{6}z_{7}}+\frac{\alpha ^{\prime }t_{34}q_{2}}{
(1+z_{6})^{2}\hat{q}_{1}}+\frac{\alpha ^{\prime }t_{35}+\alpha ^{\prime
}t_{36}+\alpha ^{\prime }t_{45}+\alpha ^{\prime }t_{46}}{z_{6}(1+z_{6})}\,.
\label{de1r}
\end{eqnarray}
This expression is the same as
\begin{eqnarray}
\frac{f}{z_{7}}\Biggl(\frac{\alpha ^{\prime }t_{5678}+1-\kappa
^{2}-\alpha ^{\prime }t_{78}}{z_{6}^{2}}+\frac{\alpha ^{\prime
}t_{3478}+1+\kappa ^{2}-\alpha ^{\prime
}t_{78}}{(1+z_{6})^{2}}\Biggl)  \nonumber \\ +\frac{\alpha
^{\prime }t_{12}-\alpha ^{\prime }t_{78}+2}{z_{6}(1+z_{6})}-
\frac{\kappa ^{2}}{z_{7}}\Biggl(\frac{\alpha ^{\prime
}t_{5678}+1}{ z_{6}^{2}(1+z_{6})}+\frac{\alpha ^{\prime
}t_{3478}+1}{z_{6}(1+z_{6})^{2}}
\Biggl)-\frac{\hat{q}_{1}\hat{q}_{2}}{z_{6}^{2}(1+z_{6})^{2}z_{7}}
\nonumber
\\
+\frac{\alpha ^{\prime }t_{78}\kappa ^{2}}{z_{7}}\Biggl(\frac{1}{
z_{6}^{2}(1+z_{6})}+\frac{1}{z_{6}(1+z_{6})^{2}}\Biggl)-\frac{(-\kappa
^{2})^{2}}{z_{6}^{2}(1+z_{6})^{2}z_{7}}+\frac{\alpha ^{\prime }t_{56}\hat{q}
_{1}}{z_{6}^{2}\hat{q}_{2}}+\frac{\alpha ^{\prime }t_{34}\hat{q}_{2}}{
(1+z_{6})^{2}\hat{q}_{1}}\,.  \label{de2r}
\end{eqnarray}

Further, the terms proportional to $t_{78}$ are integrated by
parts over $ z_{7}$ to remove this factor $t_{78}$. Analogously
the term $\sim t_{12}$ is integrated by parts over $f$ to remove
the factor $t_{12}$. The third term is integrated by parts over
$z_{6}$ to remove both nominators $(\alpha ^{\prime }t_{5678}+1)$
and $(\alpha ^{\prime }t_{3478}+1)$ in the corresponding
contributions. After these transformations we obtain (\ref
{sintu}). If we shall integrate by parts the first term in eq.
(\ref{de2r}) over $z_{6}$ is possible to reduce eq. (\ref{sintu})
to the expression
\begin{eqnarray}
I(t_{5678},\kappa ^{2},t_{3478},t_{34},t_{56},t_{12},t_{78})=(\kappa
^{2})^{-\alpha ^{\prime }t_{34}-\alpha ^{\prime }t_{56}-2}\int
dz_{7}dz_{6}dfe^{-f-z_{7}}f^{-\alpha ^{\prime }t_{12}-1}  \nonumber \\
\times z_{7}^{-\alpha ^{\prime }t_{78}-1}(1+z_{6})^{-\alpha ^{\prime
}t_{3478}}z_{6}^{-\alpha ^{\prime }t_{5678}-1}\hat{q}_{1}^{\alpha ^{\prime
}t_{34}}\hat{q}_{2}^{\alpha ^{\prime }t_{56}}\Biggl[-\frac{\alpha ^{\prime
}t_{3478}}{(1+z_{6})^{2}}+\frac{\alpha ^{\prime }t_{34}+\alpha ^{\prime
}t_{56}}{(1+z_{6})}+\frac{z_{7}\alpha ^{\prime }t_{34}}{(1+z_{6})\hat{q}_{1}}
\nonumber \\
+\frac{f\alpha ^{\prime }t_{56}}{(1+z_{6})\hat{q}_{2}}+\frac{z_{6}}{
(1+z_{6})^{2}}-\frac{f+z_{7}}{(1+z_{6})}\Biggl]\,.  \label{situ}
\end{eqnarray}
One can introduce the variable $z$ instead of $z_{6}$ according to the
relation
\begin{equation}
z_{6}=z/(1-z)\,,  \label{rede}
\end{equation}
and redenote $z_{7}=y$. After it (\ref{situ}) can be presented as follows
\begin{eqnarray}
I(t_{5678},\kappa
^{2},t_{3478},t_{34},t_{56},t_{12},t_{78})=(\kappa ^{2})^{-\alpha
^{\prime }t_{34}-\alpha ^{\prime }t_{56}-2}\int
dydzdfe^{-f-y}f^{-\alpha ^{\prime }t_{12}-1}y^{-\alpha ^{\prime
}t_{78}-1}\times  \nonumber \\ \times z^{-\alpha ^{\prime
}t_{5678}-1}(1-z)^{-\alpha ^{\prime }t_{3456}+\alpha ^{\prime
}t_{12}+\alpha ^{\prime }t_{78}}(f+yz-\kappa ^{2}(1-z)-i\epsilon
)^{\alpha ^{\prime }t_{34}}  \nonumber \\ \times (y+fz-\kappa
^{2}(1-z)-i\epsilon )^{\alpha ^{\prime }t_{56}}\Biggl[ -\alpha
^{\prime }t_{3478}(1-z)+\alpha ^{\prime }t_{34}+\alpha ^{\prime
}t_{56}  \nonumber \\ +\frac{\alpha ^{\prime
}t_{34}y(1-z)}{f+yz-\kappa ^{2}(1-z)-i\epsilon }+ \frac{\alpha
^{\prime }t_{56}f(1-z)}{y+fz-\kappa ^{2}(1-z)-i\epsilon }
+z-(f+y)\Biggl]\,.  \label{sit1u}
\end{eqnarray}

\section{Vanishing of impact factors for planar diagrams}

\setcounter{equation}{0}

The impact factor for the vector particle scattering can be
calculated from the asymptotics of Fig1b in the region where
$s_1=\pm s_7\to\infty$ while $ s_3$, $s_4$, $s_5$ and $s_6 $ are
finite. The impact factors for the states with the masses
$\alpha^{\prime}t_{34}=n_1$ and $\alpha^{\prime}t_{56}=n_2$ are
just proportional to the resudies in the poles at $\alpha^{
\prime}t_{34}=n_1$ and $\alpha^{\prime}t_{56}=n_2$. One can see
from expression (\ref{sstart}) that for the discussed asymptotics
the essential values of the integration variables are
\begin{equation}
z_{3}\rightarrow 0,\quad z_{7}/z_{6}\rightarrow 0\,.  \label{con1f}
\end{equation}
while $x$ and $y$ being defined by the relations $z_{3}=z_{4}+x$
and $ z_{5}=z_{6}+y$ are now comparable in their values with $z_4$
and $z_6$. However, the poles $\alpha^{\prime}t_{34}=n_1$ and
$\alpha^{ \prime}t_{56}=n_2 $ appear from the regions $x/z_4\to0$
and $y/z_6\to0$. In this kinematics one can expand the integrand
in powers of $x$ and $y$ to obtain the poles at
$\alpha^{\prime}t_{34}=n_1$ and $\alpha^{ \prime}t_{56}=n_2$. It
can be verified that the corresponding integral vanishes, and, so,
the impact factor for the planar diagram is equal to zero.

For the sake of simplicity we give the corresponding proof for the
boson string theory, assuming, that the external interaction
states are tachyons. In this case only the leading term in $x$ and
$y$ is needed and expression (\ref{sstart}) can be simplified as
it was done in eq. (\ref{next}). Furthermore, similar to the
multi-Regge limit we obtain
\begin{equation}
k_{2}(k_{3}+k_{4})\rightarrow -k_{2}(k_{5}+k_{6})\rightarrow k_{1}k\,,\quad
k_{7}(k_{3}+k_{4})\rightarrow -k_{7}(k_{5}+k_{6})\rightarrow k_{8}k,\,
\label{limt1s}
\end{equation}
but relations (\ref{limts}) for $k_3(k_7+k_8)$ and for
$k_5(k_7+k_8)$ are not valid. It is helpfull to redefine again the
variables according to eq. (\ref{change}). After calculating
integrals over $x$ and $y$ the asymptotics of $A^{(0)}$ turns out
to be
\begin{eqnarray}
A^{(0)}=T_p(\pm\alpha^{\prime}s_1)^{\alpha^{\prime}t_{12}+\alpha^{
\prime}t_{78} -\alpha^{\prime}t_{34}-\alpha^{\prime}t_{56}}
(\pm\alpha^{\prime}s)^{\alpha^{\prime}t_{34}+\alpha^{\prime}t_{56}+2}I^{(0)}
\nonumber \\
\Gamma(-\alpha^{\prime}t_{34}-1)\Gamma(-\alpha^{\prime}t_{56}-1)
\label{amp1l}
\end{eqnarray}
where $\Gamma(x)$ is the gamma function and
\begin{eqnarray}
I^{(0)}= \int \frac{dz_7\,dz}{z_7^2z}df \exp\biggl[-f-z_7\biggl]
z_7^{-\alpha^{\prime}t_{78}}
(1-z)^{-\alpha^{\prime}[t_{3456}-t_{12}-t_{78}]}
z^{-\alpha^{\prime}t_{5678}-1} f^{-\alpha^{\prime}t_{12}-2}
\nonumber \\ \times \biggl[f+z_7z- \frac{\pm
s_1}{s}\alpha^{\prime}k_3(k_7+k_8)(1-z)
\biggl]^{\alpha^{\prime}t_{34}+1} \biggl[z_7+fz- \frac{\pm s_1}{s}
\alpha^{\prime}k_5(k_7+k_8)(1-z)\biggl]^{\alpha^{\prime}t_{56}+1}
\,. \label{int1u}
\end{eqnarray}

In the calculation of $I^{(0)}$ we performed the change of the
integration variables as it was done in Appendix B:
$z_{7}\rightarrow z_{7}z_{4}z_{6}/(z_{4}-z_{6})$,
$z_{4}\rightarrow f=z_{4}-z_{6}$ and $ z_{6}\rightarrow
f\,z_{6}=f\,z/(1-z)$ (see eq.(\ref{rede})). The impact factor is
proportional to the sum over the residues of $I^{(0)}$ in the
poles $\alpha ^{\prime}t_{5678}=n$ for fixed values
$\alpha^{\prime}t_{34}= \alpha^{\prime}t_{56}=-1$. The parameter
$n=m-1$ takes integer values from $ n=-1$ up $n=\infty$. The
result contains the factor
\begin{equation}
\sum_{m=0}^\infty\frac{d^m}{m!dz^m} (1-z)^{-\alpha^{
\prime}[t_{3456}-t_{12}-t_{78}]}\,.  \label{su1m}
\end{equation}
The sum is calculated in the region $t_{3456}-t_{12}-t_{78}>0$
where the series is covergent. Then it is continued analytically
to physical values for $t_{3456}-t_{12}-t_{78}$. For $z
\rightarrow 1$ this sum is equal to $ (1-z)^{-\alpha^{
\prime}[t_{3456}-t_{12}-t_{78}]}=0$. Thus, the impact factor for
the planar diagram is zero. The vanishing of the impact factor for
the higher mass states $\alpha^{\prime}t_{34}=n_1$ and $
\alpha^{\prime}t_{56}=n_2$ is verified in a similar way. For the
superstring theory one can prove also the vanishing of the impact
factors for the planar diagrams.

\newpage

\centerline{Captions} \vspace*{0.5cm} \textbf{Fig. 1}. The ladder cut
determining the BFKL Pomeron in the $a+b\to a^{\prime}+b^{\prime}$ process.
The dotted line denotes a reggion, the solid one denotes a particle.

\vspace*{0.5cm}

\textbf{Fig. 2}. The diagram for the calculation of the BFKL kernel. The
dotted line denotes massless state; the solide one denotes the tower of
string states.

\newpage

\epsfbox{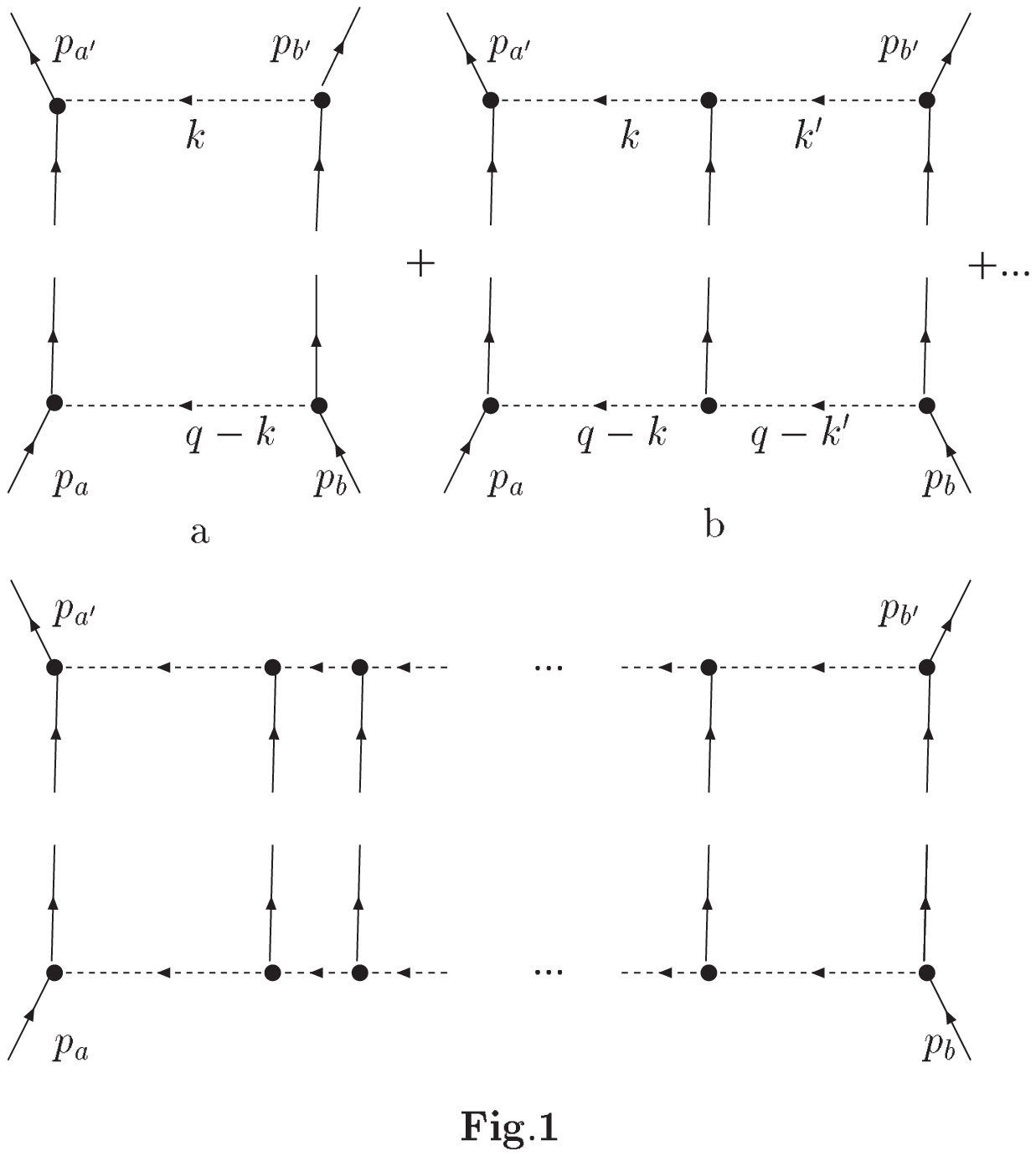}

\newpage \epsfbox{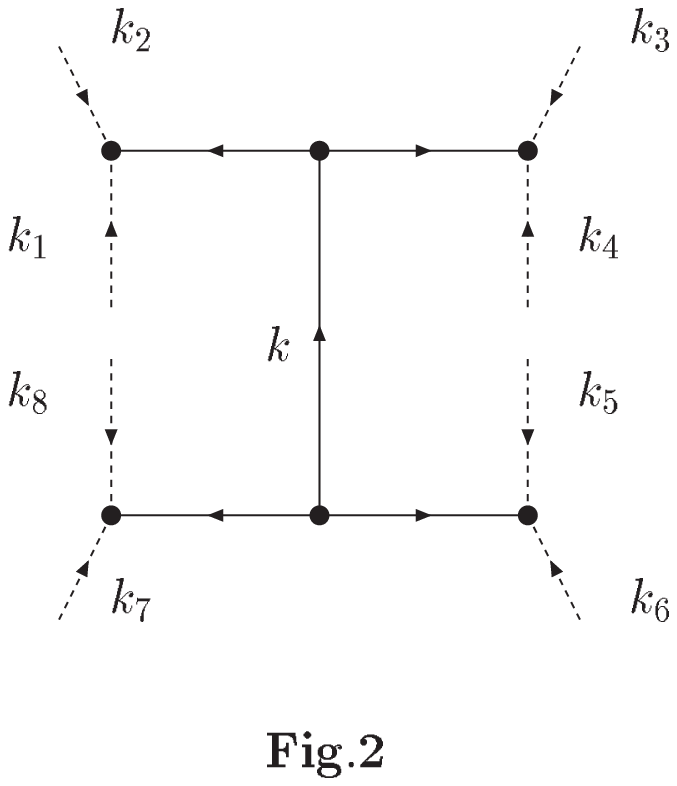}

\end{document}